\documentclass{jfm}
\usepackage{graphicx}
\usepackage{epstopdf, epsfig}
\usepackage{amsmath}
\usepackage{enumitem}
\usepackage{natbib}
\usepackage{mathtools}
\usepackage{amsfonts}
\usepackage{tikz}
\usepackage[english]{babel}
\usepackage{enumitem}
\usepackage{setspace}
\usepackage{verbatim}
\usepackage{caption}
\usepackage{subcaption}
\usepackage{wrapfig}
\usepackage{nextpage}
\usepackage[titletoc]{appendix}
\usepackage{titlesec}
\usepackage{standalone}
\usepackage{bm}
\usepackage{bibentry}
\usepackage{pdflscape}
\usepackage{rotating}
\usepackage{xcolor}
\usepackage{soul}
\usepackage{titlesec}
\usepackage{floatrow}
\usepackage{sidecap}
\usepackage{color}
\usepackage{tikz}
\usepackage{rotating}
\usetikzlibrary{shapes.misc}
\usepackage[colorlinks=true, linkcolor=blue, urlcolor=blue, citecolor=blue]{hyperref}

\allowdisplaybreaks
\tikzset{cross/.style={cross out, draw=black, minimum size=2*(#1-\pgflinewidth), inner sep=0pt, outer sep=0pt}, cross/.default={1pt}}

\shorttitle{A resolvent-based perspective on Mach waves from compressible BLs}
\shortauthor{A. Madhusudanan, G. Stroot, B. J. McKeon}

\title{A resolvent-based perspective on the generation of Mach wave radiation from compressible boundary layers}

\author{Anagha Madhusudanan\aff{1}
  \corresp{\email{anaghamadhu91@gmail.com}}
  \aunote{Current Affiliation: Dept. of Aero. Engg., Indian Institute of Science, Bangalore, India},
  Gregory Stroot\aff{2}
  \and Beverley. J. McKeon\aff{1}
    \aunote{Current Affiliation: Dept. of Mech. Eng., Stanford University, U.S.A.}}

\affiliation{\aff{1}Graduate Aerospace Laboratories, California Institute of Technology, Pasadena, CA 91125, USA
\aff{2}Department of Mechanical and Civil Engineering, California Institute of Technology, Pasadena, CA 91125, USA}

\begin{document}

\maketitle

\begin{abstract}

We identify forcing mechanisms that separately amplify subsonic and supersonic features obtained from a linearized Navier--Stokes based model for compressible parallel boundary layers. 
Resolvent analysis is used to analyse the linear model, where the non-linear terms of the linearized equations act as a forcing to the linear terms. 
Considering subsonic modes, only the solenoidal component of the forcing to the momentum equations amplify these modes. 
When considering supersonic modes, we find that these are pressure fluctuations that radiate into the freestream. 
Within the freestream, these modes closely follow the trends of inviscid Mach waves. 
There are two distinct forcing mechanisms that amplify the supersonic modes: (i) the `direct route' where the forcing to the continuity and energy equations and the dilatational component of the forcing to the momentum equations directly force the mode and (ii) the `indirect route' where the solenoidal component of the forcing to the momentum equations force a response in wall-normal velocity, and this wall-normal velocity in-turn forces the supersonic mode. 
A majority of the supersonic modes considered are dominantly forced by the direct route. 
However, when considering Mach waves that are, like in direct numerical simulations, forced from the buffer layer of the flow, the indirect route of forcing becomes significant. 
We find that these observations are also valid for a streamwise developing boundary layer.
These results are consistent with, and extend the observations in the literature regarding the solenoidal and dilatational components of velocity in compressible turbulent wall-bounded flows.

\end{abstract}

\section{Introduction}
\label{sec:introduction}

We can broadly categorise the flow features in compressible wall-bounded flows into two different kinds: (i) subsonic features that have equivalents in incompressible flows and (ii) supersonic features that have no such equivalents. 
First consider the subsonic features. 
Studies have confirmed the presence of the ubiquitous streaky structures of incompressible flows in compressible flows as well; structures such as the near-wall streaks in the buffer layer of the flow \citep{kline1967structure} and the large and very-large scale structures in the logarithmic regions of these flows (see \citet{smits2011high} and references therein). 
There is an ongoing debate on how the length scales of these structures change with increasing Mach number and wall-cooling \citep[e.g.][]{smits1989comparison, smits2006turbulent, pirozzoli2011turbulence, ganapathisubramani2006large, duan2010direct, duan2011direct, duan2011bdirect, williams2018experiments, bross2021large}. 

Now consider the second kind of flow features, i.e.\ the supersonic features that have no equivalents in incompressible flows. 
Here we consider the eddy Mach waves, which are freestream pressure fluctuations, and a majority of this manuscript will focus on these structures.  
These pressure fluctuations cause practical difficulties within wind tunnels used to measure the transition behaviour of test vehicles. 
The pressure disturbances are radiated from the boundary layers formed on the walls of the wind tunnel and thereafter impact transition measurements \citep[e.g.][]{laufer1964some, wagner1970influence, stainback1971hypersonic, pate1978dominance, schneider2001effects}. 
Studies (based on acoustic analogies) have inferred the location of the sources that radiate these Mach waves to be within the buffer layer of the boundary layer \citep{phillips1960generation, ffowcs1963noise, duan2014numerical}. 
With increasing Mach number, the intensity of these pressure radiations increases, and they also have larger propagation velocities and shallower orientation angles in the freestream \citep{laufer1964some, duan2016pressure}. 
Wall-cooling also impacts this radiation \citep{zhang2017effect}. 
Experimental measurements of these freestream radiations are notoriously challenging \citep[e.g.][]{laufer1961sound, stainback1971hypersonic, kendall1970supersonic, donaldson1995review}, and direct numerical simulations (DNS) that properly resolve these structures are expensive owing to the requirement of computational boxes with a large wall-normal extent \citep[e.g.][]{hu2006sound, duan2014numerical, duan2016pressure,  zhang2017effect}. 
It is therefore crucial to obtain models that faithfully represent these structures. 
Empirically obtained correlations such as the Pate's correlation \citep{pate1969radiated} are the typical methods by which the effects of these disturbances are currently modelled for practical purposes.

The literature described above focused on experiment-based and direct numerical simulation-based (DNS) investigations of compressible wall-bounded flows. 
The mathematical modelling of the flow is yet another method that has been employed to analyse these flows, and this will be the approach that will be pursued in the current manuscript.  
Models have been used to analyse the routes through which these flows transition to turbulence, and one such route is provided by the unstable eigenvalues that emerge from the compressible Navier--Stokes equations linearized around laminar mean profiles.  
We can categorise these unstable eigenvalues into two different kinds: (1) the first mode eigenvalues which have an equivalent in the incompressible regime and (2) the higher mode eigenvalues which do not have an equivalent in the incompressible regime \citep[e.g.][]{lees1946investigation, lees1962stability, mack1965computation, mack1975linear, mack1984boundary, malik1990numerical, ma2003receptivity, ozgen2008linear, fedorov2011high}. 
Apart from this analysis of eigenvalues, more recent studies have focused on the non-modal mechanisms that provide an additional route to transition. 
These non-modal mechanisms can be studied by either computing the optimal initial perturbations that leads to maximum transient growth \citep[e.g.][]{chang1991compressible, balakumar1992discrete, hanifi1996transient, tumin2001spatial, tumin2003optimal, zuccher2006parabolic, tempelmann2012spatial, bitter2014transient, bitter2015stability, paredes2016optimal, bugeat20193d, kamal2020application, kamal2021input} or, as will be pursued in this manuscript, by computing the optimum response of the linearized equations to a forcing \citep[e.g.][]{cook2018understanding, dwivedi2018input, dawson2019studying, dwivedi2019reattachment, bugeat20193d}. 
These studies showed that the lift-up mechanism that is responsible for amplifying the ubiquitous streaky structures in incompressible wall-bounded flows, also amplify streaky structures in the compressible counterparts of these flows \citep[e.g.][]{balakumar1992discrete, hanifi1996transient, tumin2001spatial, tumin2003optimal, zuccher2006parabolic, tempelmann2012spatial, bitter2014transient, bitter2015stability, paredes2016optimal, bugeat20193d}.  
(See \citet{fedorov2011transition} and references therein for a review regarding the transition of laminar compressible flows).

Two more recent studies that consider the modelling of these flows, and that are particularly relevant to the current work, are the studies by \citet{bugeat20193d} and \citet{bae2020resolvent} that considered laminar and turbulent compressible boundary layers, respectively. 
These studies used the resolvent analysis framework, where non-linear terms of the linearized Navier--Stokes equations are considered to be a forcing to the linear equations \citep[e.g][]{mckeon2010critical, hwang2010linear, moarref2013model, morra2021colour, towne2020resolvent, zare2017colour}. 
For compressible boundary layers, \citet{bugeat20193d} and \citet{bae2020resolvent} identified two different kinds of modes that are amplified by the resolvent operator. 
The first among these are the subsonic modes, identified as the streaks and the first modes in \citet{bugeat20193d}, and these modes have equivalents that have been studied in the incompressible regime \citep[e.g.][]{mack1984boundary, mckeon2010critical, hwang2010linear, sharma2013coherent, moarref2014low}. 
In turbulent boundary layers, \citet{bae2020resolvent} found that these subsonic modes can be scaled using the semi-local scaling of compressible flows \citep{trettel2016mean} such that they follow the trends of the incompressible modes well. 
The trends of these modes can therefore be predicted using tools developed for the incompressible regime \citep{dawson2020prediction}. 
The second among the two sets of identified modes are the supersonic modes \citep{bugeat20193d, bae2020resolvent}. 
Crucially, these modes are related to the higher order Mack modes \citep{bugeat20193d, bae2020resolvent} identified in the seminal work by \citet{mack1984boundary}, and we will explore this relationship further in the current study.  
Resolvent analysis predicts the increasing significance of these modes with increasing Mach number \citep{bae2020resolvent}, consistent with DNS \citep{duan2016pressure}. 
These trends of the subsonic and supersonic modes also hold more generally for the case of boundary layers over cooled walls with a range of wall-cooling ratios \citep{bae2020studying}. 

So far we have seen that there are both subsonic and supersonic features in compressible boundary layer flows, and that these two features are captured by the mathematical modelling technique of resolvent flow analysis where the non-linear terms act as a forcing \citep{bugeat20193d, bae2020resolvent}. 
Here, we therefore ask the following question: can we isolate the different forcing mechanisms that generate these subsonic and the supersonic modes? 
Recent studies have shown that understanding the forcing mechanisms, and thereby effectively modelling these mechanisms, is crucial for understanding and building practically useful linearized Navier--Stokes based models of flows \citep[e.g.][]{jovanovic2005componentwise, zare2017colour, morra2021colour, nogueira2021forcing, amaral2021resolvent, holford2023optimal}.
The technique of breaking the forcing into components and analysing the  parts has provided insights into various aspects of turbulent flows. 
For instance, this approach has explained the increased prevalence of channel-wide structures in Couette flows when compared to Poiseuille flows \citep{illingworth2019streamwise}; why the turbulent kinetic energy in wall-bounded flows peaks at a specific wall-normal location \citep{morra2021colour, nogueira2021forcing}; how a small fraction of the full forcing, which can be empirically modelled, generates the acoustic radiations in a jet \citep{karban2023empirical} etc.  
Here, to analyse the forcing, we will use resolvent analysis.  
Different from \citet{bugeat20193d} \& \citet{bae2020resolvent}, we concentrate on identifying the specific components of the forcing to the resolvent operator (i.e.\ the non-linear terms of the linearized equations) that are responsible for amplifying the subsonic and supersonic modes, separately. 
The Helmholtz decomposition of the forcing to the resolvent operator proves to be an instrumental tool for this purpose. 
The aim is to acquire a fundamental physical understanding of the different amplification mechanisms in the flow through identifying the mechanisms associated with individual forcing components. 
For a majority of the manuscript, we will consider the simple case of a boundary layer flow with a time-invariant mean flow that varies only in one inhomogeneous spatial (wall-normal) direction, which makes the mathematical analysis tractable and insightful. 
We also show that the conclusions drawn are applicable for the more general case of a boundary layer with a two-dimensional mean flow that varies in two inhomogeneous spatial (streamwise and wall-normal) directions. 

Here, we find that, the subsonic modes in the compressible flow are forced by the solenoidal component of the forcing alone. 
The Mach waves on the other hand have two routes through which they can be amplified: (i) the direct route where the dilatational component of the forcing to the momentum equations and the forcing to the density and temperature equations are active and (ii) the indirect route where the solenoidal forcing excites the Mach waves. 
In other words, when focusing on the indirect route, the dilatational response from a solenoidal forcing is considered. 
We find that, while the direct route is the dominant mechanism for amplifying the Mach waves, the indirect route plays a significant role for the Mach waves that are forced by the buffer layer of the flow.
While there are alternate insightful analysis techniques that have been used to probe the different mechanisms in these flows, such as for instance the examination of the interaction of vortical and acoustic mechanisms through DNS and linear stability theory in \citet{unnikrishnan2019interactions}, to the best of our knowledge, these techniques do not explain the two distinct routes of forcing the Mach waves and the different regions of the flow where these mechanisms operate.

The organisation of the rest of this paper is as follows. 
We will start with a description of the resolvent analysis of the linearized Navier--Stokes equations and its numerical implementation in \S\ref{sec:methods}. 
Section \S\ref{sec:helmholtz decomposition} will then discuss Helmholtz decomposition, a technique that will be frequently employed in this manuscript. 
In \S\ref{sec:comparing compressible and incompressible resolvent operators} the subsonic modes will be considered, and the forcing mechanisms that amplify the subsonic modes will be the topic of \S\ref{sec:forcing to the subsonic modes: helmholtz decomposition of resolvent forcing}. 
In \S\ref{sec:Mach wave radiation from the resolvent operator} we will then shift our focus to the supersonic resolvent modes, i.e.\ the resolvent Mach waves. 
The freestream contribution of these modes and the effect of viscosity on these modes will be discussed in \S\ref{sec:contribution to the boundary layer and freestream} and \S\ref{sec:the effect of viscosity on resolvent mach wave amplification}, respectively. 
The two routes of amplifying the Mach waves will be discussed in \S\ref{sec:two routes of forcing resolvent Mach wave radiation} and the contribution of the two routes across a wide parameter regime will be considered in \S\ref{sec:contribution of the direct and indirect forcing mechanisms}. 
In \S\ref{sec:resolvent mach waves and mach waves from dns: a discussion}, we will discuss the resolvent Mach waves alongside trends of these waves that are known from DNS. 
Finally, in \S\ref{sec:mach wave radiation from a streamwise developing boundary layer}, we show that the conclusions drawn are also valid for a boundary layer flow with a two-dimensional mean profile, before concluding the manuscript in \S\ref{sec:Conclusions}. 
Although most of the study focuses on a Mach $4$ and friction Reynolds number $400$ turbulent boundary layer over an adiabatic wall, the discussions are more generally applicable to laminar compressible boundary layers (appendix \S\ref{sec:laminar boundary layer}) as well as turbulent boundary layers both over adiabatic and cooled walls, for a range of Mach numbers (\S\ref{sec:contribution to the boundary layer and freestream}). 

\section{Methods}
\label{sec:methods}

\subsection{Linear Model}
\label{sec:linear model}

We consider a compressible boundary layer with the streamwise, wall-normal and spanwise directions given by $x$, $y$ and $z$, respectively. 
Although the development of the boundary layer in the streamwise direction is an important parameter to consider \citep[e.g.][]{bertolotti1992linear, govindarajan1995stability, ma2003receptivity, ran2019stochastic, ruan2021direct}, as a first approximation here we invoke the parallel flow assumption where we assume that this streamwise development is slow, and therefore neglect its effects. 
We briefly consider the impact of the streamwise development on the discussions here in \S\ref{sec:mach wave radiation from a streamwise developing boundary layer} and hope to report on this in more detail in the near future \citep{stroot2022resolvent, stroot2023generation}. 
Under this parallel flow assumption, along with the spanwise, the streamwise is also a homogeneous direction, and the mean streamwise velocity $\overline{U}(y)$, temperature $\overline{\Theta}(y)$, density $\overline{\rho}(y)$ and pressure $\overline{P}(y)$ are functions of the wall-normal direction alone. 
Additionally, under this assumption the mean wall-normal $\overline{V}(y)$ and spanwise  $\overline{W}(y)$ velocities are zero. 
Fluctuations are defined with respect to these mean quantities, where $u$, $v$ and $w$ represent the velocity fluctuations in the streamwise, wall-normal and spanwise directions, respectively and $\rho$, $\theta$ and $p$ represent the density, temperature and pressure fluctuations, respectively. 
A subscript `$\infty$' denotes freestream quantities and a subscript `$w$' denotes quantities at the wall. 
The velocities are non-dimensionalised by $U_\infty$, the length scales by the boundary layer thickness $\delta$ and temperature by $\Theta_\infty$. 
A superscript `$+$' indicates normalization of velocities and length scales by the friction velocity $u_\tau$ and the friction length scale $\mu_w/u_\tau$, respectively. 
Here $\mu$ is the first coefficient of viscosity. 

The non-dimensional numbers that define the problem are: 
(1) the Reynolds number defined as $Re=\rho_\infty U_\infty \delta/\mu_\infty$, 
(2) the freestream Mach number defined as $Ma=U_\infty/(\gamma\mathcal{R}\Theta_\infty)^{1/2}$ where $\gamma$ is the specific heat ratio and $\mathcal{R}$ is the universal gas constant and 
(3) the Prandtl number $Pr=\mu_\infty c_p/\kappa_\infty$ defined using specific heat ratio $c_p$ and the thermal conductivity $\kappa$. 
A friction Reynolds number is also defined as $Re_\tau=\rho_w u_\tau \delta/\mu_w$. 
Throughout this study $Pr=0.72$ and $\gamma=1.4$ are kept fixed. 
Both flows over adiabatic walls as well as over cooled walls are considered, and boundary layers over cooled walls are characterised by the ratio $\Theta_w/\Theta_{ad}$ where $\Theta_{w}$ is the wall-temperature and $\Theta_{ad}$ is the wall-temperature in the case of the flow over adiabatic walls. 
For most of this work we consider a $Ma=4$, $Re_\tau=400$ turbulent boundary layer over an adiabatic wall. 
However, we will show that the substance of the discussion here is applicable to turbulent boundary layers over adiabatic as well as cooled walls and over a range of Mach numbers (in \S\ref{sec:contribution to the boundary layer and freestream}), as well as to laminar boundary layers (in appendix \S\ref{sec:laminar boundary layer}). 

We linearize the Navier--Stokes equation around the mean state $(\overline{U}(y), 0, 0, \overline{\rho}(y), \overline{\Theta}(y))$ and obtain the equations for the fluctuations as:
\begin{subequations}
\begin{align}
\begin{split}
\overline{\rho}\frac{\partial u_i}{\partial t} &= 
- \overline{\rho} \overline{U} \frac{\partial u_i}{\partial x} 
- \overline{\rho} \frac{d \overline{U}}{d y} v \widehat{i} 
-\frac{1}{\gamma Ma^2} \left[ \overline{\Theta} \frac{\partial \rho}{\partial x_i} 
+ \frac{d \overline{\Theta}}{d y} \rho \widehat{j} 
+ \overline{\rho} \frac{\partial \theta}{\partial x_i}
+  \frac{d \overline{\rho}}{d y} \theta \widehat{j} \right] \\
&+ \frac{1}{Re} \left[ \frac{\partial \overline{\mu}}{\partial \overline{\Theta}} \frac{d \overline{U}}{d y} \left( \frac{\partial \theta}{\partial y} \widehat{i} 
+ \frac{\partial \theta}{\partial x} \widehat{j} \right)
+ \frac{\partial^2 \overline{\mu}}{\partial \overline{\Theta}^2} \frac{d \overline{\Theta}}{d y} \frac{d \overline{U}}{d y} \theta \widehat{i} +
\frac{\partial \overline{\mu}}{\partial y} \left( \frac{\partial u_i}{\partial y}
+ \frac{\partial v}{\partial x_i} \right) \right. \\
&+ \frac{\partial \overline{\lambda}}{\partial y} \frac{\partial u_k}{\partial x_k} \widehat{j} 
+ \left. \frac{\partial \overline{\mu}}{\partial \overline{\Theta}} \frac{d^2 \overline{U}}{d y^2} \theta \widehat{i} + 
\overline{\mu} \frac{\partial^2 u_i}{\partial x_j\partial x_j} + 
(\overline{\mu}+\overline{\lambda}) \frac{\partial^2 u_j}{\partial x_j\partial x_i} \right] + f_{u_i},
\end{split}\\
\begin{split}
\overline{\Theta} \frac{\partial \rho}{\partial t} &= 
- \overline{\Theta} \overline{U} \frac{\partial \rho}{\partial x}
- \overline{\Theta} \frac{d \overline{\rho}}{d y} v
- \frac{\partial u_i}{\partial x_i} + f_\rho, 
\end{split}\\
\begin{split}
\overline{\rho}\frac{\partial \theta}{\partial t} &= 
- \overline{\rho} \overline{U} \frac{\partial \theta}{\partial x} 
- \overline{\rho} \frac{d \overline{\Theta}}{d y} v
- (\gamma-1) \frac{\partial u_j }{\partial x_j} \\
&+ \frac{\gamma}{PrRe} \left[ 2 \frac{\partial \overline{\mu}}{\partial y} \frac{\partial \theta}{\partial y} 
+ \frac{\partial^2 \overline{\mu}}{\partial \overline{\Theta}^2} \left( \frac{d \overline{\Theta}}{d y} \right)^2 \theta 
+ \overline{\mu} \frac{\partial^2 \theta}{\partial x_j \partial x_j} 
+ \frac{\partial \overline{\mu}}{\partial \overline{\Theta}} \frac{d^2 \overline{\Theta}}{d y^2} \theta \right] \\
&+ \frac{\gamma(\gamma-1)Ma^2}{Re} \left[ 2 \overline{\mu} \frac{d \overline{U}}{d y} \left( \frac{\partial u}{\partial y} 
+ \frac{\partial v}{\partial x} \right)
+ \frac{\partial \overline{\mu}}{\partial \overline{\Theta}} \left(\frac{d \overline{U}}{d y}\right)^2 \theta
\right] 
+ f_\theta.
\end{split}
\end{align}
\label{eqn:NS_comp}
\end{subequations}\\
Here all the non-linear terms of the equation are represented by $\bm{f}=(f_u,f_v,f_w,f_\rho,f_\theta)$ where $f_u$, $f_v$ and $f_w$ represent the non-linear terms in the momentum equations, and $f_\rho$ and $f_\theta$ represent the non-linear terms in the continuity and the energy equations, respectively. 
In \eqref{eqn:NS_comp} $(u_1,u_2,u_3)$ represents $(u,v,w)$ and $(x_1,x_2,x_3)$ represents $(x,y,z)$. 
(It should be noted that, for this linearization, we have not assumed the fluctuations to be small, and instead they can assume any arbitrary value). 
Unit vectors along $x$, $y$ and $z$ are $\widehat{i}$, $\widehat{j}$ and $\widehat{k}$, respectively. 
In addition to the equations in \eqref{eqn:NS_comp}, we also have the linearized equation of state $p=\overline{\rho} \theta + \overline{\Theta} \rho$. 
The equations are scaled such that the mean pressure $\overline{P}=1$, and therefore the mean density is related to the mean temperature as $\overline{\rho}=1/\overline{\Theta}$. 
The mean viscosity is obtained as a function of temperature using the Sutherland formula $\overline{\mu} = \overline{\Theta}^{3/2} (1+C)/(\overline{\Theta}+C)$ where $C = 110.4K/\Theta_\infty$.  
The second coefficient of viscosity is given as $\lambda=-2/3\mu$. 
The subsonic lift-up mechanism \citep{landahl1980note} and critical-layer mechanism \citep{mckeon2010critical}, as well as the supersonic Mach wave generation mechanism \citep{mack1984boundary}, are easily expressed in terms of the primitive variables $(u,v,w,\rho,\theta)$ used here. 
In the future, it would however be interesting to see how the discussions here are impacted with a different choice of variables \citep{karban2020ambiguity}. 

\subsection{Resolvent operator}
\label{sec:resolvent operator}

We use the linearized equations in \eqref{eqn:NS_comp} to derive the resolvent operator for the flow. 
For this, $\bm{u}$, $\rho$, $\theta$ and $\bm{f}$ are considered in terms of their Fourier transforms in the homogeneous streamwise and spanwise directions, as well as in time:
\begin{equation}
\begin{split}
l(x,y,z,t) &= \int_{-\infty}^{\infty} \int_{-\infty}^{\infty} \int_{-\infty}^{\infty} \widehat{l}(y;k_x,k_z,\omega) e^{(ik_xx+ik_zz+i\omega)}dk_x dk_z d\omega. 
\end{split}
\end{equation}
Here $l$ represents $\bm{u}$, $\rho$, $\theta$ or $\bm{f}$, and $\widehat{\cdot}$ represents their Fourier transforms. 
($k_x,k_z$) are the streamwise and spanwise wavenumbers, ($\lambda_x,\lambda_z$) are the corresponding wavelengths and $\omega$ is the temporal frequency. 
The wavenumbers are non-dimensionalized by $(1/\delta)$ and the wavelengths by $\delta$.  
The temporal frequency $\omega$ can be written in terms of a phase-speed $c$ as $\omega=-c k_x$. 
In terms of these Fourier transforms, \eqref{eqn:NS_comp} are written as: 
\begin{equation}
\begin{split}
i\omega\bm{\widehat{q}} &= \bm{A}(k_x,k_z) \widehat{\bm{q}} + \widehat{\bm{f}}.
\label{eqn:state_space}
\end{split}
\end{equation}
The matrix $\bm{A}$ contains the finite-dimensional discrete approximations of the linearized momentum, continuity and energy equations from \eqref{eqn:NS_comp} in terms of the Fourier transforms where the derivatives $(\partial/\partial_x,\partial/\partial_y,\partial/\partial_z)$ become $(ik_x,\partial/\partial_y,ik_z)$ (for the different terms of the matrix $\bm{A}$ see \citet{dawson2019studying}). 
The vector $\widehat{\bm{q}}=(\widehat{u},\widehat{v},\widehat{w},\widehat{\rho},\widehat{\theta})$ contains the state variables and $\widehat{\bm{f}}=(\widehat{f}_u,\widehat{f}_v,\widehat{f}_w,\widehat{f}_\rho,\widehat{f}_\theta)$ the non-linear terms of the equations. 

To analyse \eqref{eqn:state_space} we need to choose a norm, and here we employ the commonly-adopted Chu norm $E$ defined as \citep{chu1965energy,hanifi1996transient}:
\begin{equation}
E = \frac{1}{2}\int_0^{\infty} \bar{\rho} (\widehat{u}^*\widehat{u}+\widehat{v}^*\widehat{v}+\widehat{w}^*\widehat{w}) + \frac{\overline{\Theta}}{\gamma \bar{\rho} Ma^2} \widehat{\rho}^*\widehat{\rho} + \frac{\bar{\rho}}{\gamma(\gamma-1) \overline{\Theta} Ma^2} \widehat{\theta}^*\widehat{\theta} \mbox{ } dy,
\label{eqn:chu}
\end{equation}
where $\cdot^*$ represents a complex conjugate. 
The Chu norm, as well as the weights corresponding to the non-uniform grid used here, are incorporated within a weight matrix $\bm{W}$. 
The discrete inner product used becomes $\langle \widehat{\bm{q}}_1, \widehat{\bm{q}}_2 \rangle = \widehat{\bm{q}}_1^* \bm{W} \widehat{\bm{q}}_2 $.
The equation in \eqref{eqn:state_space} can now be re-written as:
\begin{equation}
\widehat{\bm{q}} = \underbrace{\left[ \bm{W}^{1/2}\left( i\omega \bm{I} - \bm{A}(k_x,k_z) \right)^{-1} \bm{W}^{-1/2} \right]}_{\bm{H}(k_x,k_z,\omega)} \widehat{\bm{f}}, 
\label{eqn:resolvent}
\end{equation}
where $\bm{I}$ is the identity matrix. 
The transfer kernel $\bm{H}(k_x,k_z,\omega)$ is the resolvent operator of the flow and it maps the non-linear terms $\widehat{\bm{f}}$ to the state variables $\widehat{\bm{q}}$. 

One of the benefits of using resolvent analysis is the ability to `mask' the resolvent. 
Whereas the full resolvent admits a response and forcing in the entire spatial domain considered, the masked resolvent, when the masking is in the response, restricts the response to lie within a specific wall-normal region. 
On the other hand, if the masking is in the forcing, the forcing in the model is restricted to lie within a specific wall-normal region.
For instance, we can consider the resolvent where the response lies solely in the freestream, or where the forcing lies exclusively in the buffer layer of the flow (see \S\ref{sec:resolvent mach waves and mach waves from dns: a discussion}). 
For this masking, matrices $\bm{B}$ and $\bm{C}$ are introduced to \eqref{eqn:state_space} such that:
\begin{equation}
\begin{split}
i\omega\bm{\widehat{p}} &= \bm{A}(k_x,k_z) \widehat{\bm{p}} + \bm{B} \widehat{\bm{f}}, \qquad \bm{\widehat{q}} = \bm{C} \bm{\widehat{p}}.
\label{eqn:state_space_masked}
\end{split}
\end{equation}
If $\bm{B}$ and $\bm{C}$ equal identity, we get back \eqref{eqn:state_space}. 
To restrict the forcing or response to defined wall-normal regions, weightings such as those introduced in \citet{nogueira2020resolvent} can be incorporated in $\bm{B}$ or $\bm{C}$. 
The masked resolvent operator $\bm{H}_{mask}(k_x,k_z,\omega)$ then becomes
\begin{equation}
\widehat{\bm{q}} = \underbrace{ \bm{C} \left[ \bm{W}^{1/2}\left( i\omega \bm{I} - \bm{A}(k_x,k_z) \right)^{-1} \bm{W}^{-1/2}  \right] \bm{B}}_{\bm{H}_{mask}(k_x,k_z,\omega) } \widehat{\bm{f}}, 
\label{eqn:resolvent_masked}
\end{equation}

\subsection{Singular value decomposition of the resolvent operator}
\label{sec:Singular Value decomposition of the resolvent}

To analyse the resolvent operator in \eqref{eqn:resolvent} we perform a singular value decomposition 
\begin{equation}
\bm{H}(k_x,k_z,c) = \sum\limits_{i=1}^{5N} \bm{\psi}_i(y) \sigma_i \bm{\phi}_i(y).
\label{eqn:svd}
\end{equation} 
Here $N$ represents the number of grid-points used to discretize the wall-normal direction. 
The singular values $\sigma_i$ are arranged such that $\sigma_i \geq \sigma_{i+1}$. 
The left singular vectors $\bm{\psi}_i(y)$ are the resolvent response modes and the right singular vectors $\bm{\phi}_i(y)$ are the resolvent forcing modes. 
Therefore a forcing to the resolvent operator along $\bm{\phi}_i$ will give a response along $\bm{\psi}_i$ amplified by a factor of $\sigma_i$. 
The most sensitive forcing direction is $\bm{\phi}_1$ that is associated with the largest singular value $\sigma_1$, and the corresponding most amplified response direction is $\bm{\psi}_1$. 
If we assume that the forcing $\widehat{\bm{f}}$ in \eqref{eqn:resolvent} is unit-amplitude and broadband across $(k_x,k_z)$, then the regions of the wavenumber space where $\sigma_1$ is high represents structures that are energetic. 

The right and left singular vectors form a complete basis. Therefore any forcing $\widehat{\bm{f}}$, and any response $\widehat{\bm{q}}$ can be expressed in terms of these basis vectors as
\begin{equation}
\widehat{\bm{f}} = \sum_i \chi_i \bm{\phi}_i \quad \mbox{and} \quad \widehat{\bm{u}} = \sum_i \chi_i \sigma_i \bm{\psi}_i. 
\label{eqn:forcing_response_reconstruction}
\end{equation}
Let us assume the forcing is approximately stochastic and therefore does not have any preferred direction.
Then, in scenarios where $\sigma_1 \gg \sigma_{i\neq 1}$, it is possible that a rank-1 model where $\widehat{\bm{f}} \approx \chi_1 \bm{\phi}_1$ and $\widehat{\bm{u}} \approx \chi_1 \sigma_1 \bm{\psi}_1$ captures the flow reasonably well \citep{beneddine2016conditions, towne2018spectral}. 
This is indicative of the existence of a dominant physical mechanism that is giving rise to the resolvent amplification (such as for example the critical layer mechanism \citep{mckeon2010critical}).
To analyse if such a rank-1 approximation is valid, \citet{moarref2014low} introduced the metric $\mbox{LR} = \sigma_1^2/\sum_i \sigma_i^2$ which denotes the fraction of energy that is captured by the first resolvent mode alone. 
$\mbox{LR}$ is bounded between $0$ and $1$ and the region of the $(\lambda_x,\lambda_z)$ space where $\mbox{LR}$ is high indicates the region where a rank-1 approximation of the resolvent operator is valid.
The resolvent operator remains low-rank in the wavenumber space where, from DNS and experiments of incompressible flows, we know most of the turbulent kinetic energy resides in the flow \citep[e.g.][]{moarref2014low, bae2020resolvent}. 

\subsection{Numerical set up for the resolvent operator}
\label{sec:Numerical set up for the resolvent operator}

A summation-by-parts finite difference scheme with $N=401$ grid points is used to discretize the linear operator $\bm{A}$ \eqref{eqn:state_space} in the wall-normal direction \citep{mattsson2004summation, kamal2020application}. 
To properly resolve the wall-normal direction we employ a grid stretching technique that gives a grid that goes from $0$ to at least $y_{\mbox{\footnotesize{max}}} = 4\delta$, with half the grid points used clustered below $y_{\mbox{\footnotesize{half}}} = 1\delta$ \citep{malik1990numerical}. 
The stretched grid $y$ in terms of equidistant points $0 \leq y' \leq 1$ is given as $y = ay'/(b-y')$, with $a = y_{\mbox{\footnotesize{max}}}y_{\mbox{\footnotesize{half}}}/(y_{\mbox{\footnotesize{max}}}-2y_{\mbox{\footnotesize{half}}})$ and $b = 1+a/y_{\mbox{\footnotesize{max}}}$ \citep[e.g][]{malik1990numerical, kamal2020application}.  
Compressible boundary layer flows have pressure fluctuations that radiate into the freestream. 
These radiations are waves that have wall-normal wavelengths that are a function of their streamwise and spanwise wavenumbers $k_x$, $k_z$ and phase-speed $c$. 
For the discussions in this work it is important to properly resolve these pressure fluctuations. 
Therefore it is important to consider their wall-normal wavelengths $l$, and this $l$ can be analytically approximated as a function of $(k_x,k_z,c)$ (see \eqref{eqn:mach_wave} and \eqref{eqn:inclination_angle}). 
Since there is a large range of $l$ that exists in the flow, it would be challenging to resolve all of the waves using a fixed $y_{\mbox{\footnotesize{max}}}$ and any reasonable number of wall-normal grid points $N$. 
Therefore, for these modes we use a $y_{\mbox{\footnotesize{max}}}$ that varies with $(k_x,k_z,c)$ such that if $y_{\mbox{\footnotesize{max}}} = 4\delta$ is not sufficient to resolve at least $3l$, $y_{\mbox{\footnotesize{max}}}$ is increased to be $3l$. (In appendix \ref{sec:Grid convergence} we include a discussion on the grid convergence obtained). 
To keep the forcing to the resolvent consistent across $(k_x,k_z,c)$, all modes are forced only till $3\delta$, with a weighting as introduced in \citet{nogueira2020resolvent} used to set the forcing beyond $3\delta$ to zero.

Following \citet{mack1984boundary} and \citet{malik1990numerical}, the boundary conditions enforced at the wall are $\widehat{u}(0) = \widehat{v}(0) = \widehat{w}(0) = \widehat{\theta}(0) = 0$. 
The wall-normal momentum equation at the wall is used to get the boundary condition on density, which along with the temperature boundary condition determines boundary condition for pressure. 
Since at the freestream we can assume that the equations are inviscid, Thompson boundary conditions derived from the inviscid equations are enforced here \citep{thompson1987time, kamal2020application}. 
Additionally, a damping-layer is also required at the freestream to remove spurious numerical oscillations that arise from the finite difference operator \citep{appelo2009high} (see appendix \ref{sec:Grid convergence} for a discussion regarding this damping layer). 
Mean profiles that are required as input to the linear model in \eqref{eqn:NS_comp} are obtained from the DNS studies of \citet{ bernardini2011wall, duan2014numerical, duan2016pressure, zhang2017effect}. 

There is some arbitrariness to the choice of wall-normal extent $y_{\mbox{\footnotesize{max}}}$ and the details of the damping layer used for the wall-normal grid. 
However, the results presented here are reasonably insensitive to variations in these choices. 
This has been discussed in detail in appendix \ref{sec:Grid convergence}.

\subsection{Helmholtz decomposition}
\label{sec:helmholtz decomposition}

As the final topic in the methods section, let us briefly look at  Helmholtz decomposition, a technique that will be used frequently in this manuscript. 
The Helmholtz decomposition can be performed on any vector field. Consider a vector $\bm{q} = (q_x, q_y, q_z)$. 
The Helmholtz decomposition of $\bm{q}$ gives two components such that $\bm{q}=\bm{q}^s+\bm{q}^d$. The two components are: (i) the solenoidal component $\bm{q}^s = (q^s_x, q^s_y, q^s_z)$ which is divergence-free, i.e.\ $\nabla \cdot \bm{q}^s=0$, and (ii) the dilatational component $\bm{q}^d = (q^d_x, q^d_y, q^d_z)$ which is curl-free $\nabla \times \bm{q}^d=0$. 
Helmholtz decomposition is only unique with a defined boundary condition, and the boundary conditions that are imposed here are $q^s_y(y=0)=0$ and $q^d_x(y=0)=q^d_z(y=0)=0$ \citep{bhatia2012helmholtz}. 
As an example, consider the velocity field $\bm{u}$ from an incompressible flow. 
Since the flow is divergence free, for this case $\bm{u}=\bm{u}^s$ and $\bm{u}^d=0$. 

In \S\ref{sec:Helmholtz decomposition of velocity} we will use Helmholtz decomposition of the first resolvent response mode $\bm{\psi}_1$ to find the solenoidal and dilatational component of this mode. 
In \S\ref{sec:forcing to the subsonic modes: helmholtz decomposition of resolvent forcing} and \S\ref{sec:two routes of forcing resolvent Mach wave radiation}, we will instead focus on the Helmholtz decomposition of the first resolvent forcing mode $\bm{\phi}_1$. 
This thereafter enables us to look at the response to the solenoidal $\bm{\phi}_1^s$ and the dilatational $\bm{\phi}_1^d$ component of this forcing mode, separately. 
The response to $\bm{\phi}_1^s$ can be obtained as $H\bm{\phi}_1^s$ and to $\bm{\phi}_1^d$ as $H\bm{\phi}_1^d$.

\section{Comparing compressible and incompressible resolvent operators}
\label{sec:comparing compressible and incompressible resolvent operators}

\begin{figure}[t]
\captionsetup[subfigure]{labelformat=empty,skip=-30pt}
\begin{subfigure}[b]{\textwidth}
\centering
\includegraphics[width=\textwidth, trim={0.9cm 1cm 1.6cm 0.7cm}, clip]{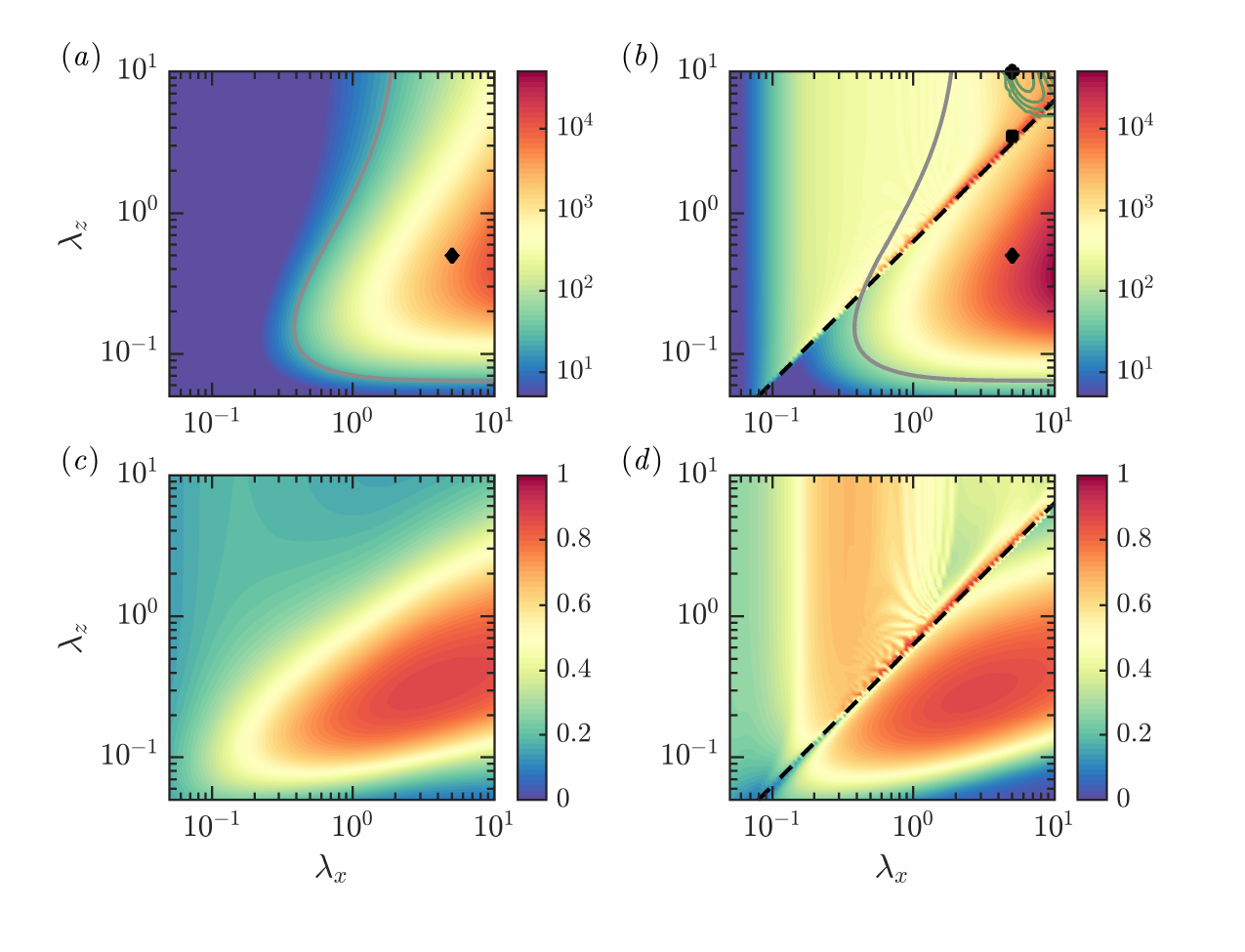}
\caption{}
\label{fig:energy_incomp}
\end{subfigure}%
\begin{subfigure}[b]{\textwidth}
\caption{}
\label{fig:energy_comp}
\end{subfigure}
\begin{subfigure}[b]{\textwidth}
\caption{}
\label{fig:LR_incomp}
\end{subfigure}
\begin{subfigure}[b]{\textwidth}
\caption{}
\label{fig:LR_comp}
\end{subfigure}
\vspace{-0.75cm}
\caption{(\textit{a,b}) The leading resolvent gain $\sigma_1$ as well as (\textit{c,d}) the fraction of energy captured by the leading resolvent mode $\mbox{LR}=\sigma_1^2/(\sum_i\sigma_i^2)$, are shown as a function of the streamwise and spanwise wavelengths ($\lambda_x$,$\lambda_z$) at a fixed phase speed $c \approx \overline{U}(y^+ = 15)$. 
(\textit{a,c}) An incompressible boundary layer with $Re_\tau= 450$ and  (\textit{b,d}) a compressible boundary layer with $Ma=4$ and $Re_\tau= 400$ over an adiabatic wall are considered. 
The black dashed line in (\textit{b,d}) indicates the relative Mach number equal to unity. 
The green contours at the top right-hand corner for the compressible case indicates the region of the wavenumber space that is unstable. 
The grey contour line in (\textit{a}) indicates $1/3$rd of the maximum energy in the incompressible case, and the same contour (computed from the incompressible case) also appears in (\textit{b}) for comparison. 
The ({\scriptsize $\blacklozenge$}) in (\textit{a,b}) and the ({\tiny $\blacksquare$}) and ({$\bullet$}) in (\textit{b}) indicates the modes that are discussed in later figures.}
\label{fig:Intro}
\end{figure}

In this section we compare the resolvent norms obtained from the incompressible and the compressible resolvent operators. 
This comparison is similar to that in \citep{bae2020resolvent}, however, in addition to the low-rank map that was compared in \citep{bae2020resolvent}, here we also look at the leading resolvent norm, which is important for the discussions here. 
In figure \ref{fig:Intro}, a $Re_\tau= 450$ incompressible resolvent operator is shown on the left, and a $Ma=4$ compressible resolvent operator at a comparable Reynolds number ($Re_\tau= 400$) is on the right. 
In the first row of figure \ref{fig:Intro}, the leading resolvent norm $\sigma_1$ \eqref{eqn:svd} is shown with respect to the streamwise and spanwise wavelengths ($\lambda_x$, $\lambda_z$), for a fixed value of $c=\overline{U}(y^+\approx 15)$. 
The colour-scale used in the figure is logarithmic. 
While the full Chu norm \eqref{eqn:chu} is used for the compressible case, the kinetic energy norm is used for the incompressible case and this difference does not significantly impact the discussions here. 
In figure \ref{fig:Intro}(\subref{fig:energy_incomp}), the grey contour line indicates $1/3$rd of the maximum energy of the incompressible case. 
The same contour line (computed from the incompressible case) is also shown in figure \ref{fig:Intro}(\subref{fig:energy_comp}). 
The green contours at the top right-hand corner for the compressible case in figure \ref{fig:Intro}(\subref{fig:energy_comp}) indicates the region of the wavenumber space that is unstable. 
In the second row of figure \ref{fig:Intro} we compare the low-rank maps $\mbox{LR}=\sigma_1^2/(\sum_i \sigma_i^2)$ (see \S\ref{sec:Singular Value decomposition of the resolvent}) that shows the fraction of energy captured by the leading resolvent mode at each ($\lambda_x,\lambda_y,c$). 

The black dashed line in figure \ref{fig:Intro}(\subref{fig:energy_comp}) indicates the region where the freestream relative Mach number is equal to unity $\overline{Ma}(y \to \infty)=1$, where $\overline{Ma}(y)$ is a local Mach number at a particular wall-height $y$ defined for each ($\lambda_x,\lambda_y,c$) as:
\begin{equation}
\overline{Ma}(y)=\frac{Ma}{\overline{\Theta}(y)^{1/2}} \left[ \frac{k_x}{k}\left(\overline{U}(y)-c \right) \right].
\label{eqn:relative_Mach}
\end{equation}
$\overline{Ma}(y)$ is the projection of the phase-speed relative to the mean flow $(\overline{U}(y)-c)$ in the direction of the streamwise wavenumber $k_x/k$ \citep{mack1984boundary}. 
For the flows considered here, at a particular ($\lambda_x,\lambda_y,c$), the maximum value of $\overline{Ma}(y)$ is $\overline{Ma}(\infty)$. 
Above the black-dashed line in figure \ref{fig:Intro}(\subref{fig:energy_comp}), $\overline{Ma}(\infty)>1$. 
Below this line $\overline{Ma}(\infty)<1$. 
In \S\ref{sec:Mach wave radiation from the resolvent operator} we will see that the supersonic resolvent modes, which is the main subject of the discussions in the current work, can only exist when $\overline{Ma}(\infty) \geq 1$ and therefore can only exist above the black-dashed line in figure \ref{fig:Intro}(\subref{fig:energy_comp}) \citep{bae2020resolvent}. 

\begin{figure}[t]
\captionsetup[subfigure]{labelformat=empty,skip=-80pt}
\begin{subfigure}[b]{\textwidth}
\centering
\includegraphics[width=\textwidth, trim={2.75cm 7.9cm 3cm 1cm}, clip]{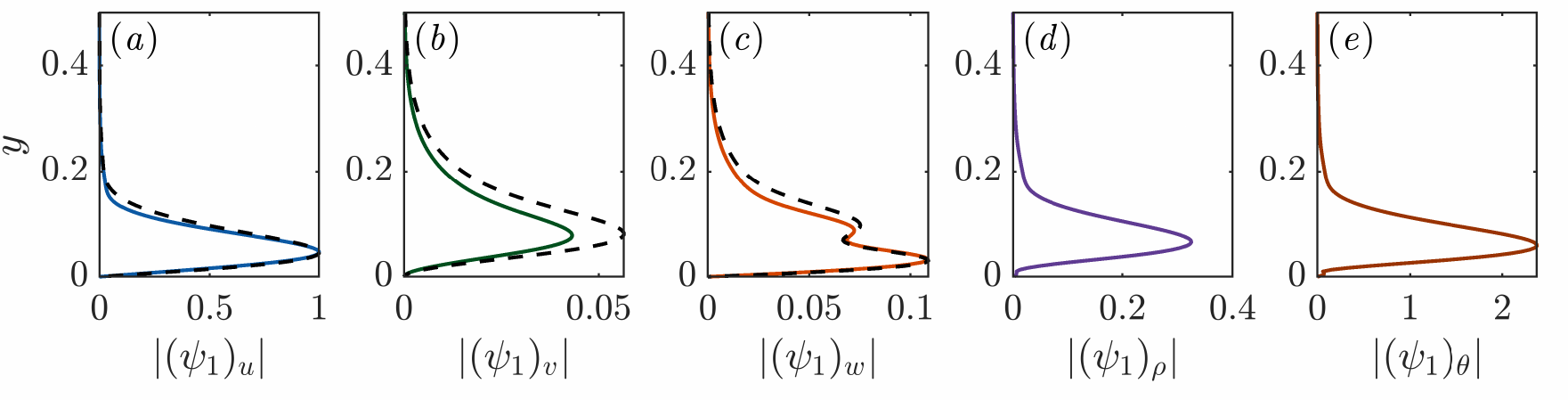}
\caption{}
\label{fig:subsonic_mode_u}
\end{subfigure}
\begin{subfigure}[b]{\textwidth}
\caption{}
\label{fig:subsonic_mode_v}
\end{subfigure}
\begin{subfigure}[b]{\textwidth}
\caption{}
\label{fig:subsonic_mode_w}
\end{subfigure}
\begin{subfigure}[b]{\textwidth}
\caption{}
\label{fig:subsonic_mode_r}
\end{subfigure}
\begin{subfigure}[b]{\textwidth}
\caption{}
\label{fig:subsonic_mode_t}
\end{subfigure}
\begin{subfigure}[b]{\textwidth}
\caption{}
\label{fig:subsonic_mode_incomp_u}
\end{subfigure}
\begin{subfigure}[b]{\textwidth}
\caption{}
\label{fig:subsonic_mode_comp_u}
\end{subfigure}
\vspace{-0.75cm}
\caption{The leading resolvent response for the modes indicated by the diamonds ({\scriptsize $\blacklozenge$}) in figure \ref{fig:Intro}(\textit{a,b}) are shown. 
The mode corresponds to $\lambda_x=5$, $\lambda_z=0.5$ and $c=\overline{U}(y^+\approx 15)$. 
The wall-normal profile of the (\textit{a}) streamwise, (\textit{b}) wall-normal and (\textit{c}) spanwise velocities are shown. 
The solid lines in (\textit{a}-\textit{c}) represents the mode from a compressible boundary layer with $Ma=4$ and $Re_\tau= 400$ and the black dashed lines represents the mode from an incompressible boundary layer with $Re_\tau= 450$. 
The profiles of (\textit{d}) density and (\textit{e}) temperature are also shown for the compressible case. }
\label{fig:CompVSIncomp}
\end{figure}

Before considering these Mach waves in detail, let us briefly consider the region below the black-dashed line, i.e.\ the subsonic modes. 
We note that the most amplified subsonic modes are linearly stable, i.e.\ do not fall within the green contour lines in figure \ref{fig:Intro}(\subref{fig:energy_comp}). 
To compare these subsonic modes to the incompressible case,  
between figures \ref{fig:Intro}(a) and \ref{fig:Intro}(b) we compare the regions of high amplification and between figures \ref{fig:Intro}(c) and \ref{fig:Intro}(d) we compare the regions where the resolvent operator is low-rank. 
Below the black-dashed line, the trends of the incompressible and the compressible resolvent operators are similar. 
This suggests that, within this region of the wavenumber space of the compressible flow, the mechanisms that are active are the same as in incompressible flows. 

For further comparison, consider the structure with $\lambda_x=5$, $\lambda_z=0.5$ and $c=\overline{U}(y^+\approx 15)$ (indicated by $({\scriptsize \blacklozenge})$ in figures \ref{fig:Intro}(\subref{fig:energy_incomp}) and \ref{fig:Intro}(\subref{fig:energy_comp})). 
The leading resolvent mode for the structure is shown in figure \ref{fig:CompVSIncomp}. 
The three components of velocity are compared in figures (\subref{fig:subsonic_mode_u}-\subref{fig:subsonic_mode_w}), with the solid coloured lines representing the compressible mode and the black dashed line representing the incompressible mode. 
For both compressible and incompressible cases, the modes are localised and reside within the boundary layer (the y-axis terminates at $0.5\delta$), and there are no significant differences between the incompressible and compressible structures. 
This similarity between the compressible and incompressible resolvent operators was explored in \citet{bae2020resolvent}, where they showed that when the compressible modes are scaled using the semi-local scaling of compressible flows \citep{trettel2016mean}, they collapse well onto the modes from the incompressible flow. 
Of course, in the case of the compressible flow, there are the additional components of temperature and density and these are shown in figures \ref{fig:CompVSIncomp}(\subref{fig:subsonic_mode_r}) and \ref{fig:CompVSIncomp}(\subref{fig:subsonic_mode_t}). 
Temperature and density also show localised profiles within the boundary layer. 

\subsection{Helmholtz decomposition of the resolvent response}
\label{sec:Helmholtz decomposition of velocity}

\begin{figure}[t]
\captionsetup[subfigure]{labelformat=empty,skip=-20pt}
\begin{subfigure}[b]{\textwidth}
\centering
\includegraphics[width=\textwidth, trim={0.5cm 0.8cm 2.6cm 8cm}, clip]{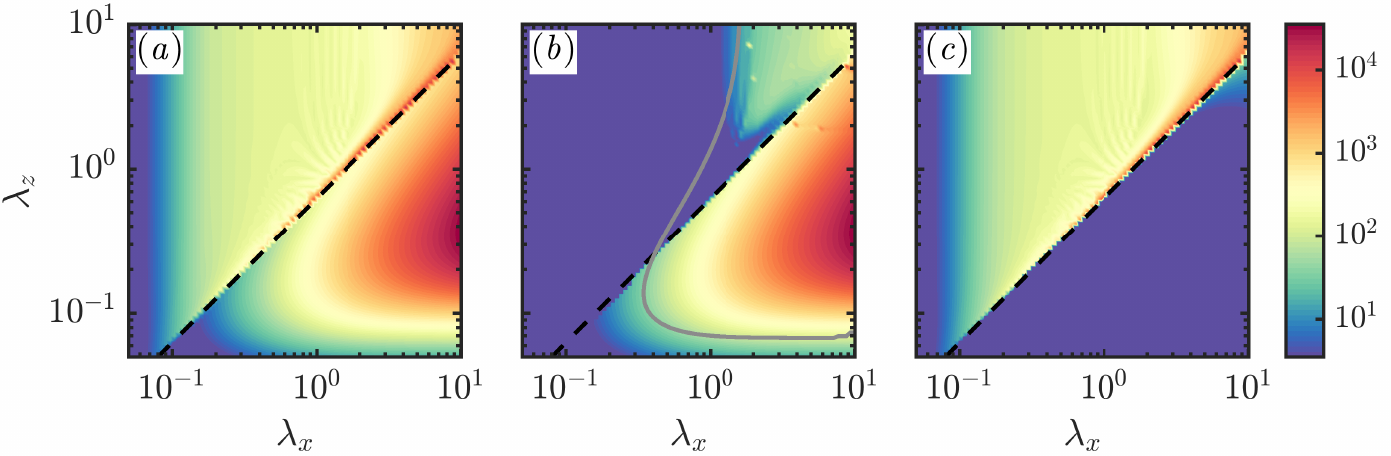}
\caption{}
\label{fig:usud_full}
\end{subfigure}%
\begin{subfigure}[b]{\textwidth}
\caption{}
\label{fig:us}
\end{subfigure}
\begin{subfigure}[b]{\textwidth}
\caption{}
\label{fig:ud}
\end{subfigure}
\vspace{-0.75cm}
\caption{The kinetic energy of (\textit{a}) the leading resolvent response $\bm{\psi}_1$ and of (\textit{b}) the solenoidal and (\textit{c}) the dilatational component of the velocity from $\bm{\psi}_1$ is shown as a function of the streamwise and spanwise wavelengths ($\lambda_x$, $\lambda_z$) and a fixed phase speed $c = \overline{U}(y^+\approx 15)$.  
The grey contour line in (\textit{b}) is the same as that shown in figure \ref{fig:Intro}(\subref{fig:energy_incomp}) and shows the kinetic energy for the incompressible case at $1/3$rd of the maximum. }
\label{fig:usud}
\end{figure}

We can attempt to isolate these incompressible-like subsonic modes from the compressible effects in the flow by doing a Helmholtz decomposition of the velocity components $\widehat{\bm{u}} = (\widehat{u}, \widehat{v}, \widehat{w})$ of the leading resolvent mode $\bm{\psi}_1$. 
As described in \S\ref{sec:helmholtz decomposition} the Helmholtz decomposition gives: (i) a solenoidal component $\widehat{\bm{u}}^s$ and (ii) a dilatational component $\widehat{\bm{u}}^d$ (in an incompressible flow, $\widehat{\bm{u}}^d=0$).  
In figure \ref{fig:usud}, the full kinetic energy in figure \ref{fig:usud}(\subref{fig:usud_full}) is compared with the kinetic energy of $\widehat{\bm{u}}^s$ in figure \ref{fig:usud}(\subref{fig:us}) and of $\widehat{\bm{u}}^d$ in figure \ref{fig:usud}(\subref{fig:ud}). 
The grey contour in figure \ref{fig:usud}(\subref{fig:us}) is the same as that shown in figure \ref{fig:Intro}(\subref{fig:energy_incomp}) and indicates the region of the wavenumber space where the incompressible resolvent is reasonably amplified.

The solenoidal response in figure \ref{fig:usud}(\subref{fig:us}) looks similar to the response of the incompressible flow in figure \ref{fig:Intro}(\subref{fig:energy_incomp}). 
Subsonic modes, i.e.\ the modes below the $\overline{Ma}(\infty)=1$ line, have solenoidal velocity, consistent with observations from DNS \citep{yu2019genuine}. 
Notably, in figure \ref{fig:usud}(\subref{fig:us}), $\widehat{\bm{u}}^s$ captures some of the energy in modes above the $\overline{Ma}(\infty)=1$ where supersonic mechanisms are active. 
The grey-contour line indicates that this occurs in the region where subsonic mechanisms are also active. 
These modes therefore have both the subsonic as well as supersonic mechanisms active. 
The co-existence of two amplification mechanisms provides an explanation for the observed decrease in the low-rank behaviour of these modes in figure \ref{fig:Intro}(\subref{fig:LR_comp}). 
However, the most interesting consequence of this co-existence of mechanisms is that it provides an additional route for amplifying the supersonic resolvent modes. 
This amplification route will be explained in \S\ref{sec:indirect route of forcing resolvent Mach wave radiation} and its potential importance to the real flow will be discussed in \S\ref{sec:resolvent mach waves and mach waves from dns: a discussion}. 

\subsection{Forcing to the subsonic modes: Helmholtz decomposition of resolvent forcing}
\label{sec:forcing to the subsonic modes: helmholtz decomposition of resolvent forcing}

In this section we take the comparison between the compressible and incompressible operators one step further. 
For incompressible resolvent operators, if we consider Helmholtz decomposition of the forcing to the momentum equations $\widehat{\bm{f}}_{\bm{u}}$, it is known that only the solenoidal component of the forcing $\widehat{\bm{f}}_{\bm{u}}^s$ has any active influence in amplifying resolvent modes \citep{rosenberg2019efficient, morra2021colour}. 
Although $\widehat{\bm{f}}_{\bm{u}}^d$ is not zero in the real flow (i.e.\ the divergence of the full forcing $\nabla \cdot \widehat{\bm{f}}_{\bm{u}} \neq 0$), this component cannot directly excite a response in velocity \citep{rosenberg2019efficient, morra2021colour}. 
Here we ask if this property of incompressible flows carries over to the subsonic modes of compressible flows.  
$\widehat{\bm{f}}_{\bm{u}}$ will be taken from the leading resolvent forcing mode $\bm{\phi}_1$ (the suboptimal modes are considered in appendix \ref{sec: Sub-optimal modes}). 
A Helmholtz decomposition of this resolvent forcing mode gives the solenoidal component $\widehat{\bm{f}}_{\bm{u}}^s = (\widehat{f}^s_u, \widehat{f}^s_v, \widehat{f}^s_w)$ and the dilatational component $\widehat{\bm{f}}_{\bm{u}}^d= (\widehat{f}^d_u, \widehat{f}^d_v, \widehat{f}^d_w)$. 
We will consider the response to the solenoidal component alone through $\widehat{\bm{f}}_1=(\widehat{f}^s_u, \widehat{f}^s_v, \widehat{f}^s_w, 0, 0)$, and the response to the dilatational component along with the forcing to the density $\widehat{f}_\rho$ and temperature $\widehat{f}_\theta$ equations through $\widehat{\bm{f}}_2=(\widehat{f}^d_u, \widehat{f}^d_v, \widehat{f}^d_w, \widehat{f}_\rho, \widehat{f}_\theta)$. 
The response to $\widehat{\bm{f}}_1$ is obtained as $H \widehat{\bm{f}}_1$ and to $\widehat{\bm{f}}_2$ as $H \widehat{\bm{f}}_2$. 

\begin{figure}[t]
\captionsetup[subfigure]{labelformat=empty,skip=-5pt}
\begin{subfigure}[b]{\textwidth}
\centering
\includegraphics[width=\textwidth, trim={2.7cm 1.2cm 3cm 0cm}, clip]{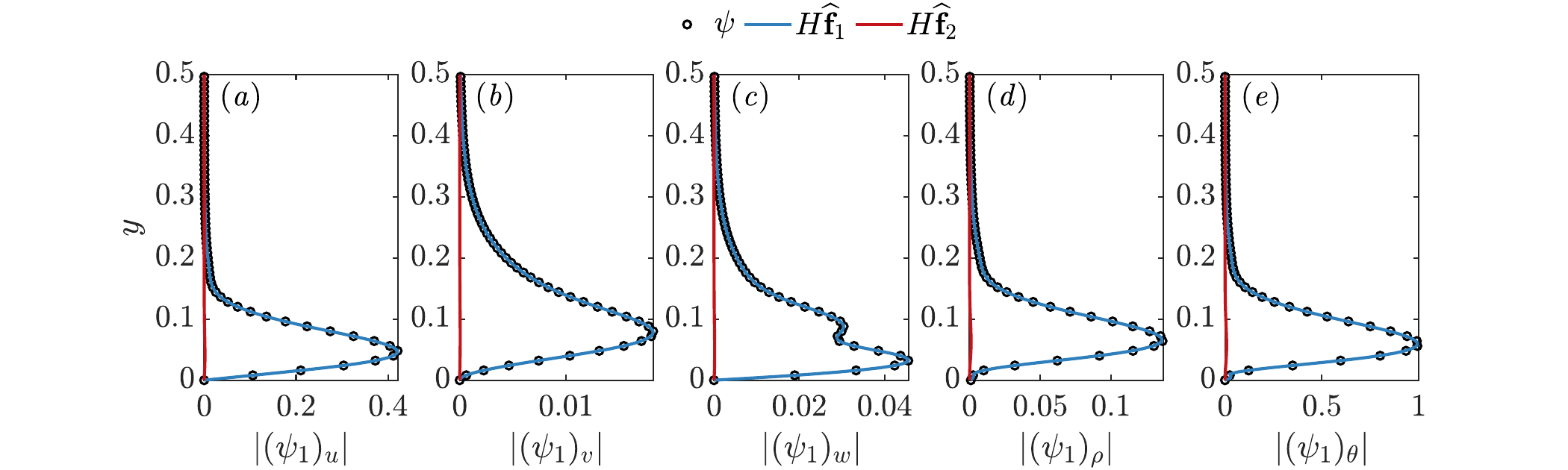}
\caption{}
\label{fig:fHelmholtz_mode_subsonic_u}
\end{subfigure}%
\begin{subfigure}[b]{\textwidth}
\caption{}
\label{fig:fHelmholtz_mode_subsonic_v}
\end{subfigure}%
\begin{subfigure}[b]{\textwidth}
\caption{}
\label{fig:fHelmholtz_mode_subsonic_w}
\end{subfigure}%
\begin{subfigure}[b]{\textwidth}
\caption{}
\label{fig:fHelmholtz_mode_subsonic_r}
\end{subfigure}%
\begin{subfigure}[b]{\textwidth}
\caption{}
\label{fig:fHelmholtz_mode_subsonic_t}
\end{subfigure}
\begin{subfigure}[b]{\textwidth}
\includegraphics[width=\textwidth, trim={2.7cm 0.05cm 3cm 1.2cm}, clip]{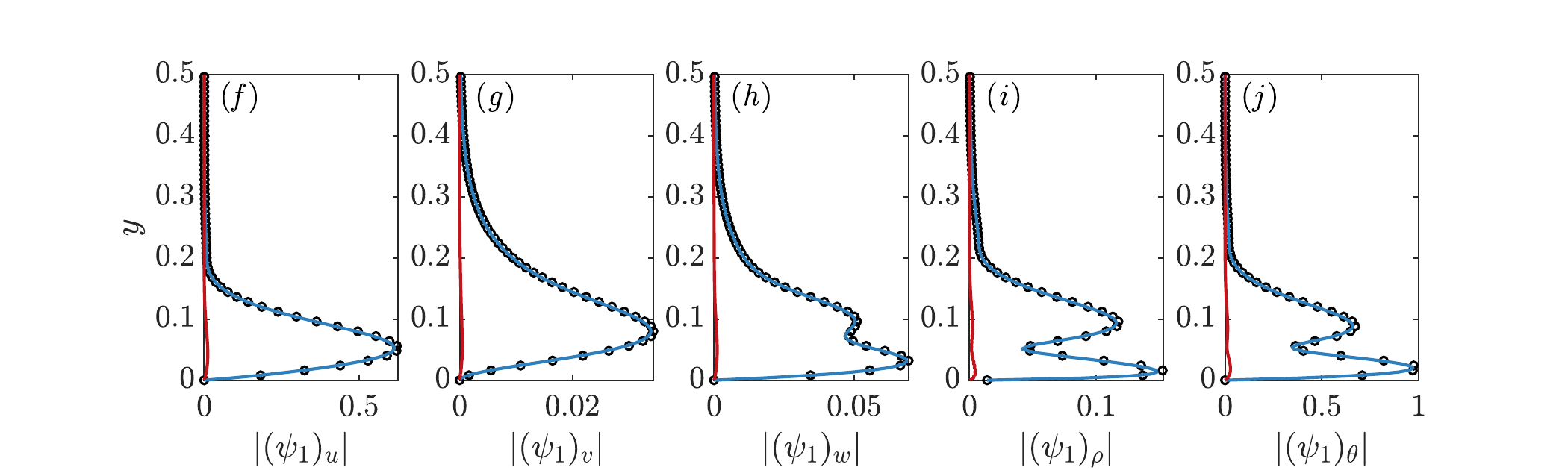}
\caption{}
\label{fig:fHelmholtz_mode_subsonic_cooled_u}
\end{subfigure}%
\begin{subfigure}[b]{\textwidth}
\caption{}
\label{fig:fHelmholtz_mode_subsonic_cooled_v}
\end{subfigure}%
\begin{subfigure}[b]{\textwidth}
\caption{}
\label{fig:fHelmholtz_mode_subsonic_cooled_w}
\end{subfigure}%
\begin{subfigure}[b]{\textwidth}
\caption{}
\label{fig:fHelmholtz_mode_subsonic_cooled_r}
\end{subfigure}%
\begin{subfigure}[b]{\textwidth}
\caption{}
\label{fig:fHelmholtz_mode_subsonic_cooled_t}
\end{subfigure}
\vspace{-0.75cm}
\caption{The response of the resolvent operator to the full leading resolvent forcing mode $\bm{\phi}_1$ (black) as well as the response to the two components of the forcing $\widehat{\bm{f}}_1$ (blue) and $\widehat{\bm{f}}_2$ (red) are shown. 
Subsonic modes with $(\lambda_x,\lambda_z,c)=(5,0.5,\overline{U}(y^+\approx 15))$ for two compressible boundary layers are shown: (\textit{a}-\textit{e}) $Ma=4$, $Re_\tau=400$ over an adiabatic wall (the mode indicated by the ({\scriptsize $\blacklozenge$}) in figure \ref{fig:Intro}(\subref{fig:energy_comp})) and (\textit{f}-\textit{j}) $Ma=6$, $Re_\tau=450$ with $\Theta_w/\Theta_{ad}=0.25$. 
The (\textit{a,f}) streamwise, (\textit{b,g}) wall-normal and (\textit{c,h}) spanwise velocities as well as the (\textit{d,i}) density and (\textit{e,j}) temperature are shown. }
\label{fig:fHelmholtz_mode_subsonic}
\end{figure}

In figures \ref{fig:fHelmholtz_mode_subsonic}(\subref{fig:fHelmholtz_mode_subsonic_u}-\subref{fig:fHelmholtz_mode_subsonic_t}), for the same structure as in figure \ref{fig:CompVSIncomp}, the five components of the leading resolvent response mode in black is compared to the response to $\widehat{\bm{f}}_1$ in blue and to $\widehat{\bm{f}}_2$ in red.  
We see that the response is almost entirely captured by the solenoidal component of the forcing. 
To see that this is more generally true, in the second row of figure \ref{fig:fHelmholtz_mode_subsonic}, the same comparison is shown for a $Re_\tau=450$ boundary layer with a higher Mach number of $Ma=6$, over a cooled wall with wall-cooling ratio $\Theta_w/\Theta_{ad}=0.25$. 
Here again we see that the solenoidal component of the forcing captures a majority of the response. 
Therefore, like in the incompressible case, the subsonic resolvent modes are actively forced only by the solenoidal component of the forcing; and this is true both for flows over adiabatic and cooled walls. 

\section{Mach wave radiation from the resolvent operator}
\label{sec:Mach wave radiation from the resolvent operator}

\begin{figure}[t]
\captionsetup[subfigure]{labelformat=empty,skip=-35pt}
\begin{subfigure}[b]{\textwidth}
\centering
\includegraphics[width=\textwidth, trim={0cm 8.3cm 0cm 0.4cm}, clip]{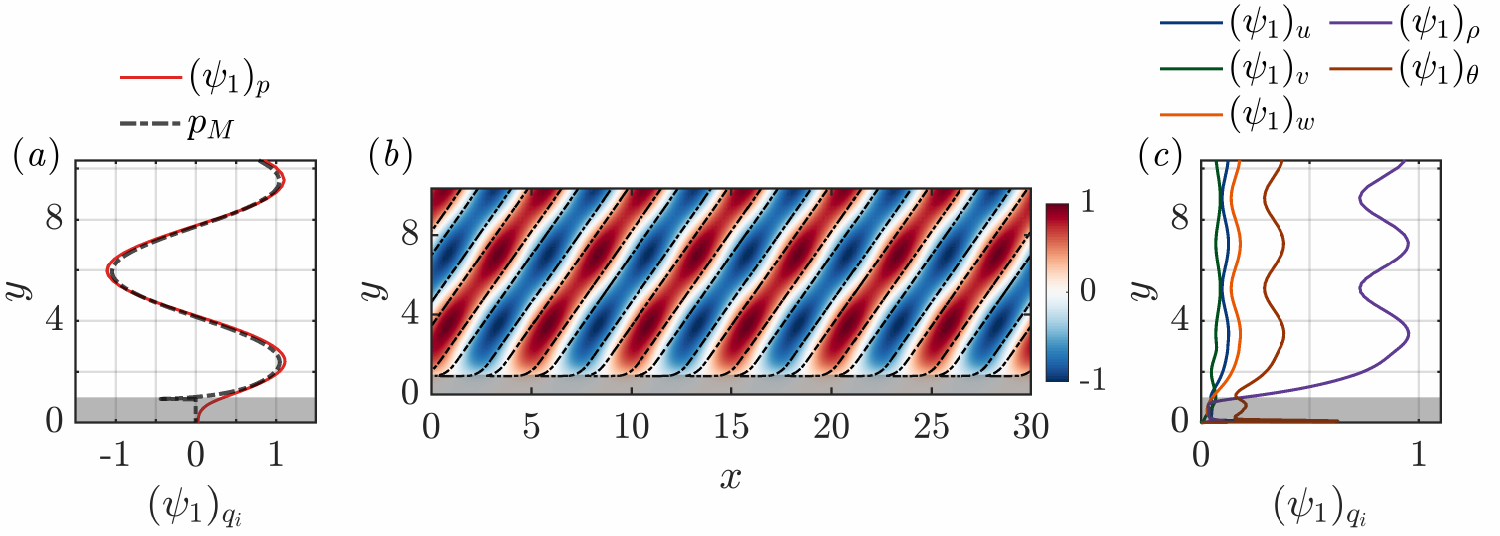}
\caption{}
\label{fig:supersonic_mode_comp_pline}
\end{subfigure}
\begin{subfigure}[b]{\textwidth}
\caption{}
\label{fig:supersonic_mode_comp_pfield}
\end{subfigure}
\begin{subfigure}[b]{\textwidth}
\caption{}
\label{fig:supersonic_mode_comp_u}
\end{subfigure}
\vspace{-0.75cm}
\caption{The leading resolvent response for the mode indicated by the ({\tiny $\blacksquare$}) in figure \ref{fig:Intro}(\textit{b}). 
The mode corresponds to $\lambda_x=5$, $\lambda_z=3.5$ and $c=\overline{U}(y^+\approx 15)$ for a compressible boundary layer with $Ma=4$ and $Re_\tau= 400$ over an adiabatic wall. 
(\textit{a}) The wall-normal profile of the real part of the pressure from the resolvent mode (red) is compared with the pressure fluctuations from the inviscid Mach wave (black dashed) given by the equation \eqref{eqn:mach_wave}.
(\textit{b}) The pressure fluctuations in a $x-y$ plane is also shown with red (positive) and blue (negative) contours showing the real part of the pressure fluctuation from the resolvent mode and the black contour lines indicating the pressure fluctuations from the inviscid Mach wave at $\pm0.5$ of the maximum. 
(\textit{c}) The wall-normal profile of the streamwise (blue), wall-normal (green) and spanwise (orange) velocities, as well as the density (purple) and temperature (brown) are also shown. 
The grey shaded regions indicate the boundary layer thickness. }
\label{fig:supersonic_mode}
\end{figure}

We will now focus on the region of the wavenumber space that falls above the $\overline{Ma}(\infty)=1$ line in figure \ref{fig:Intro}(\subref{fig:energy_comp}). 
The amplification in this region is largely due to supersonic resolvent modes.  In this section, the connection between these resolvent modes and the relative Mach number defined in \eqref{eqn:relative_Mach} is first established. We will then focus on the contribution of these modes to the freestream fluctuations of the boundary layer in \S\ref{sec:contribution to the boundary layer and freestream} and the effect of viscosity on these modes in \S\ref{sec:the effect of viscosity on resolvent mach wave amplification}. 
Thereafter, from \S\ref{sec:two routes of forcing resolvent Mach wave radiation} onward, we will concentrate on the forcing mechanisms that amplify these modes. 

These supersonic modes above the $\overline{Ma}(\infty)=1$ line are pressure fluctuations that radiate into the freestream and they occur due to the Mach wave radiation mechanism as explained in \citet{mack1984boundary}. 
We will briefly follow the derivation of the Mach wave radiation in \citet{mack1984boundary}, with the difference of keeping the non-linear terms as a forcing forcing $\bm{f}$, rather than setting them to zero as in \citet{mack1984boundary}. 
This will enable us to later isolate the two routes by which the Mach waves can cause high resolvent amplification: (i) the direct route (\S\ref{sec:direct route of forcing resolvent Mach wave radiation}) and (ii) the indirect route (\S\ref{sec:indirect route of forcing resolvent Mach wave radiation}). 

Like done in \citet{mack1984boundary}, let us for simplicity consider the inviscid form of the governing equations Fourier transformed in $(x,y,t)$ \eqref{eqn:state_space}. 
As in \eqref{eqn:NS_comp}, the non-linear terms in the equations and any external disturbances to the flow are denoted as the unknown forcing $\widehat{\bm{f}}$, therefore giving
\begin{subequations} \label{eqn:NS_inviscid}
\begin{align}
\begin{split}
i\overline{\rho} (k_x \overline{U}-\omega) \widehat{u}_i &= 
- \overline{\rho} \frac{d \overline{U}}{d y} \widehat{v} \mbox{ } \widehat{\bm{i}}
-\frac{1}{\gamma Ma^2} \frac{\partial}{\partial x_i} \widehat{p} + \widehat{f}_{u_i} 
\end{split} \label{eqn:NS_inviscid_u}\\
\begin{split}
i\overline{\rho} (k_x \overline{U}-\omega) \widehat{\theta} &= 
\phantom{-} \overline{\Theta} \frac{\partial \overline{\rho}}{\partial y} \widehat{v}
- (\gamma-1) \nabla \cdot \bm{u} + \widehat{f}_\theta
\end{split} \label{eqn:NS_inviscid_t}\\
\begin{split}
i \overline{\Theta} (k_x \overline{U}-\omega) \widehat{\rho} &=
- \overline{\Theta} \frac{\partial \overline{\rho}}{\partial y} \widehat{v} 
- \nabla \cdot \bm{u} + \widehat{f}_\rho 
\end{split} \label{eqn:NS_inviscid_r}
\end{align}
\end{subequations}

Here $\bm{u}=(u_1,u_2,u_3)=(u,v,w)$ and we have used $\overline{\rho}\overline{\Theta}=1$ to write $- \overline{\rho} \partial \overline{\Theta}/\partial y = \overline{\Theta} \partial \overline{\rho}/\partial y$.  
Pressure fluctuation $\widehat{p}$ can be written using the linearized equation of state as $\widehat{p} = \overline{\Theta} \widehat{\rho} + \overline{\rho} \widehat{\theta}$. 
Using \eqref{eqn:NS_inviscid_t} and \eqref{eqn:NS_inviscid_r}, and computing $\nabla \cdot \bm{u}$ from \eqref{eqn:NS_inviscid_u} we rewrite this equations for $\widehat{p}$ as
\begin{equation}
\begin{split}
-\frac{\partial^2 \widehat{p}}{\partial y^2}
+ k^2(1-\overline{Ma}^2)\widehat{p}  & = 
i\gamma Ma^2 \left[2 k_x \overline{\rho} \frac{d\overline{U}}{dy} 
+ \frac{d\overline{\rho}}{dy} (k_x \overline{U}-\omega) \right] \widehat{v} \\
&
- \gamma Ma^2 \nabla \cdot \widehat{\bm{f}}_{\bm{u}}
+ i Ma^2 \overline{\rho} (k_x \overline{U}-\omega) \left( \widehat{f}_\theta + \widehat{f}_\rho \right) 
.\\
\end{split}
\label{eqn:inviscid_pressure}
\end{equation}
Here $k^2=k_x^2+k_z^2$ and $\widehat{\bm{f}}_{\bm{u}} = (\widehat{f}_u, \widehat{f}_v, \widehat{f}_w)$. 
(Please see \citet{mack1984boundary} for a more complete derivation).
Note that in \eqref{eqn:inviscid_pressure} $Ma$ denotes the freestream Mach number and $\overline{Ma}$ denotes the relative Mach number $\overline{Ma}(y)$ defined in \eqref{eqn:relative_Mach}. 
Following \citet{mack1984boundary}, let us consider the unforced equation within the freestream. 
In this case, the first term on the right-hand side of \eqref{eqn:inviscid_pressure} is zero since the terms within the square brackets are zero in the freestream. 
The second and third terms are zero since we are considering the unforced equations. 
Therefore, the right hand side of \eqref{eqn:inviscid_pressure} goes to zero. 
For this equation, when $\overline{Ma}(y)>1$ we obtain a wave equation with solutions of the form
\begin{equation}
\widehat{p}_M = i \gamma \mbox{Ma}^2 \left(\frac{k_x\overline{U} - \omega}{k} \right) \exp{\left( -k \left( 1-\overline{Ma}^2 \right)^{1/2} y\right)}.  
\label{eqn:mach_wave}
\end{equation}
These are the Mach waves of the flow derived in \citet{mack1984boundary} as solutions of the inviscid unforced freestream equations when $\overline{Ma}(\infty)>1$. 
From \eqref{eqn:mach_wave} we get both upstream and downstream inclining waves, but only the downstream inclining waves satisfy the boundary conditions and therefore become valid solutions \citep{mack1984boundary}. 

Going back to figure \ref{fig:Intro}(\subref{fig:energy_comp}), the supersonic resolvent modes are amplified only when $\overline{Ma}(\infty)\geq 1$ suggesting that these resolvent modes are Mach waves \citep{bae2020resolvent}. 
To show that this is indeed the case, let us compare resolvent modes with \eqref{eqn:mach_wave}. 
We pick the $(\lambda_x,\lambda_y)$ marked by ({\tiny $\blacksquare$}) in figure \ref{fig:Intro}(\subref{fig:energy_comp}) and the pressure $(\psi_1)_p$ from the leading resolvent mode of this structure is computed using the density $(\psi_1)_\rho$ and the temperature $(\psi_1)_\theta$ components as $(\psi_1)_p = \overline{\Theta}(\psi_1)_\rho + \overline{\rho}(\psi_1)_\theta$. 
The real part of $(\psi_1)_p$ is shown as the red solid line in figure \ref{fig:supersonic_mode}(\subref{fig:supersonic_mode_comp_pline}). 
In the same figure the real part of the Mach wave obtained using \eqref{eqn:mach_wave} is also shown in black (the amplitude and phase of the Mach wave is fixed to be the same as the resolvent mode at some arbitrary wall-height, here $y=5$). 
To further compare, figure \ref{fig:supersonic_mode}(\subref{fig:supersonic_mode_comp_pfield}) shows the negative (blue) and the positive (red) pressure fluctuations from the resolvent mode $(\psi_1)_p$ in a streamwise wall-normal ($x$-$y$) plane at a fixed spanwise location (here $z=0$). 
The black line-contours represent the Mach wave from \eqref{eqn:mach_wave} (the contours represent $\pm0.5$ of the maximum value). 
From these figures, first we observe that, far from the wall, these resolvent modes are outgoing waves that radiate into the freestream. 
Additionally, we also note that, within the freestream the resolvent modes closely follow the behaviour of the inviscid Mach wave. 
This further suggests that the Mach waves are responsible for the high resolvent amplification in the wavenumber region where the freestream relative Mach number is greater than $1$, i.e.\ $\overline{Ma}(\infty)\geq 1$. 
For the sake of completeness, the velocity, density and temperature components of this mode are also shown in \ref{fig:supersonic_mode}(\subref{fig:supersonic_mode_comp_u}), and we see that these components also oscillate in the freestream. 

So far, $\overline{Ma}(y)\geq1$ just provides a condition for which we get modes that radiate into the freestream. 
The mechanism by which these modes cause high resolvent amplification is not obvious, and this will be discussed in \S\ref{sec:two routes of forcing resolvent Mach wave radiation}. 
Before that we will briefly consider: (i) the contribution of these modes to the freestream disturbances of the boundary layer in \S\ref{sec:contribution to the boundary layer and freestream} and (ii) the impact of viscosity on these resolvent modes in \S\ref{sec:the effect of viscosity on resolvent mach wave amplification}. 

\subsection{Contribution to the boundary layer and freestream}
\label{sec:contribution to the boundary layer and freestream}

\begin{figure}[t]
\captionsetup[subfigure]{labelformat=empty,skip=-10pt}
\begin{subfigure}[b]{\textwidth}
\centering
\includegraphics[width=0.8\textwidth, trim={0.5cm 0.8cm 8.6cm 8cm}, clip]{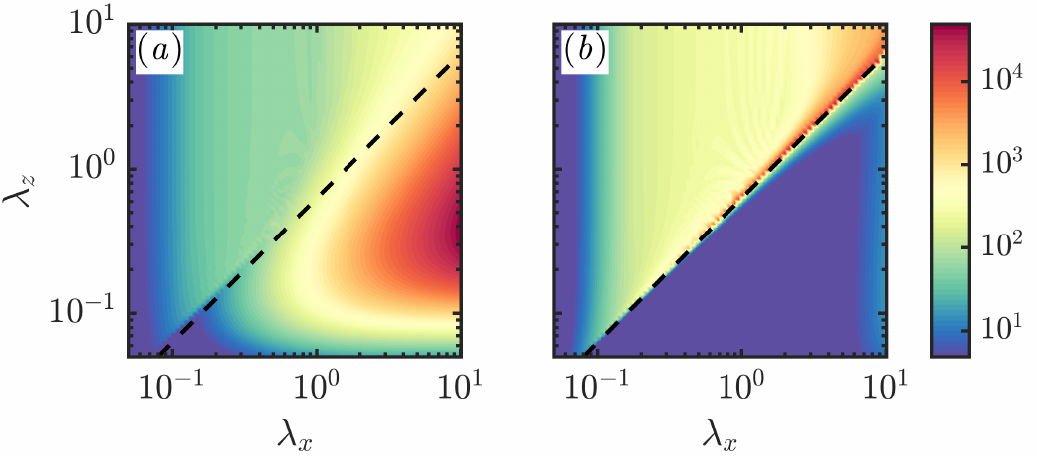}
\caption{}
\label{fig:BL}
\end{subfigure}%
\begin{subfigure}[b]{\textwidth}
\caption{}
\label{fig:FS}
\end{subfigure}
\vspace{-0.75cm}
\caption{The resolvent operator is masked to give a response (\textit{a}) within the boundary layer (i.e.\ $y\leq d_{99}$) and (\textit{b}) within the freestream (i.e.\ $y>d_{99}$), separately. 
The leading resolvent gain $\sigma_1$ from the masked operators is shown as a function of the streamwise and spanwise wavelengths $\lambda_x$ and $\lambda_z$ and for a fixed phase speed $c = \overline{U}(y^+\approx 15)$.  
The dashed black line represents the relative Mach equal to unity line.}
\label{fig:BLFS}
\end{figure}
A crucial detail about these Mach waves is that, within the wavelength range considered here, these resolvent Mach waves are the the only modes that contribute to the response in the freestream. 
To illustrate this, in figure \ref{fig:BLFS} the response to the masked resolvent is considered. 
The masking is done by introducing a weighting to the resolvent operator as done in \citet{nogueira2020resolvent} (see \S\ref{sec:resolvent operator}). 
To obtain figure \ref{fig:BLFS}(\subref{fig:BL}) the resolvent is masked such that it gives a response only within the boundary layer, i.e.\ in the region $y \leq \delta_{99}$, and for figure \ref{fig:BLFS}(\subref{fig:FS}) the response is restricted to the freestream, i.e.\ $y > \delta_{99}$. 
First, from figure \ref{fig:BLFS}(\subref{fig:BL}) we observe that the subsonic modes lie within the boundary layer, consistent with the observations in \S\ref{sec:comparing compressible and incompressible resolvent operators}. 
The contribution to the freestream in figure \ref{fig:BLFS}(\subref{fig:FS}) therefore comes solely from the Mach wave resolvent modes. 
Therefore the resolvent Mach waves are the only contributors to the freestream fluctuations from the resolvent, at least within the wavenumber space considered here. 
 
\subsection{The effect of viscosity on resolvent Mach wave amplification}
\label{sec:the effect of viscosity on resolvent mach wave amplification}

\begin{figure}[t]
\begin{subfigure}[b]{\textwidth}
\centering
\includegraphics[width=0.6\textwidth, trim={0cm 0cm 0.5cm 0cm}, clip]{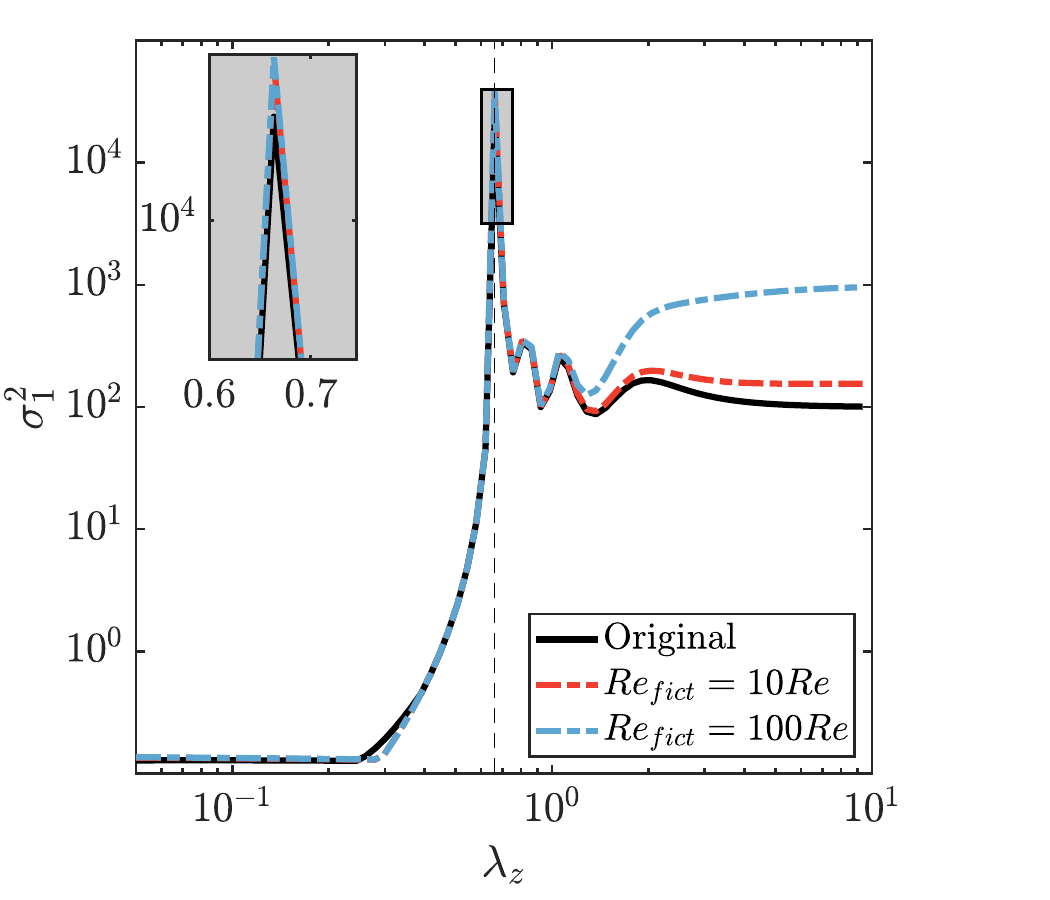}
\label{fig:RedVisc}
\end{subfigure}
\vspace{-0.75cm}
\caption{The leading resolvent amplification for modes with $\lambda_x=1$ and $c \approx \overline{U}(y^+ = 15)$ is shown as a function of $\lambda_z$. 
The solid line is obtained from the regular resolvent operator masked to give a response only within the freestream. 
The dashed lines are obtained from the fictional resolvent operators with viscosity artificially decreased to $1/10$ (red line) and $1/100$ (blue line) times the original value. 
The inset represents a zoomed in version of the boxed region in the plot. 
The dashed line represents $\overline{Ma}(\infty)=1$. }
\label{fig:RedVisc}
\end{figure}

From figure \ref{fig:Intro}(\subref{fig:energy_comp}), we see that the most linearly amplified supersonic modes are those that lie close to the $\overline{Ma}(\infty)=1$ line, and here we explore why this is the case. 
For the flows considered here, the maximum $\overline{Ma}(y)$ for the modes at the $\overline{Ma}(\infty)=1$ line is $1$ at $y=\delta_{99}$. 
Therefore, these modes have $\overline{Ma}(y)<1$ within the boundary layer, i.e.\ $\overline{Ma}(y<\delta_{99})<1$. 
Mach waves only start radiating from the wall-height where $\overline{Ma}(y) \geq 1$. 
Hence these modes near the $\overline{Ma}(\infty)=1$ line only start radiating at or near $y=\delta_{99}$, and they therefore do not have a presence within the boundary layer. 
These modes therefore exist in a region of the flow where the effect of viscosity is negligible. 
In contrast, supersonic modes that have $\overline{Ma}(\infty)>1$, i.e. fall much above the $\overline{Ma}(\infty)=1$ line, start radiating from within the boundary layer. 
These modes therefore experience the dampening effects of viscosity. 
This difference could explain why the most amplified supersonic modes lie close to the $\overline{Ma}(\infty)=1$ line. 

To illustrate this, in figure \ref{fig:RedVisc} we compare the responses from the regular resolvent operator (black solid line) to fictional resolvent operators with viscosity artificially decreased to $1/10$ (red dashed-dot line) and $1/100$ (blue dashed-dot line) times the original value (note that these operators are not physical and are used here only for illustrative purposes). 
We saw in \S\ref{sec:contribution to the boundary layer and freestream} that only the Mach waves contribute to the freestream response from the resolvent. 
Therefore, to focus solely on the Mach waves, here we `mask' the resolvent such that the response from the resolvent is restricted to lie only in the freestream (see \S\ref{sec:resolvent operator}). 
The vertical black dashed line represents $\overline{Ma}(\infty)=1$ and we are only interested in the region to the right of this line. 
First let us address the jagged response from the regular resolvent. 
Although mathematically we can have Mach waves at a continuous range of temporal frequencies \eqref{eqn:mach_wave}, because of our discretized operators, we only resolve a discrete set of these frequencies in our linear operator. 
The jagged response therefore comes about because of the proximity (or remoteness) of the selected $\omega$ to the nearest resolved frequency in the linear operator \citep{bae2020resolvent}.

From this figure, we see that the amplification of the mode at the $\overline{Ma}(\infty)=1$ line (see inset of the figure which zooms in on this mode) is not significantly effected by the reduction in viscosity. 
However, when $\overline{Ma}(\infty) > 1$, reducing the dampening effect of viscosity results in more amplified modes. 
This suggests that, the negligible effect of viscosity on the supersonic modes that lie at or close to the $\overline{Ma}(\infty)=1$ line is the reason why these modes are the most linearly amplified. 
These observations also suggests that, if in the future, the linearized equations augmented with an eddy-viscosity profile is used to study these freestream fluctuations \citep[e.g.][]{del2006linear, pujals2009note, hwang2010linear}, the amplification of modes with $\overline{Ma}(\infty) > 1$ could potentially be impacted, although their mode shapes within the freestream will likely not be affected (since eddy-viscosity in the freestream will be zero). 
However analysing the eddy-viscosity based operator is beyond the scope of the current manuscript and is left for the future. 

\section{Two routes of forcing resolvent Mach wave radiation}
\label{sec:two routes of forcing resolvent Mach wave radiation}

$\overline{Ma}(y)\geq1$ just provides a condition for which we get modes that radiate into the freestream. 
The mechanism by which these modes cause high resolvent amplification is not obvious. 
There are two routes, (i) direct and (ii) indirect, through which these Mach waves can cause high resolvent amplification, and here we will consider these two routes separately. 
In this section, two different supersonic resolvent modes from a $Ma=4$, $Re_\tau\approx400$ turbulent boundary layer over an adiabatic wall are taken as examples: 
(i) `MW1 (Mach wave 1)' with $\lambda_y=3.5$ in \S\ref{sec:direct route of forcing resolvent Mach wave radiation} (the {\tiny $\blacksquare$} in figure \ref{fig:Intro}(\subref{fig:energy_comp})) and 
(ii) `MW2' with $\lambda_y=10$ in \S\ref{sec:indirect route of forcing resolvent Mach wave radiation} (the $\bullet$ in figure \ref{fig:Intro}(\subref{fig:energy_comp})). 
Both the modes have the same $\lambda_x=5$ and $c \approx U(y^+=15)$. 
MW1 is chosen such that it falls close to the $\overline{Ma}(\infty)=1$ line, while MW2 is taken such that it lies further away from the $\overline{Ma}(\infty)=1$ line. 
This choice has two significant consequences.  
First, modes that lie close to the $\overline{Ma}(\infty)=1$ line have higher amplifications relative to other supersonic modes (see figure \ref{fig:Intro}(\subref{fig:energy_comp}) and \S\ref{sec:the effect of viscosity on resolvent mach wave amplification}). 
Therefore, MW1 has a higher resolvent amplification when compared to MW2. 
Second, the Mach wave radiates only when $\overline{Ma}(y) \geq 1$. 
For MW1, the freestream relative Mach number $\overline{Ma}(\infty)$ (i.e. the maximum value of $\overline{Ma}(y)$) is close to $1$, and therefore the mode starts radiating from near $y=\delta_{99}$ (see \S\ref{sec:the effect of viscosity on resolvent mach wave amplification}). 
On the other hand, MW2 has $\overline{Ma}(\infty)$ greater than $1$, and therefore the mode starts radiating from well within the boundary layer (see \S\ref{sec:the effect of viscosity on resolvent mach wave amplification}). 
We will see that this difference impacts the forcing mechanisms active for these modes. 
In this section: (1) the `direct route' of forcing the Mach waves will be discussed in \S\ref{sec:direct route of forcing resolvent Mach wave radiation} using MW1 and (2) the `indirect route' in \S\ref{sec:indirect route of forcing resolvent Mach wave radiation} using MW2. 
In \S\ref{sec:contribution of the direct and indirect forcing mechanisms}, we will look at the relative contributions of these two forcing routes across $(\lambda_x,\lambda_y,c)$. 

\subsection{Direct route of forcing resolvent Mach wave radiation}
\label{sec:direct route of forcing resolvent Mach wave radiation}

To probe the direct forcing route, consider the equations for the Mach waves \eqref{eqn:inviscid_pressure} in the freestream. 
Consider the inverse of the resolvent operator $L=H^{-1}$, where $L\widehat{\bm{q}}=\widehat{\bm{f}}$ \eqref{eqn:resolvent}. 
The SVD of this operator gives $L = \sum_{i=1}^{5N} \bm{\phi'}_i \sigma'_i \bm{\psi'}_i$. 
For the inviscid case, solutions from \eqref{eqn:mach_wave} $\widehat{p}_M(y)$ are solutions of L. 
Another way of saying this is that a singular vector of $L$, with a corresponding singular value of $\sigma'_i=0$, has pressure equal to $\widehat{p}_M(y)$. 

For the case of finite Reynolds numbers considered here, viscosity will damp these modes. 
Therefore, singular vectors with pressure $\widehat{p}_M(y)$ will have singular values $\sigma'_i$ close to $0$, but not $0$. 
Let us now consider the SVD of the resolvent operator $H$ in terms of the SVD of $L=H^{-1}$. 
This gives $H = \sum_{i=1}^{5N} \bm{\psi'}_i (1/\sigma'_i) \bm{\phi'}_i$. 
There is $1/\sigma'_i$ that appears in the SVD of $H$. 
Therefore, singular vectors of $L$ with pressure $\widehat{p}_M(y)$ which have $\sigma'_i$ close to $0$, will now appear as highly amplified singular vectors of $H$. 
And this is how the Mach waves $\widehat{p}_M(y)$ cause high resolvent amplification. 
Note that this mechanism can amplify $(\lambda_x,\lambda_y)$ modes that are linearly stable. 
In figure \ref{fig:Intro}(\subref{fig:energy_comp}) the green contour lines show the region of the $(\lambda_x,\lambda_y)$ space that is linearly unstable, and we observe that there are supersonic modes outside of this region, i.e.\ that are linearly stable, but still have high resolvent amplification. 

\begin{figure}[t]
\captionsetup[subfigure]{labelformat=empty,skip=-5pt}
\begin{subfigure}[b]{\textwidth}
\centering
\includegraphics[width=\textwidth, trim={2.7cm 1.2cm 3cm 1.1cm}, clip]{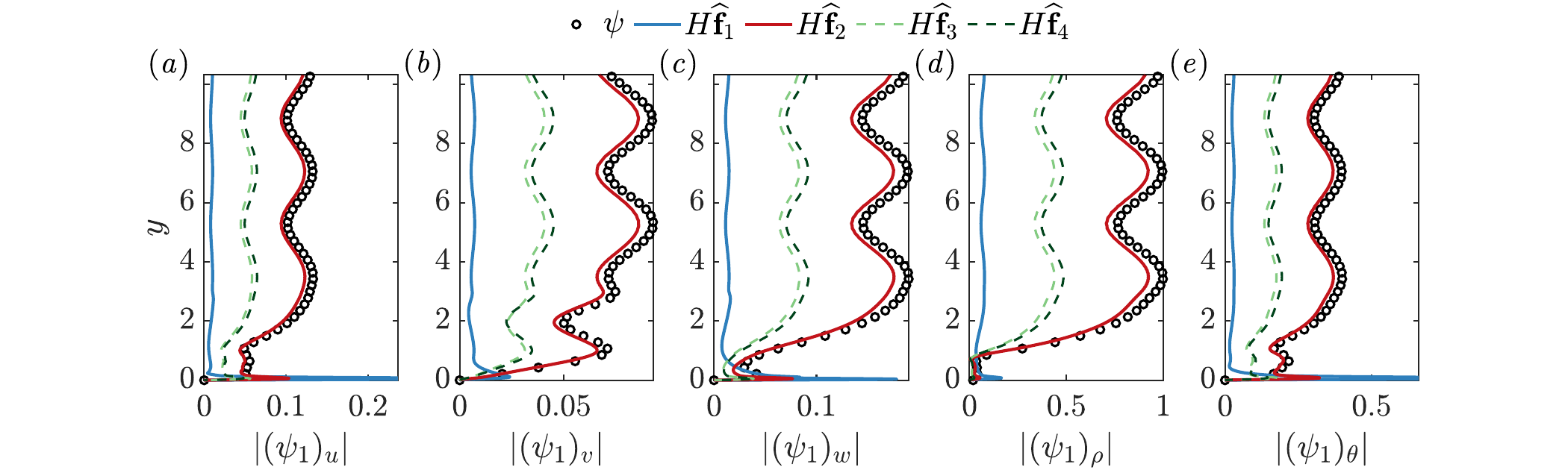}
\caption{}
\label{fig:fHelmholtz_mode_supersonic_u}
\end{subfigure}%
\begin{subfigure}[b]{\textwidth}
\caption{}
\label{fig:fHelmholtz_mode_supersonic_v}
\end{subfigure}%
\begin{subfigure}[b]{\textwidth}
\caption{}
\label{fig:fHelmholtz_mode_supersonic_w}
\end{subfigure}%
\begin{subfigure}[b]{\textwidth}
\caption{}
\label{fig:fHelmholtz_mode_supersonic_r}
\end{subfigure}%
\begin{subfigure}[b]{\textwidth}
\caption{}
\label{fig:fHelmholtz_mode_supersonic_t}
\end{subfigure} 
\begin{subfigure}[b]{\textwidth}
\includegraphics[width=\textwidth, trim={2.7cm 0.05cm 3cm 1.1cm}, clip]{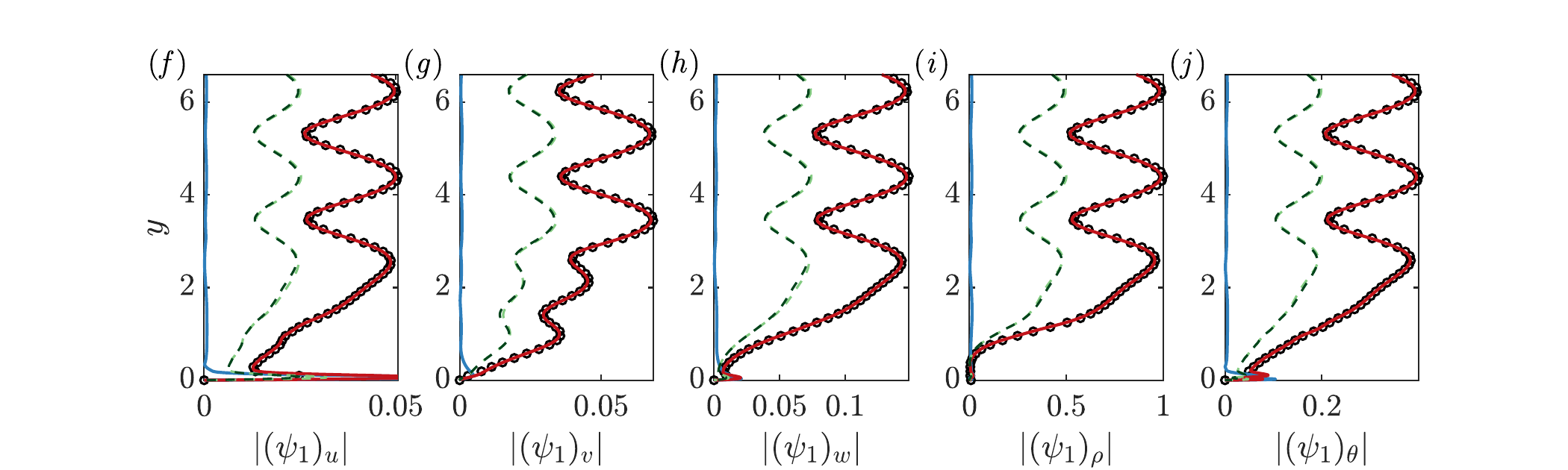}
\caption{}
\label{fig:fHelmholtz_mode_supersonic_cooled_u}
\end{subfigure}%
\begin{subfigure}[b]{\textwidth}
\caption{}
\label{fig:fHelmholtz_mode_supersonic_cooled_v}
\end{subfigure}%
\begin{subfigure}[b]{\textwidth}
\caption{}
\label{fig:fHelmholtz_mode_supersonic_cooled_w}
\end{subfigure}%
\begin{subfigure}[b]{\textwidth}
\caption{}
\label{fig:fHelmholtz_mode_supersonic_cooled_r}
\end{subfigure}%
\begin{subfigure}[b]{\textwidth}
\caption{}
\label{fig:fHelmholtz_mode_supersonic_cooled_t}
\end{subfigure}
\vspace{-0.75cm}
\caption{The response of the resolvent operator to the full leading resolvent forcing mode $\widehat{\bm{f}}=\bm{\phi}_1$ (black) as well as the response to the two components of the forcing $\widehat{\bm{f}}_1$ (blue) and $\widehat{\bm{f}}_2$ (red) are shown. 
Further, the contribution of $\widehat{\bm{f}}_{\bm{u}}^d$ alone is shown in a lighter shade of green and that of ($f_\rho$,$f_\theta$) is shown in a darker shade of green. 
Two compressible boundary layers are considered: (\textit{a}-\textit{e}) $Ma=4$, $Re_\tau=400$ over an adiabatic wall and (\textit{f}-\textit{j}) $Ma=6$, $Re_\tau=450$ with $\Theta_w/\Theta_{ad}=0.25$. 
Modes that fall close to the relative Mach equal to unity lines are shown: mode $(\lambda_x,\lambda_z,c)=(5,3.5,\overline{U}(y^+\approx 15))$ for the $Ma=4$ case (the mode indicated by the ({\tiny $\blacksquare$}) in figure \ref{fig:Intro}(\subref{fig:energy_comp})) and mode $(\lambda_x,\lambda_z,c)=(5,1.6,\overline{U}(y^+\approx 15))$ for the $Ma=6$ case. 
The (\textit{a,f}) streamwise, (\textit{b,g}) wall-normal and (\textit{c,h}) spanwise velocities as well as the (\textit{d,i}) density and (\textit{e,j}) temperature are shown. }
\label{fig:fHelmholtz_mode_supersonic}
\end{figure}

From \eqref{eqn:inviscid_pressure}, we see that, within the freestream, the pressure cannot be forced by $\widehat{v}$ (since all the terms within the square brackets are zero in the freestream). 
Therefore, only $\nabla \cdot \widehat{\bm{f}}_{\bm{u}}$, $\widehat{f}_\theta$ or $\widehat{f}_\rho$ can excite the waves. 
Now consider the Helmholtz decomposition of $\widehat{\bm{f}}_{\bm{u}}$ that gives the solenoidal $\widehat{\bm{f}}_{\bm{u}}^s$ and dilatational $\widehat{\bm{f}}_{\bm{u}}^d$ components. 
Since, by the definition of the Helmholtz decomposition, $\nabla \cdot \widehat{\bm{f}}_{\bm{u}}^s = 0$, only the dilatational component $\widehat{\bm{f}}_{\bm{u}}^{d}$ can force these modes. 
To excite the Mach waves in the freestream, we therefore need a forcing in $\widehat{\bm{f}}_{\bm{u}}^{d}$, $\widehat{f}_\theta$ or $\widehat{f}_\rho$.  

To illustrate this, in figures \ref{fig:fHelmholtz_mode_supersonic}(\subref{fig:fHelmholtz_mode_supersonic_u}-\subref{fig:fHelmholtz_mode_supersonic_t}) we consider the supersonic mode MW1 (modes MW1 and MW2 are defined at the beginning of \S\ref{sec:two routes of forcing resolvent Mach wave radiation}).  
The five components of the full leading resolvent response $\bm{\psi}_1$ are shown in black (suboptimal modes are considered in appendix \ref{sec: Sub-optimal modes}). 
The forcing $\widehat{\bm{f}}$ to this resolvent mode is the leading resolvent forcing mode $\bm{\phi}_1$. 
In red is the response to $\widehat{\bm{f}}_{\bm{u}}^d$, $\widehat{f}_\rho$ and $\widehat{f}_\theta$, i.e.\ the forcing to the resolvent is $\widehat{\bm{f}}_2=(\widehat{f}^d_u, \widehat{f}^d_v, \widehat{f}^d_w, \widehat{f}_\rho, \widehat{f}_\theta)$. 
In blue is the response to the remaining components of the forcing $\widehat{\bm{f}}_{\bm{u}}^s$, i.e.\ the resolvent forcing is $\widehat{\bm{f}}_1=(\widehat{f}^s_u, \widehat{f}^s_v, \widehat{f}^s_w, 0, 0)$. 
From figure \ref{fig:fHelmholtz_mode_supersonic}(\subref{fig:fHelmholtz_mode_supersonic_u}-\subref{fig:fHelmholtz_mode_supersonic_t}), we see that $\widehat{\bm{f}}_2$ is responsible for capturing the majority of the energy, and $\widehat{\bm{f}}_1$ plays an insignificant role for these modes. 
(The reason for why the contribution of $\widehat{\bm{f}}_1$ in figures \ref{fig:fHelmholtz_mode_supersonic}(\subref{fig:fHelmholtz_mode_supersonic_u}-\subref{fig:fHelmholtz_mode_supersonic_t}), although small, is still non-zero, is explained in \S\ref{sec:indirect route of forcing resolvent Mach wave radiation}). 
Further, the contribution of $\widehat{\bm{f}}_{\bm{u}}^d$ alone is shown in a lighter shade of green and that of ($f_\rho$,$f_\theta$) is shown in a darker shade of green. 
For the mode MW1, we note that both these components contribute almost equally to the response. 

To access the more general applicability of the discussions here, in figures \ref{fig:fHelmholtz_mode_supersonic}(\subref{fig:fHelmholtz_mode_supersonic_cooled_u}-\subref{fig:fHelmholtz_mode_supersonic_cooled_t}), we also look at a supersonic mode for a $Re_\tau=450$ boundary layer flow at a higher Mach number of $Ma=6$ and over a cooled wall with $\Theta_w/\Theta_{ad}=0.25$. 
Here again we choose a mode that falls close to the relative Mach equal to unity line which for $c \approx U(y^+=15)$ and $\lambda_x=5$ corresponds to $\lambda_z=1.6$. 
We observe similar trends as in figure \ref{fig:fHelmholtz_mode_supersonic}(\subref{fig:fHelmholtz_mode_supersonic_u}-\subref{fig:fHelmholtz_mode_supersonic_t}) where $\widehat{\bm{f}}_2$ amplifies the mode. Therefore wall cooling does not affect the trends discussed here. 

\subsection{Indirect route of forcing resolvent Mach wave radiation}
\label{sec:indirect route of forcing resolvent Mach wave radiation}

\begin{figure}[t]
\captionsetup[subfigure]{labelformat=empty,skip=-5pt}
\begin{subfigure}[b]{\textwidth}
\centering
\includegraphics[width=\textwidth, trim={2.7cm 0.05cm 3cm 1.1cm}, clip]{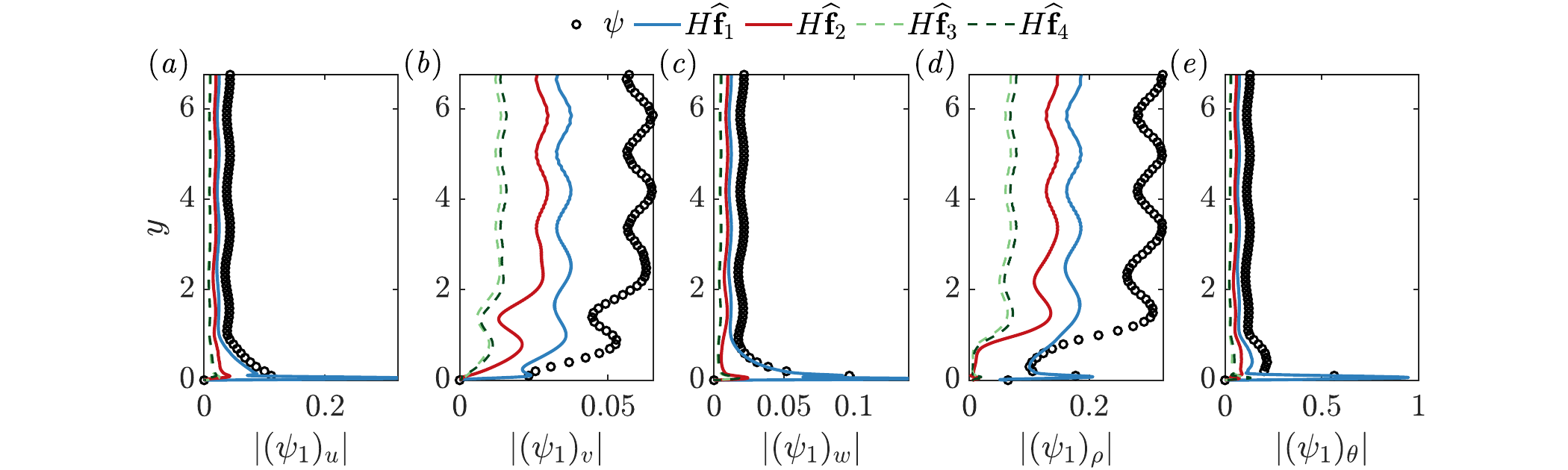}
\caption{}
\label{fig:fHelmholtz_mode_supersonic_sol_u}
\end{subfigure}%
\begin{subfigure}[b]{\textwidth}
\caption{}
\label{fig:fHelmholtz_mode_supersonic_sol_v}
\end{subfigure}%
\begin{subfigure}[b]{\textwidth}
\caption{}
\label{fig:fHelmholtz_mode_supersonic_sol_w}
\end{subfigure}%
\begin{subfigure}[b]{\textwidth}
\caption{}
\label{fig:fHelmholtz_mode_supersonic_sol_r}
\end{subfigure}%
\begin{subfigure}[b]{\textwidth}
\caption{}
\label{fig:fHelmholtz_mode_supersonic_sol_t}
\end{subfigure} 
\vspace{-0.75cm}
\caption{The response of the resolvent operator to the full leading resolvent forcing mode $\widehat{\bm{f}}=\bm{\phi}_1$ (black) as well as the response to the two components of the forcing $\widehat{\bm{f}}_1$ (blue) and $\widehat{\bm{f}}_2$ (red) are shown. 
Further, the contribution of $\widehat{\bm{f}}_{\bm{u}}^d$ alone is shown in a lighter shade of green and that of ($f_\rho$,$f_\theta$) is shown in a darker shade of green. 
A supersonic mode with $(\lambda_x,\lambda_z,c)=(5,10,\overline{U}(y^+\approx 15))$ for a compressible boundary layer over an adiabatic wall with $Ma=4$, $Re_\tau=400$ is considered (the mode indicated by the ({$\bullet$}) in figure \ref{fig:Intro}(\subref{fig:energy_comp})). 
The (\textit{a}) streamwise, (\textit{b}) wall-normal and (\textit{c}) spanwise velocities as well as the (\textit{d}) density and (\textit{e}) temperature are shown. }
\label{fig:fHelmholtz_mode_supersonic_sol}
\end{figure}

So far we considered mode MW1 and found that mainly the forcing components $\widehat{\bm{f}}_{\bm{u}}^{d}$, $\widehat{f}_\theta$ or $\widehat{f}_\rho$ excites the mode. 
Let us now consider mode MW2 where this is not the case. 
Figure \ref{fig:fHelmholtz_mode_supersonic_sol} is the equivalent of figures \ref{fig:fHelmholtz_mode_supersonic}(\subref{fig:fHelmholtz_mode_supersonic_u}-\subref{fig:fHelmholtz_mode_supersonic_t}), but for mode MW2. 
The five components of the full leading resolvent mode of MW2 is shown in black, along with the response to $\widehat{\bm{f}}_1$ in blue and $\widehat{\bm{f}}_2$ in red (the definitions of $\widehat{\bm{f}}_1$ and $\widehat{\bm{f}}_2$ are the same as in \S\ref{sec:direct route of forcing resolvent Mach wave radiation}). 
Unlike for MW1, for MW2 we find that $\widehat{\bm{f}}_1$ (blue line), and therefore the solenoidal component of the forcing to the momentum equations $\widehat{\bm{f}}_{\bm{u}}^s$, plays a significant role in capturing the mode. 
The question then is, why does $\widehat{\bm{f}}_{\bm{u}}^s$ excite MW2, but not MW1? 

To understand this, first, from \eqref{eqn:inviscid_pressure}, we note that wall-normal velocity $\widehat{v}$ can excite the resolvent Mach waves. 
Second, from figure \ref{fig:Intro}(\subref{fig:energy_comp}), we note that both MW1 and MW2 
lie within the grey contour line, which represents the region where incompressible mechanisms are reasonably active. 
Therefore, a possible route through which $\widehat{\bm{f}}_{\bm{u}}^s$ can excite Mach waves is: incompressible-like mechanisms which causes $\widehat{\bm{f}}_{\bm{u}}^s$ to excite a response in $\widehat{v}$, and this $\widehat{v}$ then exciting the Mach waves. 
Here we will call this the indirect route of forcing the Mach waves. 
However, this mechanism still does not explain the differences between the trends in the forcing to MW1 and MW2, specifically why the indirect route is prominent for MW2 and not for MW1. 

\begin{figure}[t]
\captionsetup[subfigure]{labelformat=empty,skip=-20pt}
\begin{subfigure}[b]{\textwidth}
\centering
\includegraphics[width=0.85\textwidth, trim={8.5cm 3.9cm 1.2cm 7.6cm}, clip]{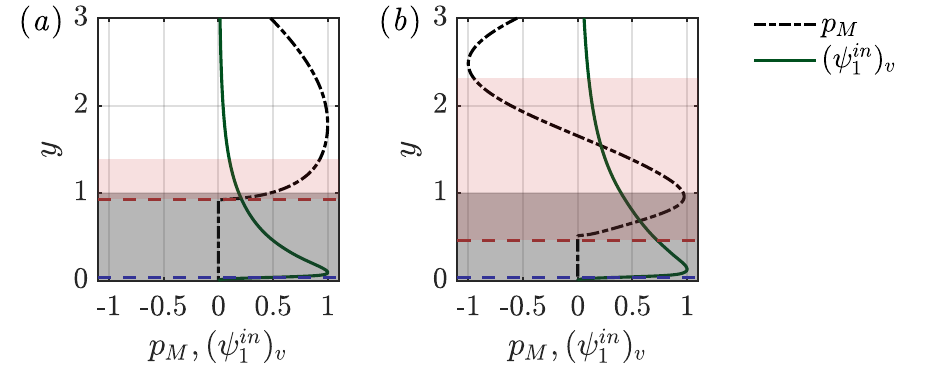}
\caption{}
\label{fig:NormalVsNonnormal_nonnormal_mode1}
\end{subfigure}
\begin{subfigure}[b]{\textwidth}
\caption{}
\label{fig:NormalVsNonnormal_nonnormal_mode2}
\end{subfigure}
\vspace{-0.75cm}
\caption{The analytical Mach wave $\widehat{p}_M(y)$  \eqref{eqn:inviscid_pressure} and $(\psi^{in}_1)_{v}(y)$ the wall-normal velocity obtained from the incompressible resolvent with the compressible mean $H^{in}$ are shown for the modes (\textit{a}) MW1 and (\textit{b}) MW2. 
The grey shaded region represents the region within the boundary layer. 
The red-shaded region represents the overlap between $\widehat{p}_M(y)$ and $(\psi^{in}_1)_{v}(y)$, where $\overline{Ma}(y) \geq 1$ and therefore $\widehat{p}_M(y)$ exists, and $(\psi^{in}_1)_{v}(y)$ is at least $10\%$ of it's maximum value. }
\label{fig:NormalVsNonnormal}
\end{figure}

To probe this, in figure \ref{fig:NormalVsNonnormal}(\subref{fig:NormalVsNonnormal_nonnormal_mode1}) we consider MW1 and in figure \ref{fig:NormalVsNonnormal}(\subref{fig:NormalVsNonnormal_nonnormal_mode2}) we consider MW2. 
Two separate profiles are plotted in these figures. 
The first profile $\widehat{p}_M(y)$ is obtained from the analytic expression for the Mach wave radiation \eqref{eqn:mach_wave}, and indicates the wall-normal region where these Mach waves exist. 
The red-dashed line represents the wall-height above which the relative Mach number $\overline{Ma}(y)\geq 1$, and therefore $\widehat{p}_M(y)$ starts radiating only from this wall-height.
To compute the second profile $(\psi^{in}_1)_{v}(y)$, we consider a fictional incompressible resolvent operator $H^{in}$ (i.e.\ the resolvent operator at $Ma=0$), but with the mean velocity profile $\overline{U}$ from the compressible case. 
In effect, $H^{in}$ captures the mechanisms that would hypothetically be active in a $Ma=0$ flow that maintains the compressible mean velocity profile ($H^{in}$ is not physically relevant and is used here only for illustration). 
In other words, $H^{in}$ should be able to approximately capture the incompressible mechanisms in this flow. 
The profile $(\psi^{in}_1)_{v}(y)$ is the wall-normal velocity component of the leading resolvent mode obtained from $H^{in}$. 
Finally, the red-shaded region represents the overlap between $\widehat{p}_M(y)$ and $(\psi^{in}_1)_{v}(y)$ (in this region $\overline{Ma}(y) \geq 1$ and therefore $\widehat{p}_M(y)$ exists, and $(\psi^{in}_1)_{v}(y)$ is at least $10\%$ of it's maximum value). 
We see that, since MW2 starts radiating from well within the boundary layer, this region of overlap is significantly higher for the mode.  
The $\widehat{v}$ generated through incompressible-like mechanisms forced by $\widehat{\bm{f}}_{\bm{u}}^s$ can therefore much more efficiently drive the Mach waves in MW2, in comparison to MW1. 
This provides a possible explanation for why $\widehat{\bm{f}}_{\bm{u}}^s$ plays a significant role in exciting MW2. 
The relatively small overlap observed in the case of MW1 also explains why the solenoidal component gives a small but non-zero response (blue line) in figures \ref{fig:fHelmholtz_mode_supersonic}(\subref{fig:fHelmholtz_mode_supersonic_u}-\subref{fig:fHelmholtz_mode_supersonic_t}).  

From this we see that there are two factors that are responsible for this indirect forcing. (1) The modes should have active incompressible-like mechanisms. This is true for both MW1 and MW2 and can be measured by the leading singular value of $H^{in}$, $\sigma^{in}_{1}(\lambda_x,\lambda_y,c)$. (2) And there should be an overlap between $\widehat{p}_M(y)$ and $(\psi^{in}_1)_{v}(y)$. This is more significant for MW2 and can be measured using a projection of $(\psi^{in}_1)_{v}$ onto $\widehat{p}_M(y)$ defined as:
\begin{equation}
\eta(\lambda_x,\lambda_y,c) = \frac{ \left( \int_0^{y_{f}}\widehat{p}_M(y) (\psi^{in}_1)_{v}^*(y) \right)^2}{ \int_0^{y_{f}} \widehat{p}_M(y) \widehat{p}_M^*(y) \int_0^{y_{f}} (\psi^{in}_1)_{v}(y) (\psi^{in}_1)_{v}^*(y) }.
\end{equation}
Therefore, if this mechanism is indeed responsible for amplification, then the mode should have a higher value of $(\sigma^{in}_{1})^2 \eta$. 
The wall-height $y_f$ is the height till which we force the resolvent and here $y_f=3$ (see \S\ref{sec:Numerical set up for the resolvent operator}). 
The value of $(\sigma^{in}_{1})^2 \eta$ is approximately $102$ for MW1 and $202$ for MW2, showing that the indirect route is more active for MW2 (although computing this number shows that this route is not completely absent for MW1, but just less significant). 
We will use this metric $(\sigma^{in}_{1})^2 \eta$ in \S\ref{sec:contribution of the direct and indirect forcing mechanisms}. 

\section{Contribution of the direct and indirect forcing mechanisms}
\label{sec:contribution of the direct and indirect forcing mechanisms}

So far, we have considered a decomposition of the forcing to the resolvent into two components: (1) $\widehat{\bm{f}}_1=(\widehat{f}^s_u, \widehat{f}^s_v, \widehat{f}^s_w, 0, 0)$ which contains the solenoidal component of the forcing to the momentum equations and (2) $\widehat{\bm{f}}_2=(\widehat{f}^d_u, \widehat{f}^d_v, \widehat{f}^d_w, \widehat{f}_\rho, \widehat{f}_\theta)$ that contains the dilatational component of the forcing to the momentum equations as well as the forcing to the density and temperature equations. 
Here a Helmholtz decomposition of the leading resolvent forcing mode $\widehat{\bm{f}}_{\bm{u}} = \bm{\phi}_1$ is giving the solenoidal component $\widehat{\bm{f}}_{\bm{u}}^s = \bm{\phi}_1^s$ and the dilatational component $\widehat{\bm{f}}_{\bm{u}}^d = \bm{\phi}_1^d$ of the forcing. 
Subsonic modes are forced by $\widehat{\bm{f}}_1$.  
On the other hand, there are two routes through which the supersonic modes can be forced: the direct route where $\widehat{\bm{f}}_2$ is active and the indirect route where $\widehat{\bm{f}}_1$ plays a significant role. 
So far we have only looked at individual $(\lambda_x,\lambda_y,c)$ modes and the purpose of this section is to analyze the relative importance of $\widehat{\bm{f}}_1$ and $\widehat{\bm{f}}_2$ over a range of $(\lambda_x, \lambda_y,c)$. 

\begin{figure} 
\captionsetup[subfigure]{labelformat=empty,skip=-75pt}
\begin{subfigure}[b]{\textwidth}
\centering
\includegraphics[width=\textwidth, trim={0.25cm 0.25cm 12.7cm 0.25cm}, clip]{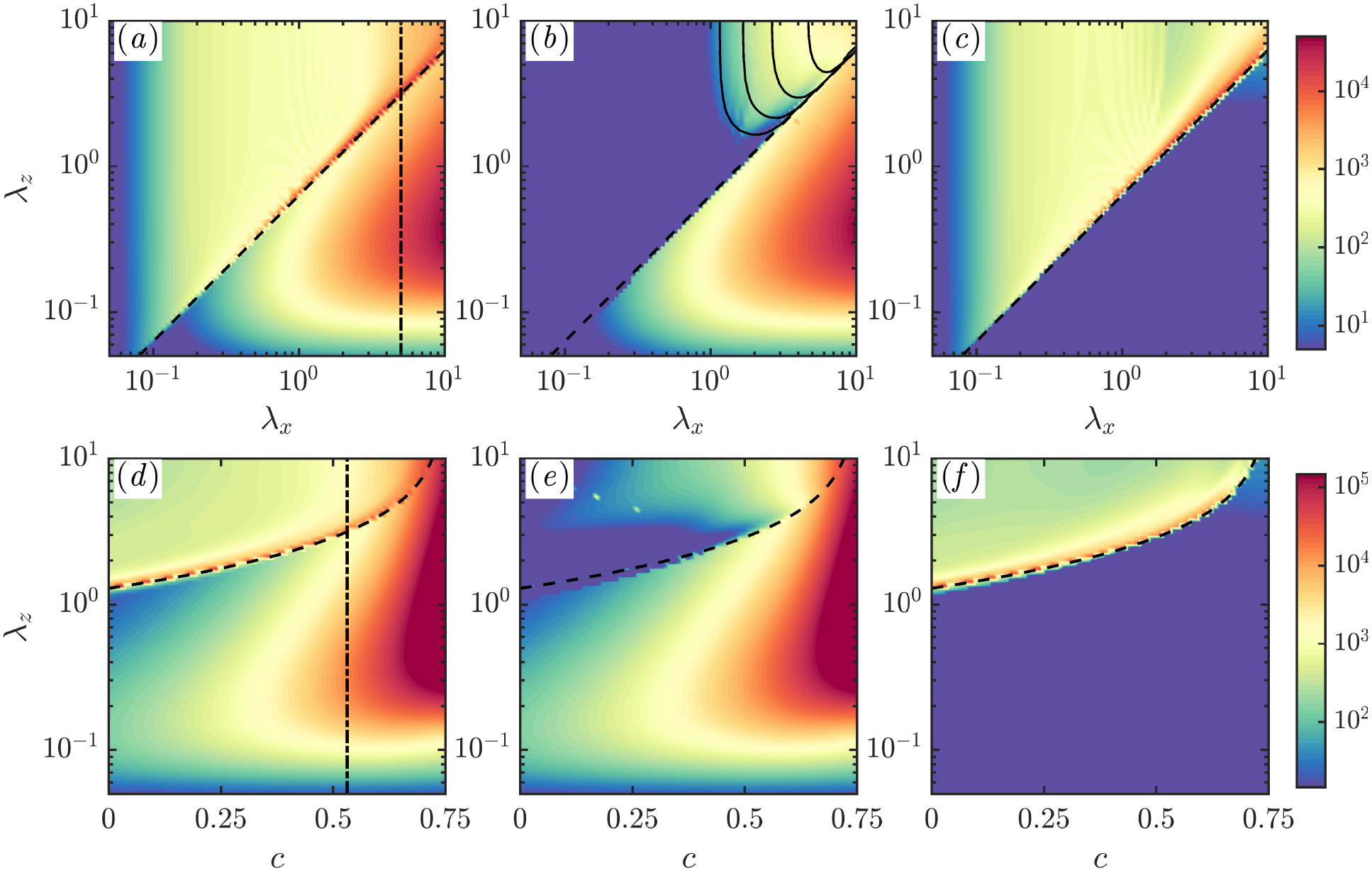}
\caption{}
\label{fig:fsfd_kxky_1}
\end{subfigure}
\begin{subfigure}[b]{\textwidth}
\caption{}
\label{fig:fsfd_kxky_2}
\end{subfigure}
\begin{subfigure}[b]{\textwidth}
\caption{}
\label{fig:fsfd_kxky_3}
\end{subfigure}
\begin{subfigure}[b]{\textwidth}
\caption{}
\label{fig:fsfd_cky_1}
\end{subfigure}
\begin{subfigure}[b]{\textwidth}
\caption{}{}
\label{fig:fsfd_cky_2}
\end{subfigure}
\begin{subfigure}[b]{\textwidth}
\caption{}
\label{fig:fsfd_cky_3}
\end{subfigure}
\caption{The Chu norm of the response of the resolvent operator to (\textit{a,d}) the full leading resolvent forcing mode $\bm{\phi}_1$ as well as the two components of the forcing (\textit{b,e}) $\widehat{\bm{f}}_1$ and (\textit{c,f}) $\widehat{\bm{f}}_2$ are shown in two ways: (\textit{a}-\textit{c}) as a function of the streamwise and spanwise wavelengths $(\lambda_x,\lambda_z)$ at a fixed phase-speed $c=\overline{U}(y^+\approx 15)$ and (\textit{d}-\textit{f}) as a function of phase-speeds and spanwise wavelengths (c,$\lambda_z$) at a fixed streamwise wavelengths $\lambda_x=5$. 
The vertical dashed-dot line in (\textit{a}) indicates $\lambda_x=5$ in (\textit{d}) indicates $c=\overline{U}(y^+\approx 15)$. 
The solid black contours in (\textit{b}) indicates contours of $-1:1:2$ contours of $\log_{10}((\sigma^{in}_{1})^2 \eta)$. 
The dashed contour lines indicate the relative Mach equal to unity. }
\label{fig:fsfd}
\end{figure}

In figure \ref{fig:fsfd}, for the case of a turbulent boundary layer at $Ma=4$ and $Re_\tau=400$, the full resolvent response in the first column (figures \ref{fig:fsfd}(\subref{fig:fsfd_kxky_1},\subref{fig:fsfd_cky_1})) is compared to responses to $\widehat{\bm{f}}_1$ in the second column (figures \ref{fig:fsfd}(\subref{fig:fsfd_kxky_2},\subref{fig:fsfd_cky_2})) and $\widehat{\bm{f}}_2$ in the third column (figures \ref{fig:fsfd}(\subref{fig:fsfd_kxky_3},\subref{fig:fsfd_cky_3})). 
The first row of the figure (figures \ref{fig:fsfd}(\subref{fig:fsfd_kxky_1}-\subref{fig:fsfd_kxky_3})) shows the responses as a function of ($\lambda_x$,$\lambda_z$) at a fixed phase-speed of $c=U(y^+\approx15)$. 
To consider a range of $c$, the second row of the figure (figures \ref{fig:fsfd}(\subref{fig:fsfd_cky_1}-\subref{fig:fsfd_cky_3})) shows the responses as a function of $c$ and $\lambda_z$, at a fixed value of $\lambda_x=5$. 
(The values of $c$ in figures \ref{fig:fsfd}(\subref{fig:fsfd_cky_1}-\subref{fig:fsfd_cky_3}) are taken only till $0.75$, since the amplification of the subsonic streaks with $c>0.75$ is too high to be clearly depicted along with the lesser amplified supersonic modes). 
The black contour lines represent $-1:1:2$ contours of $\log_{10}((\sigma^{in}_{1})^2 \eta)$ which, as seen in \S\ref{sec:indirect route of forcing resolvent Mach wave radiation}, indicates the region where (i) incompressible-like mechanisms are active and (ii) $(\psi^{in}_1)_{v}$ projects onto $\widehat{p}_M(y)$ (see \S\ref{sec:indirect route of forcing resolvent Mach wave radiation}). 
The black dashed lines in the figures (or curves in the case of figures \ref{fig:fsfd}(\subref{fig:fsfd_cky_1}-\subref{fig:fsfd_cky_3})) indicates relative Mach equal to unity $\overline{Ma}(\infty)=1$. 
Note that the colour-scales in this figure are logarithmic. 

\begin{figure} 
\captionsetup[subfigure]{labelformat=empty,skip=-75pt}
\begin{subfigure}[b]{\textwidth}
\centering
\includegraphics[width=\textwidth, trim={0.25cm 1.25cm 12.7cm 9.5cm}, clip]{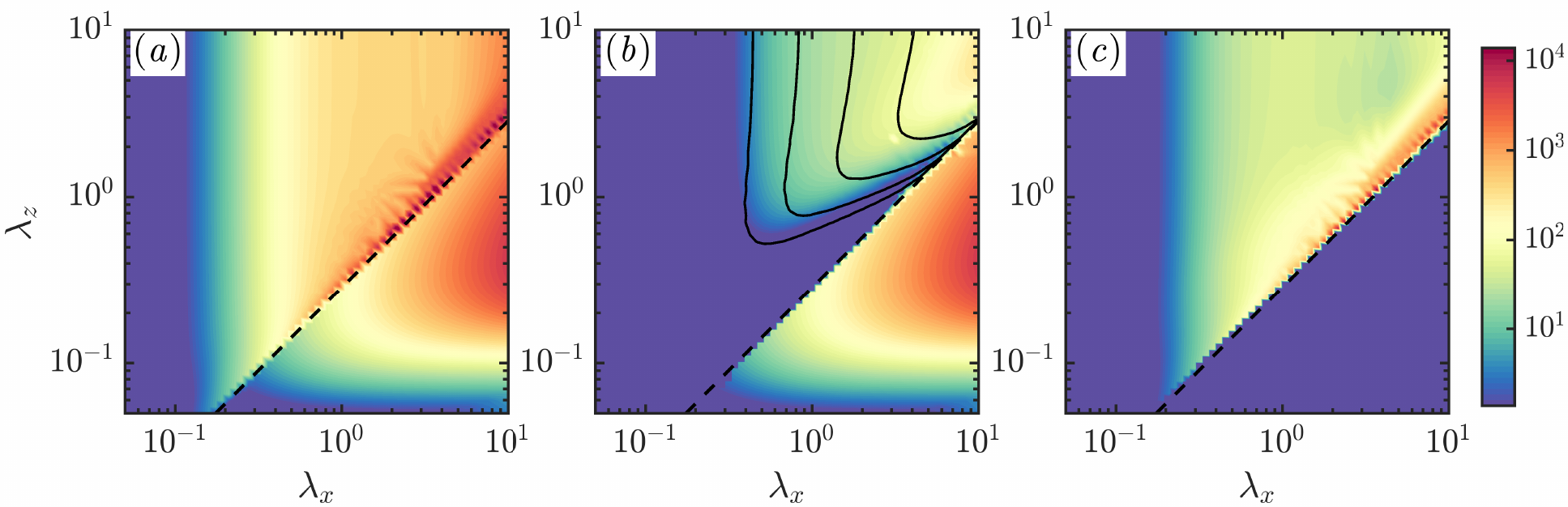}
\caption{}
\label{fig:AllCases_1}
\end{subfigure}
\begin{subfigure}[b]{\textwidth}
\caption{}
\label{fig:AllCases_2}
\end{subfigure}
\begin{subfigure}[b]{\textwidth}
\caption{}
\label{fig:AllCases_3}
\end{subfigure}
\begin{subfigure}[b]{\textwidth}
\caption{}
\label{fig:AllCases_4}
\end{subfigure}
\begin{subfigure}[b]{\textwidth}
\caption{}{}
\label{fig:AllCases_5}
\end{subfigure}
\begin{subfigure}[b]{\textwidth}
\caption{}
\label{fig:AllCases_6}
\end{subfigure}
\caption{A turbulent boundary layers with $Ma=6$, $Re_\tau=450$ and $\Theta_w/\Theta_{ad}=0.25$ are considered. The Chu norm of the response of the resolvent operator to (\textit{a}) the full leading resolvent forcing mode $\bm{\phi}_1$ as well as the two components of the forcing (\textit{b}) $\widehat{\bm{f}}_1$ and (\textit{c}) $\widehat{\bm{f}}_2$ are shown as a function of the streamwise ($\lambda_x$) and spanwise ($\lambda_z$) wavelengths for a fixed value of phase speed $c=\overline{U}(y^+\approx 15)$. 
The dashed lines indicate the relative Mach equal to unity. }
\label{fig:AllCases}
\end{figure}

From figures \ref{fig:fsfd}(\subref{fig:fsfd_kxky_2},\subref{fig:fsfd_cky_2}) we observe that $\widehat{\bm{f}}_1$ captures the responses in the subsonic modes, i.e.\ the responses in the region below the $\overline{Ma}(\infty)=1$ line. 
On the other hand, figures \ref{fig:fsfd}(\subref{fig:fsfd_kxky_3},\subref{fig:fsfd_cky_3})  shows that $\widehat{\bm{f}}_2$ captures the majority of the response in the supersonic modes, i.e.\ the responses in the region above the $\overline{Ma}(\infty)=1$ line. 
This is especially true when considering the most amplified supersonic modes that are near the $\overline{Ma}(\infty)=1$ line, consistent with observations in \S\ref{sec:two routes of forcing resolvent Mach wave radiation}.  
Also in line with \S\ref{sec:two routes of forcing resolvent Mach wave radiation}, from figures \ref{fig:fsfd}(\subref{fig:fsfd_kxky_2},\subref{fig:fsfd_cky_2}) we note that there is a small set of supersonic modes where the contribution of $\widehat{\bm{f}}_1$ is not zero. 
The black contour lines in figure \ref{fig:fsfd}(\subref{fig:fsfd_kxky_2}) follow the trends of these modes, which suggests that an incompressible-like mechanism is active, and this mechanism can force the Mach waves. 
Therefore, the indirect route of forcing (\S\ref{sec:indirect route of forcing resolvent Mach wave radiation}) is active for these modes. 

In figure \ref{fig:AllCases} we consider a $Ma=6$, $Re_\tau=450$ turbulent boundary layer over a cooled wall with wall-cooling ratio $\Theta_w/\Theta_{ad}=0.25$, with figures \ref{fig:AllCases}(\subref{fig:AllCases_1}), \ref{fig:AllCases}(\subref{fig:AllCases_2}) and \ref{fig:AllCases}(\subref{fig:AllCases_3}) showing responses to the full forcing, to $\widehat{\bm{f}}_1$ and to $\widehat{\bm{f}}_2$, respectively. 
We observe that the same trends as observed for the $Ma=4$ boundary layer is valid here, indicating that wall-cooling does not impact the trends discussed.   
These discussions are also equally valid for laminar compressible boundary layers, and Appendix \S\ref{sec:laminar boundary layer} briefly discusses this case. 
Another point to note is that, we have so far only discussed the most amplified first resolvent mode, and Appendix \S\ref{sec: Sub-optimal modes} contains a brief discussion on the effect of the different components of the forcing on the sub-optimal resolvent modes. 

From these observations, we see that within the majority of the wavenumber space above the $\overline{Ma}(\infty)=1$ line, the contribution of $\widehat{\bm{f}}_2$ is orders of magnitude higher than the contribution of $\widehat{\bm{f}}_1$ (given that the colour-scale is logarithmic). 
This is especially true for the most amplified resolvent modes that lie close to the $\overline{Ma}(\infty)=1$ line. 
However, this does not mean that the contribution of $\widehat{\bm{f}}_1$ is not significant for supersonic modes. 
In the next section we will discuss a comparison with DNS where $\widehat{\bm{f}}_1$ can be seen to play a dominant role in exciting the supersonic modes of the flow. 

\section{Resolvent Mach waves and Mach waves from DNS: A discussion}
\label{sec:resolvent mach waves and mach waves from dns: a discussion}

\begin{figure} 
\captionsetup[subfigure]{labelformat=empty,skip=-75pt}
\begin{subfigure}[b]{\textwidth}
\centering
\includegraphics[width=\textwidth, trim={0.25cm 7.5cm 12.7cm 0.25cm}, clip]{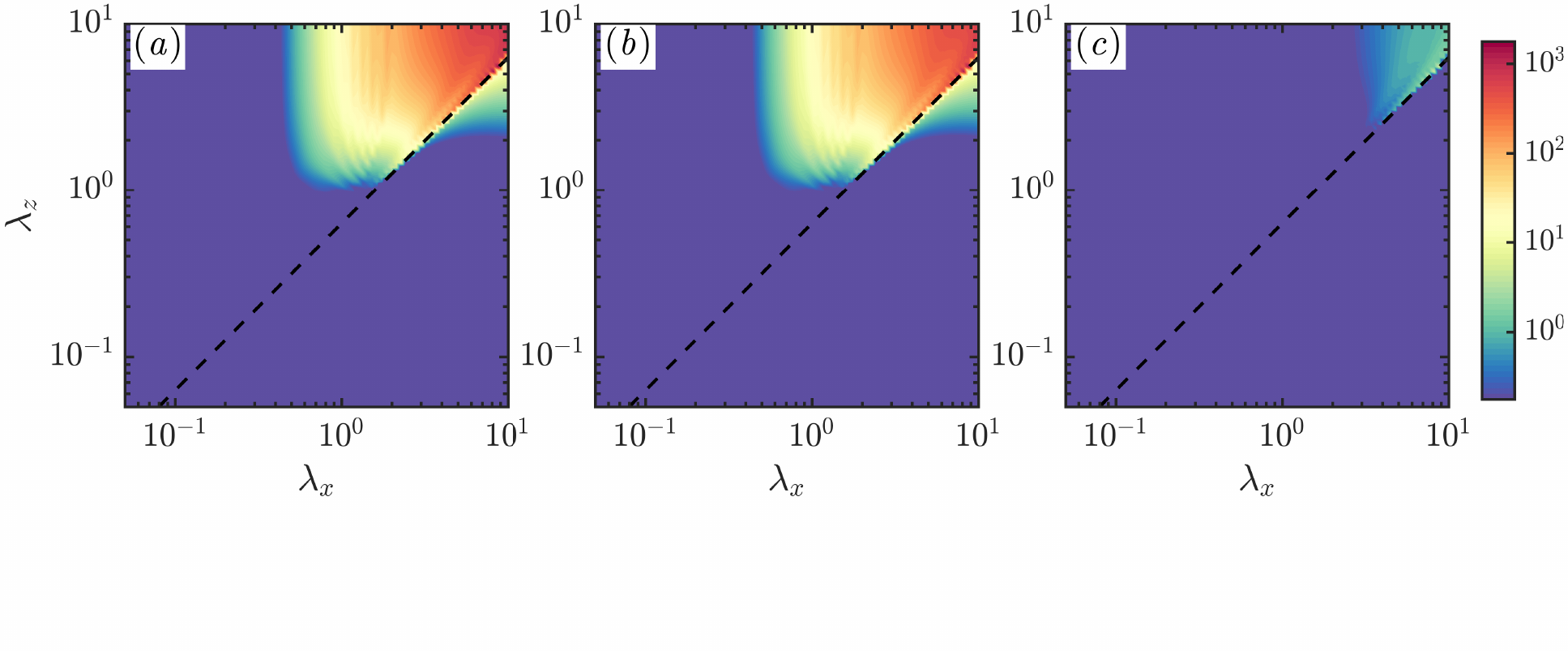}
\caption{}
\label{fig:FrcB_RspF_1}
\end{subfigure}
\begin{subfigure}[b]{\textwidth}
\caption{}
\label{fig:FrcB_RspF_2}
\end{subfigure}
\begin{subfigure}[b]{\textwidth}
\caption{}
\label{fig:FrcB_RspF_3}
\end{subfigure}
\caption{The response of the masked resolvent is shown; the response is masked to lie solely within the freestream, and the forcing to lie within $y^+<30$. 
The Chu norm of the response of the resolvent operator to (\textit{a}) the full leading resolvent forcing mode $\bm{\phi}_1$ as well as the two components of the forcing (\textit{b}) $\widehat{\bm{f}}_1$ and (\textit{c}) $\widehat{\bm{f}}_2$ are shown as a function of the streamwise ($\lambda_x$) and spanwise ($\lambda_z$) wavelengths for a fixed value of phase speed $c \approx \overline{U}(y^+=15)$. 
The $Ma=4$, $Re_\tau=400$ turbulent boundary layer over an adiabatic wall is considered. 
The dashed black line indicate the relative Mach equal to unity line. }
\label{fig:FrcB_RspF}
\end{figure}
In this section we will discuss the resolvent Mach waves alongside trends of these waves that are known from DNS \citep[e.g.][]{duan2014numerical, duan2016pressure, zhang2017effect}. 
The turbulence within the boundary layer forces Mach wave radiation. 
DNS studies show that with increasing $Ma$ the freestream inclination angles of these Mach waves decrease and their propagation velocities increase \citep{duan2016pressure}. 
On the other hand, increasing wall-cooling does not seem to impact propagation speeds of the waves, but the inclination angles are reported to be slightly steeper \citep{zhang2017effect}. 
Crucially, these studies employed acoustic analogy based arguments, and identified the sources of these waves to lie within the buffer layer of the flow. 

In order to analyse these Mach waves that are radiated by the buffer layer, we have to move away from using resolvent analysis over the entire wall-normal domain, as has been done in previous works \citep[e.g.][]{bugeat20193d, bae2020resolvent}. 
We will instead use the `masked resolvent', where the forcing and the response to the resolvent is masked, i.e.\ restricted to lie in particular wall-normal regions of the flow. 
First considering the response, Mach wave responses alone are of interest here and we have observed that these Mach waves are the sole contributors to the freestream resolvent response (see \S\ref{sec:contribution to the boundary layer and freestream}). 
Therefore the resolvent response can be masked such that it lies within the freestream alone. 
Next, considering the forcing, DNS studies have shown that the sources to these waves lie within the buffer layer of the flow \citep[e.g.][]{duan2014numerical, duan2016pressure, zhang2017effect}. 
Therefore, the resolvent forcing can be masked such that it is localised within the buffer layer. 
So we are here considering the resolvent operator such that its forcing can only lie within $y^+<30$ and the response within $y>\delta_{99}$. 
The weighting function introduced in \citet{nogueira2020resolvent} is used to mask the resolvent (see \S\ref{sec:resolvent operator}). 

Figure \ref{fig:FrcB_RspF} shows the response to such a masked resolvent, with figures \ref{fig:FrcB_RspF}(\subref{fig:FrcB_RspF_1}), \ref{fig:FrcB_RspF}(\subref{fig:FrcB_RspF_2}) and \ref{fig:FrcB_RspF}(\subref{fig:FrcB_RspF_3}) showing the response to the full forcing, to $\widehat{\bm{f}}_1$ and to $\widehat{\bm{f}}_2$, respectively. 
Here $c=\overline{U}(y^+\approx 15)$ is taken from within the buffer layer. 
Comparing figure \ref{fig:FrcB_RspF}(\subref{fig:FrcB_RspF_1}) to figure \ref{fig:fsfd}(\subref{fig:fsfd_kxky_2}), the response of the masked resolvent is concentrated in the region where we found that the indirect route of forcing is significant. 
Additionally from figures \ref{fig:FrcB_RspF}(\subref{fig:FrcB_RspF_2}) and \ref{fig:FrcB_RspF}(\subref{fig:FrcB_RspF_3}), we see that $\widehat{\bm{f}}_1$ captures almost all of the response. 
Therefore, based on the discussions in \S\ref{sec:contribution of the direct and indirect forcing mechanisms}, we can postulate that the indirect route of forcing amplifies these modes. 
Purely from a resolvent perspective, we can therefore hypothesise that the Mach wave radiation due to the turbulence within the boundary layer are generated due to the indirect route of forcing as described in \S\ref{sec:indirect route of forcing resolvent Mach wave radiation}. 

Presuming the accuracy of this hypothesis, both the indirect route of Mach wave forcing and the subsonic modes are forced by the solenoidal component of the forcing. 
This could potentially reduce the problem of modelling the full high-rank non-linear forcing in these flows to the simpler problem of modelling only the lower-rank solenoidal component of the forcing. 

\subsection{Inclination angle of the supersonic resolvent modes}
\label{sec:Inclination angle of the supersonic resolvent modes}

\begin{figure}[t]
\begin{subfigure}[b]{\textwidth}
\centering
\includegraphics[width=0.55\textwidth, trim={0.5cm 0cm 1.3cm 0cm}, clip]{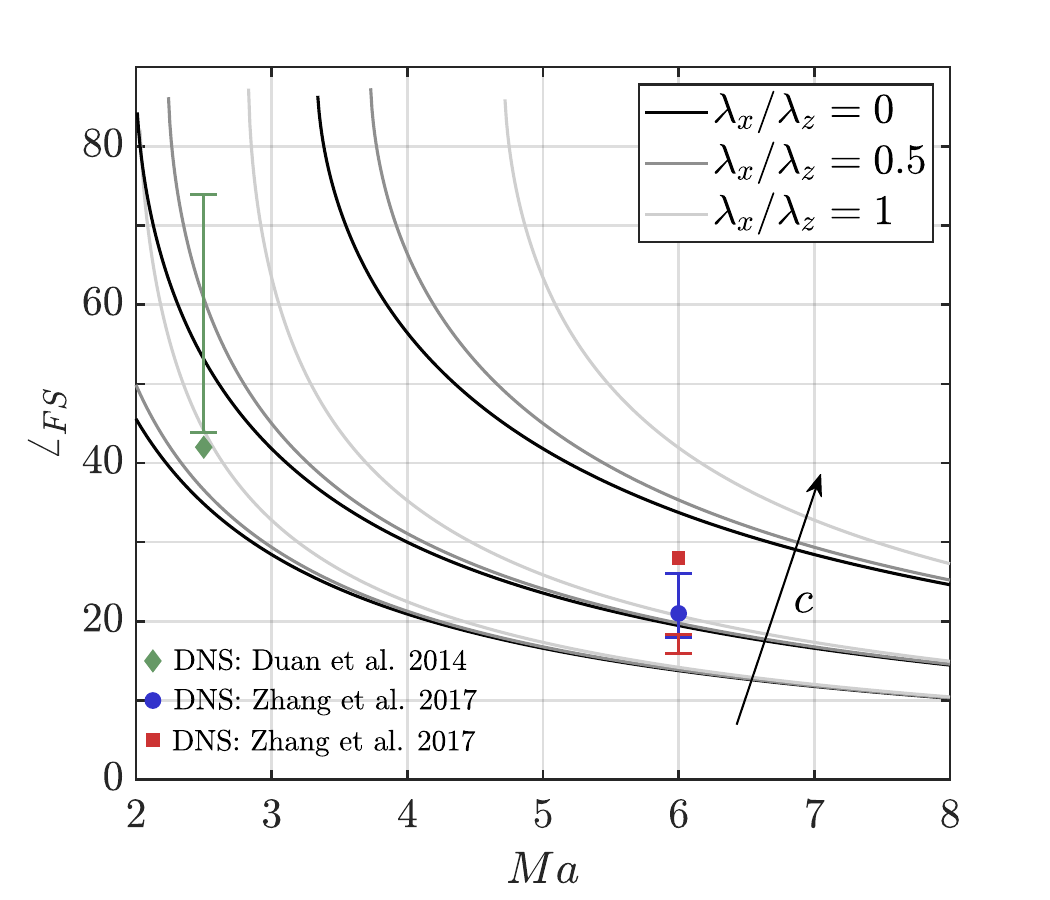}
\label{fig:inclination_angle_PS}
\end{subfigure}
\begin{subfigure}[b]{\textwidth}
\label{fig:inclination_angle_AR}
\end{subfigure}
\vspace{-0.75cm}
\caption{The inclination angle of the inviscid Mach waves given by the equation \eqref{eqn:inclination_angle} is shown with respect to Mach number for a range of aspect ratios and phase speeds. 
Three different phase-speeds of $c=0.3,0.5 \mbox{ and }0.7$ increasing in the direction of the arrow and three different values of aspect ratios $\lambda_x/\lambda_z=0,0.5 \mbox{ and }1.0$ are shown. 
The markers indicate the average inclination angles of the Mach waves from DNS reported in the literature \citep{duan2014numerical,zhang2017effect}. 
The DNS data at $Ma=6$ is for the case of cooled walls, and the two markers indicate two different ratios of $\Theta_w/\Theta_{ad}=0.76$ (blue) and $0.25$ (red). 
The intervals represent the range of inclination angles predicted by the resolvent model under two assumptions: (1) $k_z=0$ and (2) $\overline{U}(y^+ = 10) \leq c \leq \overline{U}(y^+ = 20)$. }
\label{fig:inclination_angle}
\end{figure}
DNS studies have investigated the freestream inclination angles of the Mach wave radiation generated from the boundary layer \citep[e.g.][]{duan2014numerical,zhang2017effect}. 
Although DNS datasets show some degree of variability in the freestream inclination angles observed, these studies consistently emphasise the existence of a statistically dominant angle. 
When considering the Mach waves from the resolvent model, we can use the fact that these waves are pressure oscillation of the form given by \eqref{eqn:mach_wave} and compute the freestream inclination angle of these Mach waves as:
\begin{equation}
\angle_{FS} = \tan^{-1} \left( \frac{-1j}{\sqrt{1+(\lambda_x/\lambda_z)^2-\mbox{Ma}^2 (1-c)^2}} \right)
\label{eqn:inclination_angle}
\end{equation}
At a fixed Mach number, we see that the inclination angle depends on the aspect ratio $\lambda_x/\lambda_z$ and the phase speed $c$ of the mode. 
From the resolvent, we therefore have a range of inclination angles. 
The mechanism that picks out the statistically dominant inclination angles found in DNS is still not clear. 
To probe into this mechanism, in the future there is a requirement of comparing DNS data with the resolvent model. 
More specifically, constructing a resolvent-based low-order model of the freestream radiations using DNS data from the buffer layer alone, will potentially elucidate the mechanism that is responsible for the selection of a predominant angle in DNS studies. 
This detailed comparison with DNS, however, is beyond the scope of the current manuscript, and is left as an important future direction of work.   

Although we cannot do a one-to-one comparison with DNS, we can still compare some of the trends observed in DNS to that from the resolvent model. 
For this, in figure \ref{fig:inclination_angle}, the solid lines show the inclination angles of the resolvent freestream radiation computed from \eqref{eqn:inclination_angle} as a function of Mach number for a range of aspect ratios and phase-speeds. 
The phase-speed increases in the direction of the arrow in figure \ref{fig:inclination_angle}, and for each phase-speed lighter coloured lines represent larger aspect ratios. 
First, we observe that phase-speeds have a more prominent impact on the freestream inclination angles, when compared to aspect-ratios. 
Second, from \eqref{eqn:inclination_angle}, we observe that wall-cooling does not have a direct impact on the inclination angles of the structures. 
It only has an indirect effect through a control on the values of $c$ that can be admitted for these waves (through the differences in mean profiles due to wall-cooling). 
Finally, we note that, in general the inclination angles obtained from the model decrease with increasing Mach number, and this trend is consistent with DNS \citep{duan2014numerical}. 

The markers show the values of the average freestream inclination angles that are obtained from DNS \citep{duan2014numerical, zhang2017effect} from a $Ma=2.5$ flow over an adiabatic wall and a $Ma=6$ flow with two wall-cooling ratios of $\Theta_w/\Theta_{ad}=0.76 \mbox{ and } 0.25$ (indicated by the two markers at $Ma=6$). 
The intervals plotted show the range of inclination angles predicted by the resolvent under two assumptions: (1) the phase-speed lies within the buffer layer $\overline{U}(y^+ \approx 10) \leq c \leq \overline{U}(y^+ \approx 20)$, 
and (2) the aspect-ratio is zero, since from figure \ref{fig:FrcB_RspF} the most amplified resolvent modes that can be forced from the buffer layer tend to have large $\lambda_z$ (also see \citet{bugeat20193d}). 
It could be argued that, when the effect of wall-cooling is not significant (green and blue markers), the average inclination angles obtained from DNS fall roughly within the range of inclination angles that are observed from the model. 
However, there are major discrepancies.  
Firstly, The range of inclination angles predicted for $Ma=2.5$ is too large for any meaningful comparison to be made. 
Secondly, DNS predicts an increase in the inclination angle with increasing wall-cooling, while resolvent predicts the opposite trend, i.e.\ a decrease in the inclination angle. 

These differences could be due to several reasons, and some of them are: (1) the assumption of phase-speeds and aspect-ratios that we have considered for plotting the intervals are oversimplified, or (2) the rank-1 resolvent model that we have used here is not sufficient. 
However, the most obvious restriction in comparing the inclination angles as done here is the assumption that all $(k_x,k_z,c)$ modes are forced equally. 
This cannot be true, since the dominant turbulent structures within the boundary layer will determine the extent to which these modes are forced. 
These turbulent structures can also be impacted by wall-cooling, thereby influencing the dominant angles of these radiations. 
To address this, there is again the need to construct resolvent-based low-rank models of the freestream, and this is an important future direction of work. 

\section{Mach wave radiation from a streamwise developing boundary layer}
\label{sec:mach wave radiation from a streamwise developing boundary layer}

\begin{figure}
\begin{subfigure}[b]{0.75\textwidth}
\centering
\includegraphics[width=\textwidth, trim={1cm 1.7cm 1.5cm 0.5cm}, clip]{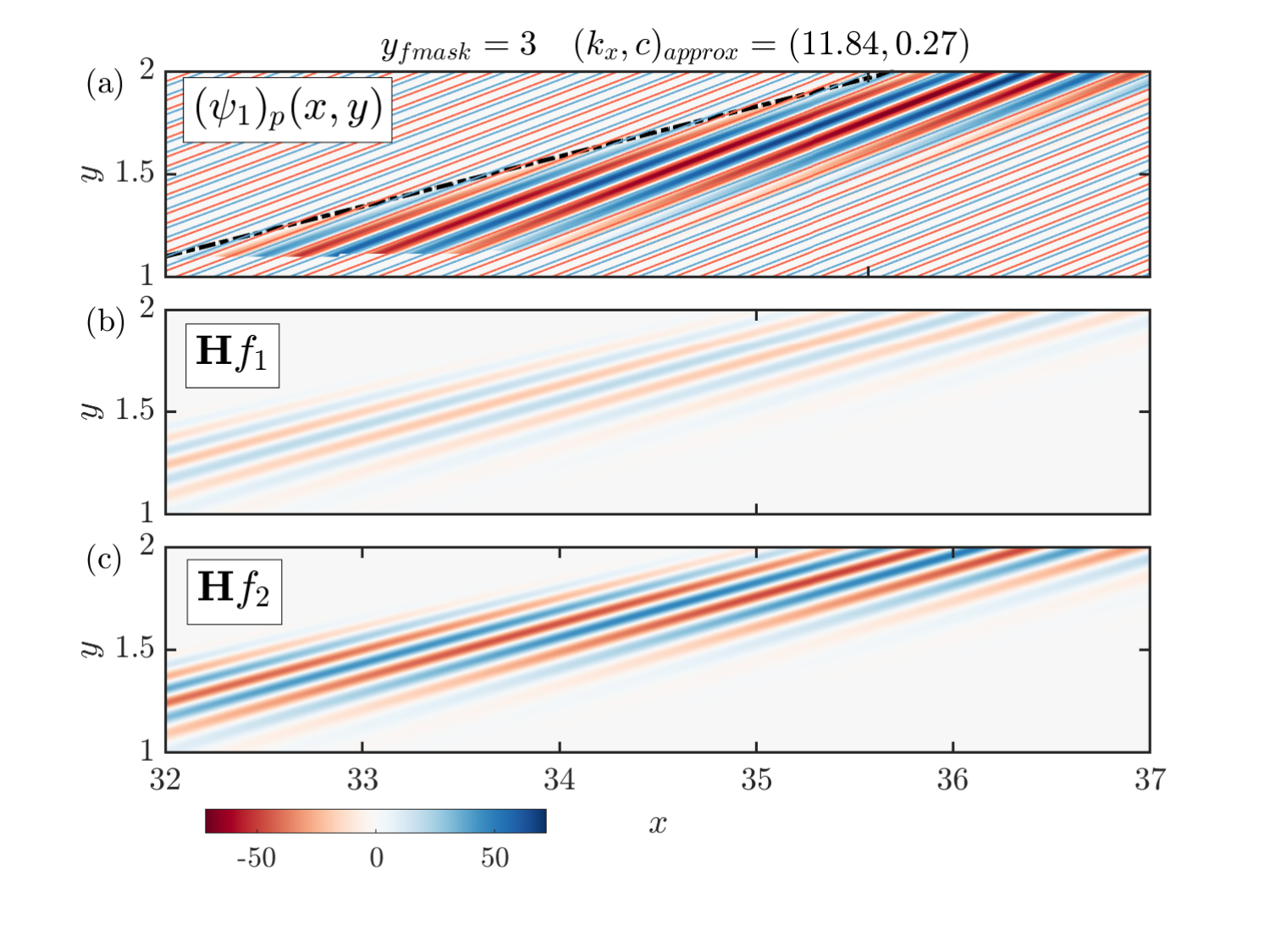}
\label{fig:ModeShape_HighMask}
\end{subfigure}
\begin{subfigure}[b]{0.75\textwidth}
\includegraphics[width=\textwidth, trim={1cm 1.7cm 1.5cm 0.5cm}, clip]{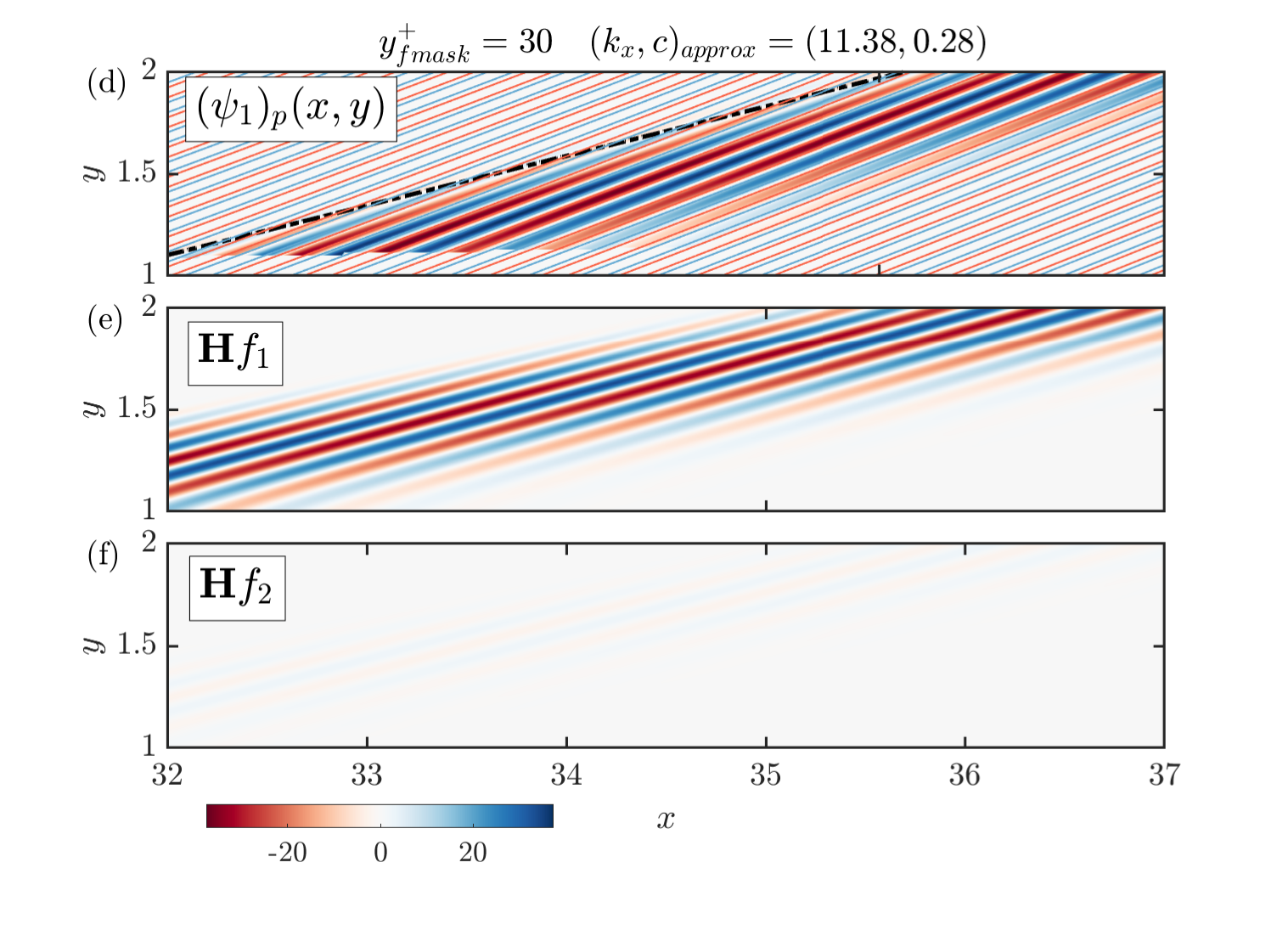}
\label{fig:ModeShape_LowMask}
\end{subfigure}
\vspace{-0.5cm}
\caption{The leading resolvent response mode obtained from the 2D resolvent analysis for spanwise wavenumber $k_z=12.62$ and temporal frequency $\omega=3.15$. 
(\textit{a,d}) The full response is shown, with the resolvent masked such that the response is solely in the freestream. 
Also shown are the responses to the forcing (\textit{b,e}) $\widehat{\bm{f}}_1$ and (\textit{c,f}) $\widehat{\bm{f}}_2$. 
The modes obtained from two different masking for the forcing are shown: (\textit{a,b,c}) the forcing lies throughout the boundary layer and part of the freestream, here till $y\leq3$ and (\textit{d,e,f}) the forcing lies within the buffer layer,  $y^+\leq30$. 
Note, for clarity, only the freestream and a subset of the streamwise domain used to compute the resolvent is shown here.  
The approximate streamwise wavenumber $k_x$ and phase-speed $c$ obtained using a Fourier transform of the mode, are also shown as the titles. 
The contour lines in (\textit{a,d}) show the analytical solution obtained by solving the inviscid pressure equations in the freestream, with red and blue contours representing positive and negative pressure fluctuations, respectively.} 
\label{fig:ModeShape_2D}
\end{figure}
The aim of this section is to briefly show that the mechanisms generating the Mach wave radiations, as discussed here, can be generalised to a streamwise developing boundary layer. 
In this case, the mean velocity is a function of the streamwise and wall-normal directions $\overline{U}(x,y)$. 
Unlike in the case of the 1D boundary layer, for the growing boundary layer the wall-normal mean velocity $\overline{V}(x,y)$ is not zero.  
Without the simplifying assumptions for the 1D boundary layer, the inviscid equations for pressure in \eqref{eqn:inviscid_pressure} will now become
\begin{equation}
\begin{split}
\omega^2p 
&- 2\omega\overline{U}_i \frac{\partial p}{\partial x_i} 
+ \overline{U}_i\overline{U}_j \frac{\partial p}{\partial x_i \partial x_j}
+ \overline{U}_i \frac{\partial \overline{U}_j }{\partial x_i}\frac{\partial p}{\partial x_j}
+ \gamma \omega \frac{\partial \overline{U}_j }{\partial x_j} p
+\gamma \overline{U}_i \frac{\partial \overline{U}_j }{\partial x_j} \frac{\partial p}{\partial x_i}  \\ 
&- \gamma \frac{\partial \overline{U}_i }{\partial x_j}\frac{\partial \overline{U}_j }{\partial x_i} p 
- \frac{1}{M^2} \frac{\partial \overline{\Theta} }{\partial x_j} \frac{\partial p}{\partial x_j}
- \frac{\overline{\Theta}}{M^2} \frac{\partial^2 p}{\partial x_j^2}
= 
\gamma \left[ 2\frac{\partial \overline{U}_i}{\partial x_j} \frac{\partial }{\partial x_i} 
+\frac{\partial^2 \overline{U}_i }{\partial x_i\partial x_j} \right] u_j  \\
& + \left[ \omega  +\overline{U}_i \frac{\partial }{\partial x_i}\right]
\left( \overline{\Theta}f_\rho+f_\theta \right)
- \gamma \frac{\partial f_{u_i}}{\partial x_i}.
\end{split}
\label{eqn:inviscid_pressure_2D}
\end{equation}
Here $(x_1,x_2,x_3)=(x,y,z)$, $(u_1,u_2,u_3)=(u,v,w)$, $(\overline{U}_1,\overline{U}_2,\overline{U}_3)=(\overline{U},\overline{V},0)$ and $(f_{u_1},f_{u_2},f_{u_3})=(f_x,f_y,f_z)$. 
A Fourier transform is taken in the homogeneous spanwise direction $z$. 
Within the freestream, the LHS of \eqref{eqn:inviscid_pressure_2D} can be solved using separation of variables to obtain solutions of the form \citep{mack1990inviscid}
\begin{equation}
\begin{split}
p(x,y)=A\exp(r_xx)\exp(r_yx). 
\end{split}
\label{eqn:2D_pressure_wave}
\end{equation}
Here $A$ is a constant at $(k_z,\omega)$, and $r_x$ and $r_y$ can be obtained by solving the ordinary differential equations obtained from the separation of variables. 
These wavenumbers follow one among these two sets of quadratic equations: (i) $c_1 r_x^2 - c_3r_x - l_1 = 0$ and $c_2 r_y^2 + (c_5 r_x - c_4) r_y + (l_1 + c_6) = 0$ or (ii) $c_2 r_y^2 - c_4r_y - l_2 = 0$ and $c_1 r_x^2 + (c_5 r_y - c_3) r_x + (l_2 + c_6) = 0$. 
The constants depend on the mean freestream quantities alone as: 
$c_1 = (1 - 1/M^2)$, 
$c_2 = (\overline{V}_\infty^2 - 1/M^2)$, 
$c_3 = 2\omega$, 
$c_4 = 2\omega\overline{V}_\infty$, 
$c_5 = 2\overline{V}_\infty$ and 
$c_6 = (\omega^2 - k_z^2/M^2 )$, 
where $\overline{V}_\infty$ represents the freestream wall-normal velocity normalised by $U_\infty$. 
By varying the constant $l_1$ (or $l_2$) we can get a family of solutions at a range of $(r_x,r_y)$. 
Therefore, for the streamwise developing boundary layer, we can obtain wave-like solutions within the freestream, consistent with the observations for the 1D case in the rest of the manuscript \citep{mack1990inviscid}.

Now considering the RHS of \eqref{eqn:inviscid_pressure_2D}, and comparing it with the 1D boundary layer \eqref{eqn:inviscid_pressure}, we can obtain similar observations. 
There are two routes of forcing. 
First, the dilatational part of the forcing to the momentum equations $\widehat{\bm{f}}_d$ (obtained through $\partial f_{u_i}/\partial x_i$) along with the forcing to the continuity $f_\rho$ and energy $f_\theta$ equations, can directly force the pressure waves. 
This gives the direct route of forcing as in \S\ref{sec:direct route of forcing resolvent Mach wave radiation}. 
Second, the solenoidal component of the forcing can generate a response in velocity $u_j$, which in turn can force the pressure equations within the boundary layer. 
(Although $u_j$ cannot be assumed to be zero in the freestream, the terms within the square brackets that multiply with $u_j$ can be assumed to be zero in this region.)  
This gives us the indirect route of forcing as in \S\ref{sec:indirect route of forcing resolvent Mach wave radiation}. 
Consistent with prior observations from the 1D case, the indirect route of forcing involves the mean shear profiles, and can therefore be expected to be dominant in the buffer layer where this mean shear is high.

As done in \S\ref{sec:Mach wave radiation from the resolvent operator}, to confirm if these conclusions derived from the inviscid solutions do indeed capture features of the resolvent at a finite Reynolds number, figure \ref{fig:ModeShape_2D} shows a 2D resolvent mode. 
To obtain the 2D resolvent mode, the linearized discrete Navier-Stokes equations contained in the matrix $\bm{A}$ \eqref{eqn:state_space} are now constructed using the 2D mean profiles $\overline{U}(x,y)$ and $\overline{V}(x,y)$ obtained from \citet{duan2016pressure}.  
The derivatives are now $(\partial/\partial x,\partial/\partial y,\partial/\partial z)=(\partial/\partial x,\partial/\partial y, i k_z)$. 
The boundary conditions in the freestream are the same as for the 1D resolvent (see \ref{sec:Numerical set up for the resolvent operator}). 
The inlet and outlet boundary conditions are the non-reflecting Navier-Stokes characteristic boundary conditions \citep{POINSOT1992104} with damping sponges to prevent any reflections \citep{freund1997sponges}. 
For the 2D resolvent, at a particular value of $(k_z,\omega)$, we obtain a range of modes at different $k_x$, that includes both subsonic and supersonic modes. 
To obtain just the Mach wave radiation that is of interest here, we will therefore consider the resolvent with the response masked such that it lies only in the freestream. 

Figures \ref{fig:ModeShape_2D}(a) and \ref{fig:ModeShape_2D}(d) show the pressure obtained from the 2D resolvent mode as the filled contours. 
Note, for clarity, only the freestream and a subset of the streamwise domain used to compute the resolvent is shown in the figure. 
The line contours represent the solution to the inviscid freestream pressure equation \eqref{eqn:2D_pressure_wave}.  
The constant for the quadratic equation $l_1$ (or $l_2$) is fixed such that $r_x$ (or $r_y$) matches the streamwise (or wall-normal) wavenumber of the resolvent mode in the freestream, which is obtained from a Fourier transform. 
Here red and blue represent positive and negative fluctuations, respectively. 
A $Ma=5.86$, $T_w/T_{ad}\approx0.76, T_\infty = 55 K$ boundary layer is considered here. 
A temporal frequency $\omega=3.15$ is chosen, since, from DNS, we know that this frequency is energetic in the freestream \citet{duan2016pressure}, and a spanwise wavenumber of $k_z=12.62$ is chosen. 
The dashed-dot line represents the Mach angle computed as $\sin^{-1}{(1/Ma)}$. 
Unlike in the 1D resolvent, due to the finite nature of the domain, the forcing is restricted to a finite streamwise extent, and therefore the radiation from the resulting response modes will be restricted by the Mach angle (note, as seen in the figure, the inclination angle of the mode can be different from this Mach angle). 
The forcing and response are not masked in the streamwise direction. 
Two options for masking the forcing in the wall-normal direction are used: (i) in figures \ref{fig:ModeShape_2D}(a,b,c) the forcing is allowed to lie anywhere within the boundary layer or part of the freestream, i.e.\ at $y\leq3$ and (ii) in figures \ref{fig:ModeShape_2D}(d,e,f) the forcing lies exclusively within the buffer layer of the flow, i.e.\ $y^+\leq 30$. 
The response to the forcing $\widehat{\bm{f}}_1$ is shown in figures \ref{fig:ModeShape_2D}(b) and \ref{fig:ModeShape_2D}(e), and to $\widehat{\bm{f}}_2$ in figures \ref{fig:ModeShape_2D}(c) and \ref{fig:ModeShape_2D}(f).

First, from figures \ref{fig:ModeShape_2D}(a) and \ref{fig:ModeShape_2D}(d), we see that the inviscid equations do match the resolvent modes reasonably well. 
This shows that, like we observed for the 1D resolvent, the 2D resolvent also admits wave-like solutions in the freestream, which can be analytically modelled as solutions to the inviscid freestream equations. 
Second, comparing figures \ref{fig:ModeShape_2D}(b) and \ref{fig:ModeShape_2D}(c), we see that, when the forcing is allowed reach to $y=3$, $\widehat{\bm{f}}_2$ contributes dominantly to the mode. 
In other words, in this case, through the direct route, the dilatational component of the forcing to the momentum equations and the forcing to the energy and continuity equations force the mode. 
However, from figures \ref{fig:ModeShape_2D}(e) and \ref{fig:ModeShape_2D}(f), we see that this observation is no longer true when the forcing is within the buffer layer alone. 
In this case, through the indirect route, the solenoidal component of the forcing to the momentum equations $\widehat{\bm{f}}_1$ dominates the response.

Therefore, similar to the 1D resolvent, there are two routes through which the Mach waves in the 2D resolvent can be forced: (i) the direct and (ii) the indirect route. 
The indirect route likely becomes more important when considering modes forced by the buffer layer of the flow. 
This is consistent with the results obtained from the 1D resolvent, and therefore shows that, in this case, the conclusions drawn here from the 1D resolvent analysis are valid for the trends from the 2D resolvent. 
Also consistent with the 1D resolvent, by varying $k_z$, we can obtain modes (at a range of $k_x$) with different ratios of the solenoidal to the dilatational responses (not shown here for brevity). 
It would be interesting, in the future, to understand how the pressure radiations obtained in \eqref{eqn:2D_pressure_wave} may relate to mechanisms where the spatial modulation of wavepackets leads to radiation \citep[e.g.][]{jordan2013wave, reba2010wave}. 
A detailed analysis of the 2D resolvent is therefore important and is a crucial future direction of work.


\section{Conclusions}
\label{sec:Conclusions}

We identified the forcing mechanisms that separately amplify the subsonic and the supersonic modes in a linearized Navier-Stokes equations-based model of compressible boundary layer flows.  
The resolvent analysis framework was used for this purpose, where the non-linear terms of the linearized momentum ($\widehat{\bm{f}}_{\bm{u}}$), continuity ($\widehat{f}_\rho$) and temperature ($\widehat{f}_\theta$) equations were treated as a forcing to the linearized equations. 
To identify the separate forcing mechanisms that are active, we use a Helmholtz decomposition of $\widehat{\bm{f}}_{\bm{u}}$ that gives two components: (i) the divergence-free solenoidal component $\widehat{\bm{f}}_{\bm{u}}^s$ and (ii) the curl-free dilatational component $\widehat{\bm{f}}_{\bm{u}}^d$. 

First considering the subsonic modes, these are structures localised within the boundary layer and are similar to structures in the incompressible flow (figures \ref{fig:CompVSIncomp} and figure \ref{fig:BLFS}). 
The velocity fluctuations of these structures are divergence-free (figure \ref{fig:usud}). 
Only the solenoidal component of the forcing to the momentum equations $\widehat{\bm{f}}_{\bm{u}}^s$ can amplify these modes (figure \ref{fig:fsfd}). 
All the other components of forcing, that includes the dilatational component $\widehat{\bm{f}}_{\bm{u}}^d$, as well as $\widehat{f}_\rho$ and $\widehat{f}_\theta$, play a negligible role in amplifying the subsonic modes.  
This is consistent with what has been previously observed in the incompressible regime \citep[e.g.][]{rosenberg2019efficient}. 
It is interesting that similar trends are here obtained for compressible flows as well.

Now considering the supersonic modes, i.e.\ the resolvent Mach waves, these modes are pressure fluctuations that radiate into the freestream of the boundary layer (figures \ref{fig:supersonic_mode}). 
These modes contribute to the energy in the freestream of the flow (figures \ref{fig:BLFS}). 
The most amplified resolvent Mach waves start radiating from the edge of the boundary layer, where the dampening effect of viscosity is negligible. 
Importantly, within the freestream, these resolvent modes closely follow the trends of the inviscid Mach waves \citep{mack1984boundary}. 
We identified two routes through which these modes can be amplified: (i) the `direct route' where $\widehat{\bm{f}}_{\bm{u}}^d$, $\widehat{f}_\rho$ and $\widehat{f}_\theta$ directly force these modes and (ii) the `indirect route' where $\widehat{\bm{f}}_{\bm{u}}^s$ forces a response in wall-normal velocity $\widehat{v}$ through incompressible-like mechanisms and this $\widehat{v}$ in-turn forces the supersonic mode. 
The direct route of forcing appears dominant for a majority of the supersonic modes considered. 
However, the indirect route plays the dominant role for Mach waves that are forced by the buffer layer of the flow. 
This distinction is crucial since DNS studies have found that the buffer layer of compressible boundary layers contains the sources that generate the Mach waves found in the freestream \citep[e.g.][]{duan2014numerical}. 
We also show that these observations are more generally true for a streamwise developing boundary layer as well.

These observations, as well as preliminary discussions regarding the freestream inclination angles of these resolvent modes, show that, in the future, constructing resolvent-based low-rank models of freestream disturbances in the real flow is crucial for further understanding and modelling of these structures.  
Additionally, considering the role of the solenoidal component of the forcing in exciting the subsonic modes and the supersonic modes from the buffer layer, it would be interesting to see how much of the flow can be modelled by considering this solenoidal component of the forcing alone, instead of the much higher-rank full forcing. 


\mbox{}

\noindent
\textbf{Acknowledgements}: We acknowledge support from the Air Force Office of Scientific Research grant FA9550-20-1-0173. 
We would also like to thank Prof.\ Lian Duan for providing us the 2D mean profiles used in \S\ref{sec:mach wave radiation from a streamwise developing boundary layer} and Prof.\ Anthony Leonard for helpful discussions regarding this work.
We are also grateful to the anonymous referees for their valuable suggestions and questions regarding this work. 

\mbox{}

\noindent
\textbf{Declaration of interests}: The authors report no conflicts of interest.

\appendix

\section{Suboptimal modes}
\label{sec: Sub-optimal modes}

\begin{figure}[t!]
\captionsetup[subfigure]{labelformat=empty,skip=-27pt}
\begin{subfigure}[b]{\textwidth}
\centering
\includegraphics[width=\textwidth, trim={1.8cm 5.6cm 1.7cm 5.3cm}, clip]{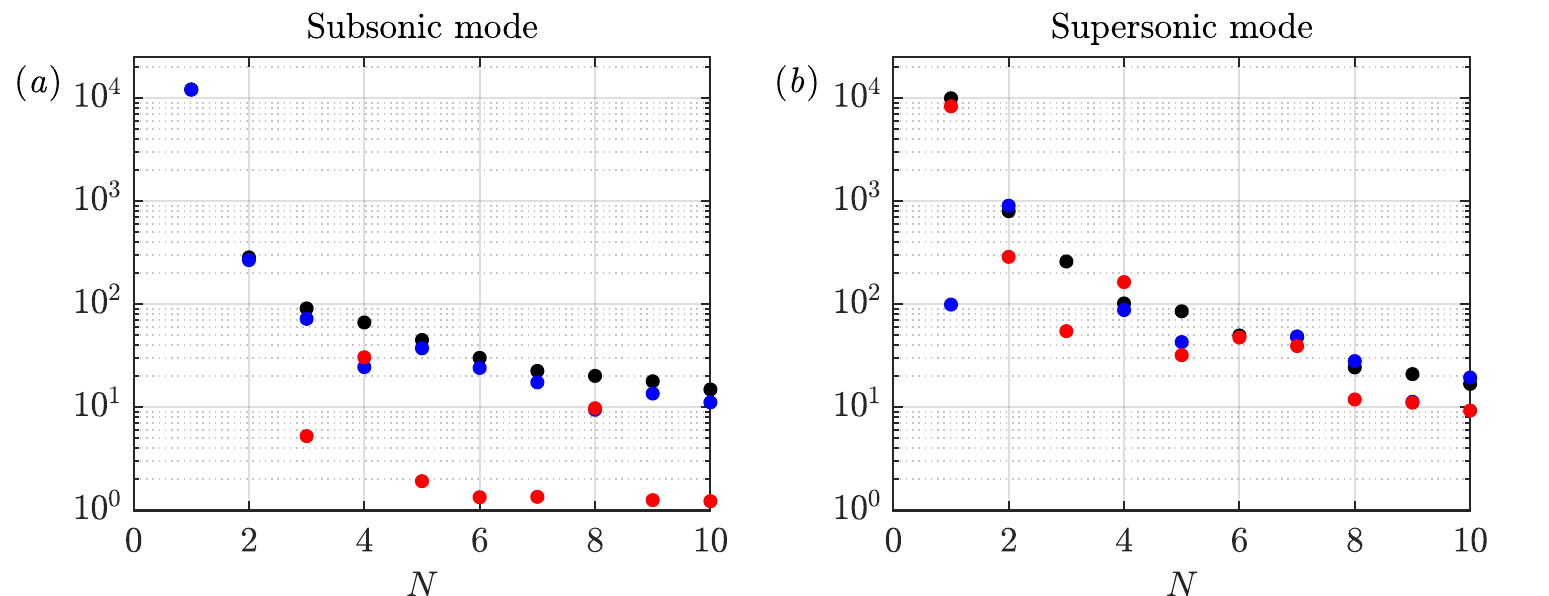}
\caption{}
\label{fig:Suboptimals_subsonic}
\end{subfigure}
\begin{subfigure}[b]{\textwidth}
\caption{}
\label{fig:Suboptimals_supersonic}
\end{subfigure}
\caption{The Chu norm of the response of the resolvent operator to the full resolvent forcing mode $\bm{\phi}_i$ (black) as well as the two components of the forcing $\widehat{\bm{f}}_1$ (blue) and $\widehat{\bm{f}}_2$ (red) are shown for the first 10 resolvent modes. 
(\textit{a}) The subsonic mode considered in figure \ref{fig:fHelmholtz_mode_subsonic} and (\textit{b}) the supersonic mode considered in figure \ref{fig:fHelmholtz_mode_supersonic} are shown. }
\label{fig:Suboptimals}
\end{figure}

So far we have only considered the leading resolvent mode. 
For the sake of completeness, in this section we will briefly consider the effect that the different forcing components have on the suboptimal modes while noting that for a more complete understanding of if, and how many, suboptimal modes should be considered, we will require resolvent-based reconstructions of DNS data. 
In figure \ref{fig:Suboptimals} the first 10 resolvent modes, i.e.\ those corresponding to $\sigma_1 - \sigma_{10}$ \eqref{eqn:svd} are shown. Two $(k_x,k_z)$ pairs are considered: (i) the subsonic mode indicated by ({\scriptsize $\blacklozenge$}) in figure \ref{fig:Intro}(\subref{fig:energy_comp}) is shown in figure \ref{fig:Suboptimals}(\subref{fig:Suboptimals_subsonic}) and (ii) the supersonic mode MW1 indicated by ({\tiny $\blacksquare$}) in figure \ref{fig:Intro}(\subref{fig:energy_comp}) is shown in figure \ref{fig:Suboptimals}(\subref{fig:Suboptimals_supersonic}). 
The Chu norm of the response to the full resolvent forcing is depicted in black, while the response to $\widehat{\bm{f}}_1$ is shown in blue and to $\widehat{\bm{f}}_2$ is shown in red.
(The energy of the full response is equal to the sum of the energies of the responses to $\widehat{\bm{f}}_1$ and $\widehat{\bm{f}}_2$ and twice the cross correlation between these responses. 
While the energies of the responses to $\widehat{\bm{f}}_1$ and $\widehat{\bm{f}}_2$ are positive, their cross-correlation can be negative. 
This is why, for certain modes such as for the $N=4$ mode in figure \ref{fig:Suboptimals}(\subref{fig:Suboptimals_supersonic}), energy of the response to $\widehat{\bm{f}}_1$ or $\widehat{\bm{f}}_2$ is higher than the full response in black). 

Considering the response of the subsonic mode in figure \ref{fig:Suboptimals}(\subref{fig:Suboptimals_subsonic}), we find that the first three modes are captured by the solenoidal component of the forcing (the red dotes are not visible for $N=1,2$ since they fall below range of $y-$axis shown in the figure). 
When considering the supersonic mode, we find that the second resolvent mode is captured by the solenoidal component of the forcing. 
Things brings us back to the observation that MW1 exists in a region of the wavenumber space where both incompressible-like and purely compressible mechanisms coexist, and also that incompressible-like mechanisms can exite compressible modes. 
The compressible mechanisms forced by the direct-route are most active for the mode considered here, and therefore appears as the first resolvent mode. 
For the current work, we leave the discussion of the suboptimal modes at this point, while noting that this topic requires further investigation, especially when constructing resolvent-based low-rank models of the freestream.   

\section{Grid convergence}
\label{sec:Grid convergence}

\begin{figure}[t] 
\captionsetup[subfigure]{labelformat=empty,skip=-25pt}
\begin{subfigure}[b]{\textwidth}
\centering
\includegraphics[width=0.75\textwidth, trim={1.0cm 0cm 4.5cm 0cm}, clip]{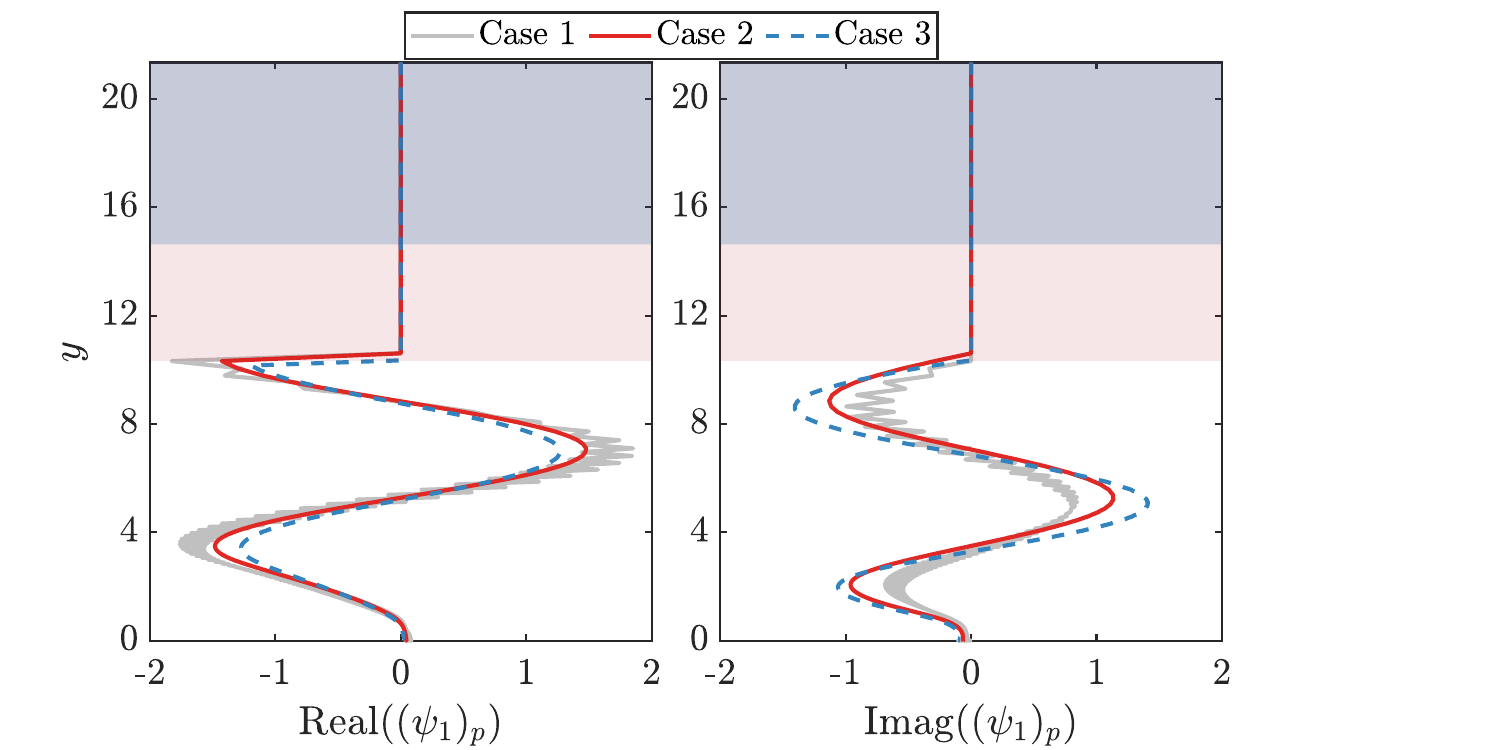}
\caption{}
\label{fig:CompMode_GridConvergence_1}
\end{subfigure}
\begin{subfigure}[b]{\textwidth}
\caption{}
\label{fig:CompMode_GridConvergence_2}
\end{subfigure}
\caption{The real and imaginary parts of the pressure from the leading resolvent response for a supersonic mode. 
The mode corresponds to $\lambda_x=5$, $\lambda_z=3.5$ and $c=\overline{U}(y^+\approx 15)$ for a compressible boundary layer over an adiabatic wall with $Ma=4$ and $Re_\tau= 400$ (the mode indicated by the ({\tiny $\blacksquare$}) in figure \ref{fig:Intro}(\textit{b})). 
The modes obtained using three different grids are shown: (i) Case 1: the grid used in this work but without a damping layer, (ii) Case 2: the grid used in this work with $N=401$, $y_{\mbox{\footnotesize{max}}}=4\delta$ and $y_{\mbox{\footnotesize{max}}}=3l$ in the subsonic and supersonic regions, respectively and (iii) Case 3: with $N=601$, $y_{\mbox{\footnotesize{max}}}=5\delta$ and $y_{\mbox{\footnotesize{max}}}=4l$ in the subsonic and supersonic regions, respectively}
\label{fig:CompMode_GridConvergence}
\end{figure}

To discretize the equations in \eqref{eqn:NS_comp} in the wall-normal direction, we use a summation-by-parts finite difference scheme with $N=401$ grid points. 
A grid stretching method is employed to properly resolve the wall-normal direction \citep{mattsson2004summation, kamal2020application} (see \S\ref{sec:Numerical set up for the resolvent operator}). 
This grid stretching introduces spurious numerical oscillations in the modes obtained. 
Following \citet{appelo2009high}, a damping-layer along with an artificial viscosity is used to remedy these spurious oscillations. 
The role of this damping layer is to slow down the waves within it and to implement it we use the damping layer and artificial viscosity defined in \citet{appelo2009high} (for the damping layer equation 4 from \citet{appelo2009high} with $p=4$, $q=4$ and $\epsilon_L=10^{-4}$ is used, and for the artificial viscosity equation 9 from the reference with $K=[0]$ and $\gamma_k=0.15$ is used). It is ensured that at least $20$ grid-points are included within the damping layer. 

\begin{figure}[t] 
\captionsetup[subfigure]{labelformat=empty,skip=-40pt}
\begin{subfigure}[b]{\textwidth}
\centering
\includegraphics[width=\textwidth, trim={1.7cm 1.1cm 4.5cm 1.75cm}, clip]{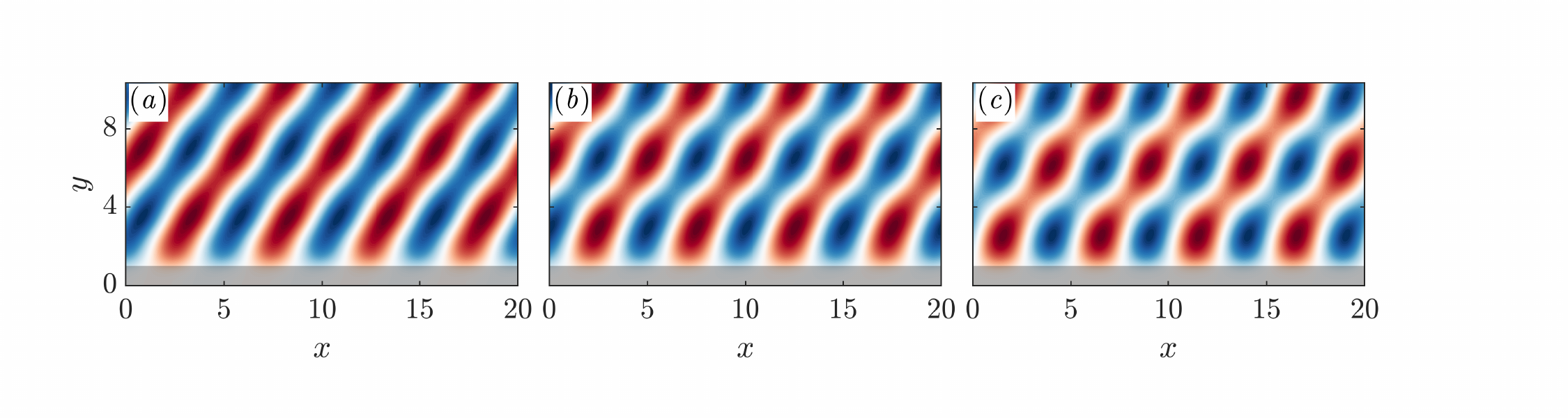}
\caption{}
\label{fig:CompMode_DampingStrength_1}
\end{subfigure}
\begin{subfigure}[b]{\textwidth}
\caption{}
\label{fig:CompMode_DampingStrength_2}
\end{subfigure}
\begin{subfigure}[b]{\textwidth}
\caption{}
\label{fig:CompMode_DampingStrength_3}
\end{subfigure}
\caption{The pressure fluctuations obtained from the leading resolvent response for a supersonic mode is shown with the red and blue contours representing positive and negative pressure fluctuations, respectively. 
The mode corresponds to $\lambda_x=5$, $\lambda_z=3.5$ and $c=\overline{U}(y^+\approx 15)$ for a compressible boundary layer over an adiabatic wall with $Ma=4$ and $Re_\tau= 400$ (the mode indicated by the ({\tiny $\blacksquare$}) in figure \ref{fig:Intro}(\textit{b})). 
The strength of the damping layer used increases from ($\textit{a}$) $0.15$ to ($\textit{b}$) $0.5$ and then to ($\textit{c}$) $1.5$.}
\label{fig:CompMode_DampingStrength}
\end{figure}

\begin{figure} 
\captionsetup[subfigure]{labelformat=empty,skip=-75pt}
\begin{subfigure}[b]{\textwidth}
\centering
\includegraphics[width=\textwidth, trim={0.25cm 1.1cm 12.7cm 0.25cm}, clip]{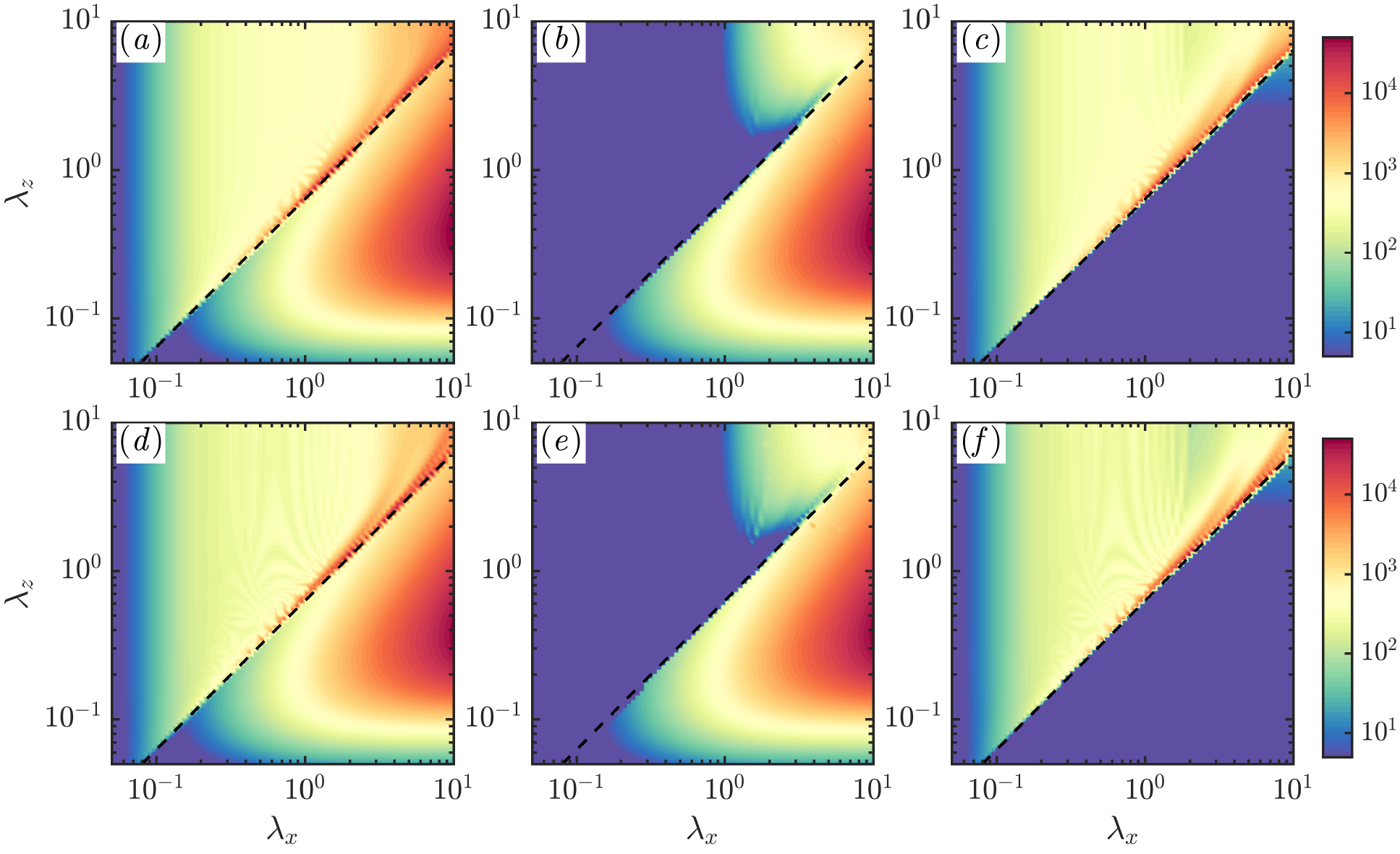}
\caption{}
\label{fig:fsfd_GridConv_1}
\end{subfigure}
\begin{subfigure}[b]{\textwidth}
\caption{}
\label{fig:fsfd_GridConv_2}
\end{subfigure}
\begin{subfigure}[b]{\textwidth}
\caption{}
\label{fig:fsfd_GridConv_3}
\end{subfigure}
\begin{subfigure}[b]{\textwidth}
\caption{}
\label{fig:fsfd_GridConv_4}
\end{subfigure}
\begin{subfigure}[b]{\textwidth}
\caption{}{}
\label{fig:fsfd_GridConv_5}
\end{subfigure}
\begin{subfigure}[b]{\textwidth}
\caption{}
\label{fig:fsfd_GridConv_6}
\end{subfigure}
\caption{The Chu norm of the response of the resolvent operator to (\textit{a,d}) the full leading resolvent forcing mode $\bm{\phi}_1$ as well as the two components of the forcing (\textit{b,e}) $\widehat{\bm{f}}_1$ and (\textit{c,f}) $\widehat{\bm{f}}_2$ are shown as a function of the streamwise ($\lambda_x$) and spanwise ($\lambda_z$) wavelengths for a fixed value of phase speed $c=\overline{U}(y^+\approx 15)$. 
The $Ma=4$, $Re_\tau=400$ turbulent boundary layer over an adiabatic wall (that was also considered in figure \ref{fig:fsfd}) is shown here. 
(\textit{a}-\textit{c}) The responses obtained using $N=601$ grid points and a $y_{\mbox{\footnotesize{max}}} = 5\delta$ in the subsonic region and $y_{\mbox{\footnotesize{max}}} = 4l$ in the supersonic region are shown.  
(\textit{d}-\textit{f}) Also shown are the responses obtained with a grid where the strength of the damping layer is increased to $\gamma_k=1.5$ (from the $\gamma_k=0.15$ used for figure \ref{fig:fsfd}). 
These responses are compared to those shown in figures \ref{fig:fsfd}(\subref{fig:fsfd_kxky_1}-\subref{fig:fsfd_kxky_3}) to illustrate the insensitivity of the obtained results to the grid used. 
The dashed lines in all figures indicate the relative Mach equal to unity line. }
\label{fig:fsfd_GridConv}
\end{figure}

It is important to consider whether the grid stretching and the introduction of the damping layer impact the conclusions drawn in the current work. 
To investigate this, in figure \ref{fig:CompMode_GridConvergence} the real and imaginary parts of the pressure obtained from the leading resolvent response for the supersonic mode $\lambda_x=5$, $\lambda_y=3.5$ and $c\approx U(y^+=15)$ are shown. 
Three grids are considered in figure \ref{fig:CompMode_GridConvergence}: (i) Case 1 (grey solid line): the same grid as that used in this work, but without a damping layer, (ii) Case 2 (red solid line): the grid used in this work with $N=401$ and $y_{\mbox{\footnotesize{max}}}=4\delta$ for the subsonic modes and $y_{\mbox{\footnotesize{max}}}=3l$ for the supersonic modes (where $l$ is the wavelength of the Mach waves) and (iii) Case 3 (blue dashed line): with $N=601$ and $y_{\mbox{\footnotesize{max}}}=5\delta$ for the subsonic modes and $y_{\mbox{\footnotesize{max}}}=4l$ for the supersonic modes. 
The red shaded regions towards the freestream indicate the extent of the damping layer for Case 2, and the blue shaded region indicates the damping layer for Case 3 (since the number of points in the damping layer is fixed to be $20$, the wall-normal extent of the damping layer changes with $N$). 
To keep the comparison consistent across the three cases, all the responses are set to zero within the red-shaded region and beyond into the freestream. 
Considering the grey lines from Case 1, where no damping layer or artificial viscosity is included, we observe sawtooth oscillations that arise due to the grid stretching. 
From the modes obtained with damping layers, we note that this damping removes the sawtooth oscillations. 
Comparing the responses from Case 2 and Case 3, we notice that there are small differences between the modes. 
This is probably expected given that the extent of the damping layer is different for the two cases. 
Importantly, these differences do not impact the conclusions drawn in the current work. 
To see this, in figures \ref{fig:fsfd_GridConv}(\subref{fig:fsfd_GridConv_1}-\subref{fig:fsfd_GridConv_3}) we reproduce figures \ref{fig:fsfd}(\subref{fig:fsfd_kxky_1}-\subref{fig:fsfd_kxky_3}) using more number of grid points and a larger $y_{\mbox{\footnotesize{max}}}$. 
In other words, the data plotted in figures \ref{fig:fsfd_GridConv}(\subref{fig:fsfd_GridConv_1}-\subref{fig:fsfd_GridConv_3}) are computed using the grid Case 3, while the data in figures \ref{fig:fsfd}(\subref{fig:fsfd_kxky_1}-\subref{fig:fsfd_kxky_3}) were computed using the grid Case 2. 
We see that there are no significant differences that arise due to an increase in the number of grid points as well as an increase in the maximum extent of the wall-normal grid. 

To investigate the effect of the damping layer a little further, in figure \ref{fig:CompMode_DampingStrength} the real part of the pressure obtained using the grid Case 2 is shown, with the strength of the artificial viscosity increasing moving from left to right. 
We notice that there is a beating in the pressure fluctuations that is made worse as the strength of this artificial viscosity increases. 
This is likely due to reflections introduced by the artificial viscosity. This shows that we should exercise caution while introducing artificial viscosity (removing the artificial viscosity is not a valid option since it will re-introduce the saw-tooth oscillations). 
However, the results in the current work are not affected by the strength of the artificial viscosity. 
To illustrate this, in figures \ref{fig:fsfd_GridConv}(\subref{fig:fsfd_GridConv_4}-\subref{fig:fsfd_GridConv_6}) we reproduce figures \ref{fig:fsfd}(\subref{fig:fsfd_kxky_1}-\subref{fig:fsfd_kxky_3}) with a grid that uses a stronger artificial viscosity (the strength is determined by the value of $\gamma_k$ which increases from $0.15$ in figures \ref{fig:fsfd}(\subref{fig:fsfd_kxky_1}-\subref{fig:fsfd_kxky_3}) to $1.5$ in figures \ref{fig:fsfd_GridConv}(\subref{fig:fsfd_GridConv_4}-\subref{fig:fsfd_GridConv_6})). 
There are no significant differences that are introduced by this change in the strength of the damping layer. 

The main conclusions drawn in the current work are therefore not significantly impacted by changes in the grid used. 

\section{Laminar boundary layer}
\label{sec:laminar boundary layer}

\begin{figure} 
\captionsetup[subfigure]{labelformat=empty,skip=-75pt}
\begin{subfigure}[b]{\textwidth}
\centering
\includegraphics[width=\textwidth, trim={0.25cm 7.5cm 12.7cm 0.25cm}, clip]{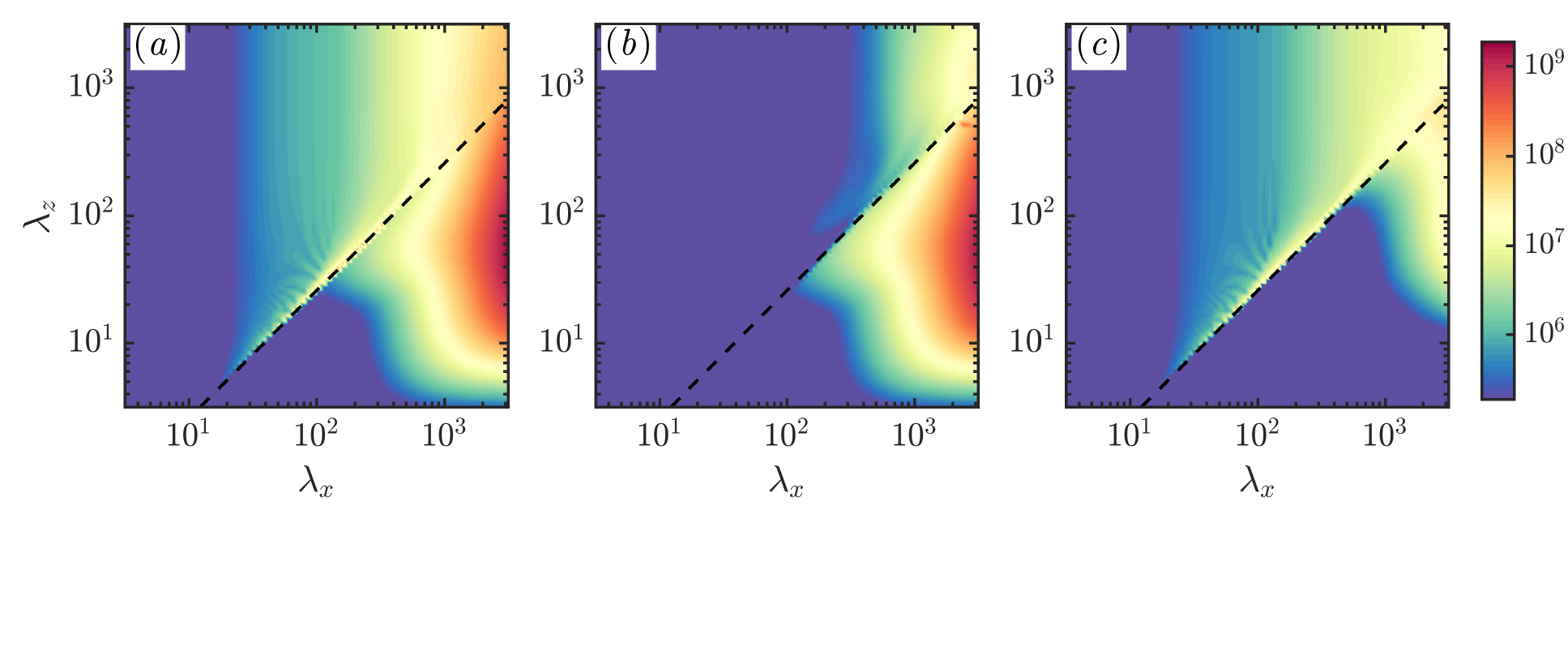}
\caption{}
\label{fig:laminar_1}
\end{subfigure}
\begin{subfigure}[b]{\textwidth}
\caption{}
\label{fig:laminar_2}
\end{subfigure}
\begin{subfigure}[b]{\textwidth}
\caption{}
\label{fig:laminar_3}
\end{subfigure}
\caption{The Chu norm of the response of the resolvent operator to (\textit{a}) the full leading resolvent forcing mode $\bm{\phi}_1$ as well as the two components of the forcing (\textit{b}) $\widehat{\bm{f}}_1$ and (\textit{c}) $\widehat{\bm{f}}_2$ are shown as a function of the streamwise ($\lambda_x$) and spanwise ($\lambda_z$) wavelengths for a fixed value of phase speed $c=0.5$. 
The $Ma=8$, $Re=1000$ laminar boundary layer over an adiabatic wall is considered. 
This figure is equivalent to figure \ref{fig:fsfd} for the case of a laminar flow. 
The dashed lines indicate the relative Mach equal to unity line. }
\label{fig:laminar}
\end{figure}
We will also briefly consider the case of a laminar compressible boundary layer and show that the trends discussed here are more generally applicable to compressible laminar boundary layers as well. 
The Reynolds number for the laminar case is defined as $Re=U_\infty l/\nu_\infty$, where the reference length scale is defined as $l=(x\nu_\infty/U_\infty)^{1/2}$. 
The mean streamwise velocity $\overline{U}$ and temperature $\overline{T}$ profiles for the compressible laminar boundary layers that are required as input to the model are obtained from the Mangler-Levy-Lees similarity profiles as in \citet{malik1990numerical}. 
In figures \ref{fig:laminar}(\subref{fig:laminar_1}-\subref{fig:laminar_3}) we reproduce figures \ref{fig:fsfd}(\subref{fig:fsfd_kxky_1}-\subref{fig:fsfd_kxky_3}) for the case of a $Ma=8$ and $Re=1000$ laminar compressible boundary layer. 
We see that, similar to the case of the turbulent flow, $\widehat{\bm{f}}_1$ and $\widehat{\bm{f}}_2$ approximately capture the subsonic and supersonic regions, separately.  

\bibliographystyle{jfm}
\bibliography{ref}

\end{document}